\documentclass[]{aastex63}
\usepackage{chemformula}
\let\ce\ch
\usepackage[caption=false]{subfig}
\usepackage{hyperref}
\usepackage{xcolor}
\definecolor{medium-blue}{rgb}{0,0,1}
\hypersetup{colorlinks, urlcolor={medium-blue}}

\received{December 18 2020}
\revised{April 5 2021}
\accepted{April 14 2021}
\published{June 8 2021}
\submitjournal{Astronomical Journal}
\shorttitle{Hybrid Gas-Dust Outbursts from 67P/Churyumov-Gerasimenko}
\shortauthors{Noonan et al.}
\begin{document}
\title{Analysis of Hybrid Gas-Dust Outbursts Observed at 67P/Churyumov-Gerasimenko}
\author[0000-0003-2152-6987]{John W. Noonan}
\affiliation{Department of Space Studies, Southwest Research Institute, Suite 300, 1050 Walnut Street, Boulder, Colorado 80302,USA}
\affiliation{Lunar and Planetary Laboratory,University of Arizona, 1629 E University Blvd, Tucson, Arizona 85721-0092, USA }

\author[0000-0002-2968-0455]{Giovanna Rinaldi}
\affiliation{Istituto di Astrofisica e Planetologia Spaziali, Istituto Nazionale di Astrofisica, via del Fosso del Cavaliere 100 00133 Rome, Italy}

\author[0000-0002-9318-259X]{Paul D. Feldman}
\affiliation{Department of Physics and Astronomy, The Johns Hopkins University, 3400 N. Charles Street, Baltimore, Maryland 21218, USA }

\author[0000-0001-5018-7537]{S. Alan Stern}
\affiliation{Department of Space Studies, Southwest Research Institute, Suite 300, 1050 Walnut Street, Boulder, Colorado 80302,USA}
 
\author[0000-0002-3672-0603]{Joel Wm. Parker}
\affiliation{Department of Space Studies, Southwest Research Institute, Suite 300, 1050 Walnut Street, Boulder, Colorado 80302,USA}

\author[0000-0003-0797-5313]{Brian A. Keeney}
\affiliation{Department of Space Studies, Southwest Research Institute, Suite 300, 1050 Walnut Street, Boulder, Colorado 80302,USA}

\author{Dominique Bockel\'ee-Morvan}
\affiliation{LESIA, Observatoire de Paris, Universit\'e PSL, CNRS, Sorbonne Universit\'e, Université de Paris, 5 place Jules Janssen, 92195 Meudon, France}

\author[0000-0002-8227-9564]{Ronald J. Vervack Jr.}
\affiliation{Johns Hopkins University Applied Physics Laboratory, 11100 Johns Hopkins Road, Laurel, Maryland 20723-6099,USA}

\author[0000-0002-5358-392X]{Andrew J. Steffl}
\affiliation{Department of Space Studies, Southwest Research Institute, Suite 300, 1050 Walnut Street, Boulder, Colorado 80302,USA}

\author[0000-0003-2781-6897]{Matthew M. Knight}
\affiliation{Astronomy Department, University of Maryland, College Park, Maryland 20742, USA}
\affiliation{Department of Physics, United States Naval Academy, 572C Holloway Rd, Annapolis MD 21402, USA}

\author{Rebecca N. Schindhelm}
\affiliation{Department of Space Studies, Southwest Research Institute, Suite 300, 1050 Walnut Street, Boulder, Colorado 80302,USA}
\affiliation{Ball Aerospace and Technology Corp., 1600 Commerce Street, Boulder, CO 80301, USA}

\author[0000-0002-4230-6759]{Lori M. Feaga}
\affiliation{Astronomy Department, University of Maryland, College Park, Maryland 20742, USA}

\author{Jon Pineau}
\affiliation{Stellar Solutions, Inc., Palo Alto, California 94306}

\author{Richard Medina}
\affiliation{Department of Space Operations, Southwest Research Institute, Suite 300, 1050 Walnut Street, Boulder, Colorado 80302,USA}

\author[0000-0003-0951-7762]{Harold A. Weaver}
\affiliation{Johns Hopkins University Applied Physics Laboratory, 11100 Johns Hopkins Road, Laurel, Maryland 20723-6099,USA}

\author{Jean-Loup Bertaux}
\affiliation{LATMOS, CNRS/UVSQ/IPSL, 11 Boulevard d'Alembert, 78280 Guyancourt, France}

\author{Michael F. A'Hearn}
\altaffiliation{Deceased}
\affiliation{Astronomy Department, University of Maryland, College Park, Maryland 20742, USA}

\correspondingauthor{John Noonan}
\email{noonan@lpl.arizona.edu}

% Abstract of the paper
\begin{abstract}
Cometary outbursts offer a valuable window into the composition of comet nuclei with their forceful ejection of dust and volatiles in explosive events, revealing the interior components of the comet. Understanding how different types of outbursts influence the dust properties and volatile abundances to better interpret what signatures can be attributed to primordial composition and what features are the result of processing is an important task best undertaken with a multi-instrument approach. The European Space Agency \textit{Rosetta} mission to 67P/Churyumov-Gerasimenko carried a suite of instruments capable of carrying out this task in the near-nucleus coma with unprecedented spatial and spectral resolution. In this work we discuss two outbursts that occurred November 7 2015 and were observed by three instruments on board: the Alice ultraviolet spectrograph, the Visual Infrared and Thermal Imaging Spectrometer (VIRTIS), and the Optical, Spectroscopic, and Infrared Remote Imaging System (OSIRIS). Together the observations show that mixed gas and dust outbursts can have different spectral signatures representative of their initiating mechanisms, with the first outburst showing indicators of a cliff collapse origin and the second more representative of fresh volatiles being exposed via a deepening fracture. This analysis opens up the possibility of remote spectral classification of cometary outbursts with future work. 

\end{abstract}

\section{Introduction}\label{Intro}
The European Space Agency's \textit{Rosetta} spacecraft escorted the comet 67P/Churyumov-Gerasimenko from   August 2014 until September 2016. For a broad range of comet and heliocentric distances the spacecraft observed changes to the comet's coma, nucleus, and plasma environment. One particularly frequent form of these changes comes from cometary outbursts. Characterizing outbursts and their impact on the near-nucleus coma is critical to understanding the relationship between outburst traits, chemical composition, and level of dissociative electron impact. So far both gas and dust outbursts have been identified by the Alice instrument, with several outbursts even having overlapping traits \citep{feldman2016nature,Steffl2015,Steffl2018}. This dichotomy is difficult to disentangle with any single instrument's dataset; a multi-instrument approach is required to make meaningful progress in proper outburst characterization \citep{Grun2016,pajola2017pristine,agarwal2017evidence}. The initiation of outbursts, gas or dust, may leave unique clues in the coma signature that could be traced with a multi-instrument technique. Reviewing the library of \textit{Rosetta} data, identifying outbursts in the data, and correlating different instrument datasets represents the next crucial step in understanding the chaotic nature and source of cometary outbursts.

Alice observations of outbursts have revealed a range of compositions and emission processes within these periods of increased activity. \ce{H2O}, \ce{CO2}, \ce{CO}, and \ce{O2} were all indirectly observed within outbursts via emission from the daughter products H,C, and O \citep{feldman2016nature}. The supervolatile species \ce{O2} is thought to be the outburst initiator based on the outburst model put forward by \cite{skorov2016model}, which modeled the similarly volatile CO, due to the abundance of \ce{O2} in outbursts \citep{feldman2016nature}. The increased emission strength of the semi-forbidden \ion{O}{1}] 1356 \AA\ feature during outbursts with respect to other atomic emissions also indicates changes to the dissociative electron impact emission environment, but whether that increase is caused by an increase in the neutral density, electron density, electron energy, or a combination of the three remains to be determined \citep{feldman2016nature}. This correlation has been among the most intriguing results from \textit{Rosetta}; the discovery of the prevalence of dissocative electron impact emission, the result of collisions between energetic electrons and neutral molecules like \ce{H2O}, \ce{CO2}, and \ce{O2}, both at large heliocentric distances when it was correlated with solar wind interaction \citep{feldman2015measurements,bodewits2016changes,galand2020far} and nearer perihelion when it was more sporadic and linked to transient events \citep{feldman2016nature,noonan2018ultraviolet}. This emission mechanism is tied to the near-nucleus coma, typically within 10 kilometers of the nucleus, and had remained undetected in any comet prior to the \textit{Rosetta} mission.  

During the perihelion passage, several outbursts were observed with the Optical, Spectrocopic and Infrared Remote Imaging System (OSIRIS) and Visible InfraRed Thermal Imaging Spectrometer (VIRTIS) onboard Rosetta \citep{Coradini2007}, in both VIRTIS-H and VIRTIS-M channels. The outbursts have very different morphologies, with narrow and collimated plumes (August 10, September 13) and broad blobs (September 14) \citep{vincent2016summer,lin2017investigating,Rinaldi2018,bockelee2017comet}. The outbursts have been characterized by sudden increases in dust scattered solar light  over a period of 5-30 minutes, without a corresponding enhancement of CO$_2$ or H$_2$O vibrational bands characteristic of the comet dust activity. This rapid onset is correlated with a change of the visible and infrared dust color from red to less red implying the presence of very small grains ($\sim$ 100 nm) in the outburst material. The sudden increase is also correlated with a large increase of the color temperature (from 300 K to up to 630 K) and large bolometric albedos ($\sim$ 0.7) indicate bright grains in the ejecta, which could either be silicatic grains, implying the thermal degradation of the carbonaceous material, or icy grains. The 3 $\mu$m absorption band from water ice is not detected in the spectra, whereas signatures of organic compounds near 3.4 $\mu$m are observed in emission. However, for the same outburst, Alice has observed a strong absorption feature around 170 nm, characteristic of water ice \citep{Steffl2015,Steffl2018}. Because the UV wavelengths are more sensitive to the small particles with respect to the IR wavelengths, the presence of the absorption feature around 170 nm and the absence at 3 $\mu$m would be consistent with the presence of very small ice particles, less than 100 nm.   

At the moment cometary outbursts are well-known but poorly understood phenomena. The aim of this work is to take advantage of the capabilities of three instruments to analyze the dust and gas coma behaviour during these transient events in the post-perihelion period when the comet was at a heliocentric distance of 1.61 au. The comparison allows us to infer possible time evolution properties of the gas and dust activity. This paper reviews data taken between 12:00 UTC and 19:18 UTC on 2015 November 7 and is intended to serve as a companion paper to an additional multi-instrument analysis undertaken by \citet{Noonan2021spatial}, which describes data taken between 21:26 UTC November 7 and 10:30 UTC November 8, 2015.

In Section \ref{Obs}, we describe the ALICE and VIRTIS-M instruments and datasets. In Sections \ref{results:ALICE}, \ref{results:VIRTIS} , and \ref{results:Other} the results from each instrument, including OSIRIS and NAVCAM, are discussed individually. Finally, in Section \ref{Discussion} we combine the three separate analyzes to show the inverse relationship in these outbursts between gas and dust production, the implication for different outburst mechanisms, and provide an argument for further investigation of the Alice data for small outbursts. A brief summary of our findings is outlined in Section \ref{Summary}.

%\begin{table*}
%\begin{center}
%\begin{tabular}{c c c c c c}
%\hline
% Number & Time Range & Comet & Sub-Spacecraft & Sub-Spacecraft \\
% of Observations & (UTC) &  Distance (km) & Latitude ($^{\circ}$) & Longitude ($^{\circ}$)  \\
%\hline
% 38 & 12:00-19:18 & 237.0 - 228.8 & -3.20 -- -0.99 & 103.6 -- -109.0 \\
%	\end{tabular}
%\caption{ Alice observations analyzed in this work. All times are on 2015 November 7. }	
%\label{table:observations}
%\end{center}    
%\end{table*}

%\begin{table*}
%\begin{center}
%\footnotesize\centering
%	\caption{VIRTIS-M observation analyzed in this work}
%	\label{tab:observations}%[htbp]
%    \begin{adjustbox}{width=\textwidth}
    
\begin{table*}
%\begin{sidewaystable}
	\centering   %\rotatebox{90}
	\caption{VIRTIS-M observations contemporary with Alice observations}
	\label{tab:VM_observations}%[htbp]
  %  \begin{adjustbox}{max width=1.\textwidth}
	\begin{tabular}{cccccccccccc}
        \hline \hline
        \\
ID&	 VIS File name & $S_{Cube}$  & R & $t_{exp}$ & $t_{start}$  & $\Delta$t & $d_{S/C}$ & $\Phi$  & Subsolar  & Subsolar  & $r_{h}$  \\
&	&	  &   &  & & &  &  & lat & long  &  \\
 &	 &  & [m$/$px] & [sec] &[UTC] & [sec]& [km] & ($\circ$) & ($\circ$) & ($\circ$)  & [au] \\
		\hline
        \\
 %       & \multicolumn{10}{c}{July 18$^{th}$ 2015 } \\
 %       \hline
  1&  	V1\_00405529400 &		256 126 432 & 45.07 & 16 & 15:04:40 & 2592 & 230 & 62.77&-16.62 &50.93		&	1.61\\
 2 A&  	V1\_00405532100&		256 126 432 & 45.07 & 16 & 15:49:41 & 2592 & 230 & 62.77 &-14.05&	28.41	&	1.61\\
 3& V1\_00405534800 &		256 126 432 & 45.07 & 16 & 16:34:40 & 2592 & 229 & 62.75 &-12.56&11.78		&	1.61\\
 4 B& 	V1\_00405537500&		256 126 432 & 45.07  & 16 & 17:19:41 & 2592& 229 & 62.67 &	-12.21 &357.43	&	1.61\\
 5& V1\_00405540200 &		256 126 432 & 45.07 & 16 & 18:04:40 & 2624 & 228 & 62.55 &	-12.95&342.61
	&	1.61\\
 6& V1\_00405542900&		256 107 432 & 45.07 & 16 & 18:49:40 & 2176  & 228 & 62.42 &	-14.68	&325.87&	1.61\\
		\hline
	\end{tabular}
%    \end{adjustbox}
  	\flushleft  
     {\bf{Note}}:
{\it{Column 1}}: Assigned letter for each image cube containing an outburst. 
{\it{Column 2}}: Observation file name.                                   
{\it{Column 3}}: Cube size in number of samples, number of scan lines and spectral bands (432 for each channel).                              
{\it{Column 4}}: Pixel size at the distance of the observation. 
{\it{Column 5}}: Exposure time for each line. 
{\it{Column 6}}: Start time of the image cube (UTC).                          
{\it{Column 7}}: Total duration time for the image cube from acquisition start to stop.                                                                       
{\it{Column 8}}: Distance of spacecraft from the comet center. {\it{Column 9}}: Observation phase angle.                                       {\it{Column 10}}: Subsolar latitude.                                         
{\it{Column 11}}: Subsolar longitude.                                          
{\it{Column 12}}: Heliocentric distance.                                  
%\end{sidewaystable}
\label{tab:observations}
\end{table*}

\section{Observations}\label{Obs}
The outbursts in this paper were identified while examining the atomic emission light curves from the Alice instrument for the companion paper. In Alice data outbursts are best characterized as sharp increases to atomic emissions relative to prior observations. To correctly identify an outburst items such as geometry changes must be ruled out. 

\subsection{Geometry and Spacecraft Pointing}\label{geometry}
On 2015 November 7, \textit{Rosetta} and 67P were moving away from the Sun after perihelion in August 2015 and had a heliocentric distance of 1.61 au. At 12:00 UTC on November 7, when the first observations used in this analysis were taken, \textit{Rosetta} was 237 km from the nucleus of 67P. Between 12:00 UTC and 19:18 UTC, when the final observations were taken prior to an observing gap, the spacecraft had decreased its comet-centric distance to 228 km. Over that same time period the phase angle decreased from 64$^{\circ}$ to 61.5$^{\circ}$. 

\subsection{Alice Instrument Description}\label{Instrument:Alice}
The Alice instrument was a low-power and light-weight imaging spectrograph on-board the \textit{Rosetta} spacecraft with the goals of constraining the comet's UV surface properties, identifying atomic emissions from the coma, and understanding nucleus-coma interactions. The Alice instrument had a bandpass of 700 \AA\--2050 \AA\ with a spectral resolution characterized in flight to be 11 \AA\ at the center of the ``dog bone" shaped slit, which had a narrow central section and wide top and bottom sections. The upper and lower portions were both $0.10^{\circ}$ (210 $\mu$m) wide, while the narrow center was $0.05^{\circ}$ (100 $\mu$m) wide in the middle $2.0^{\circ}$ of the slit. The lower and middle sections were 2.0$^{\circ}$ long, while the upper section was 1.53$^{\circ}$. In total the slit was $5.53^{\circ}$ long. The detector in the Alice instrument was a microchannel plate with 32 rows in the spatial dimension and 1024 columns in the spectral dimension. Only rows 5 through 23 (zero indexed) of the 32 spatial rows were exposed to incoming light from the aperture. Each detector row subtended $0.30^{\circ}$ on the sky. One notable detector effect that was present in the data is identified as the odd/even effect, where the Alice detector tended to push counts to odd rows over even rows \citep{feldman2011rosetta,chaufray2017rosetta}. The full details of the instrument are available in \citet{stern2007alice}.

\begin{figure}
\centering
\gridline{\rotatefig{-90}{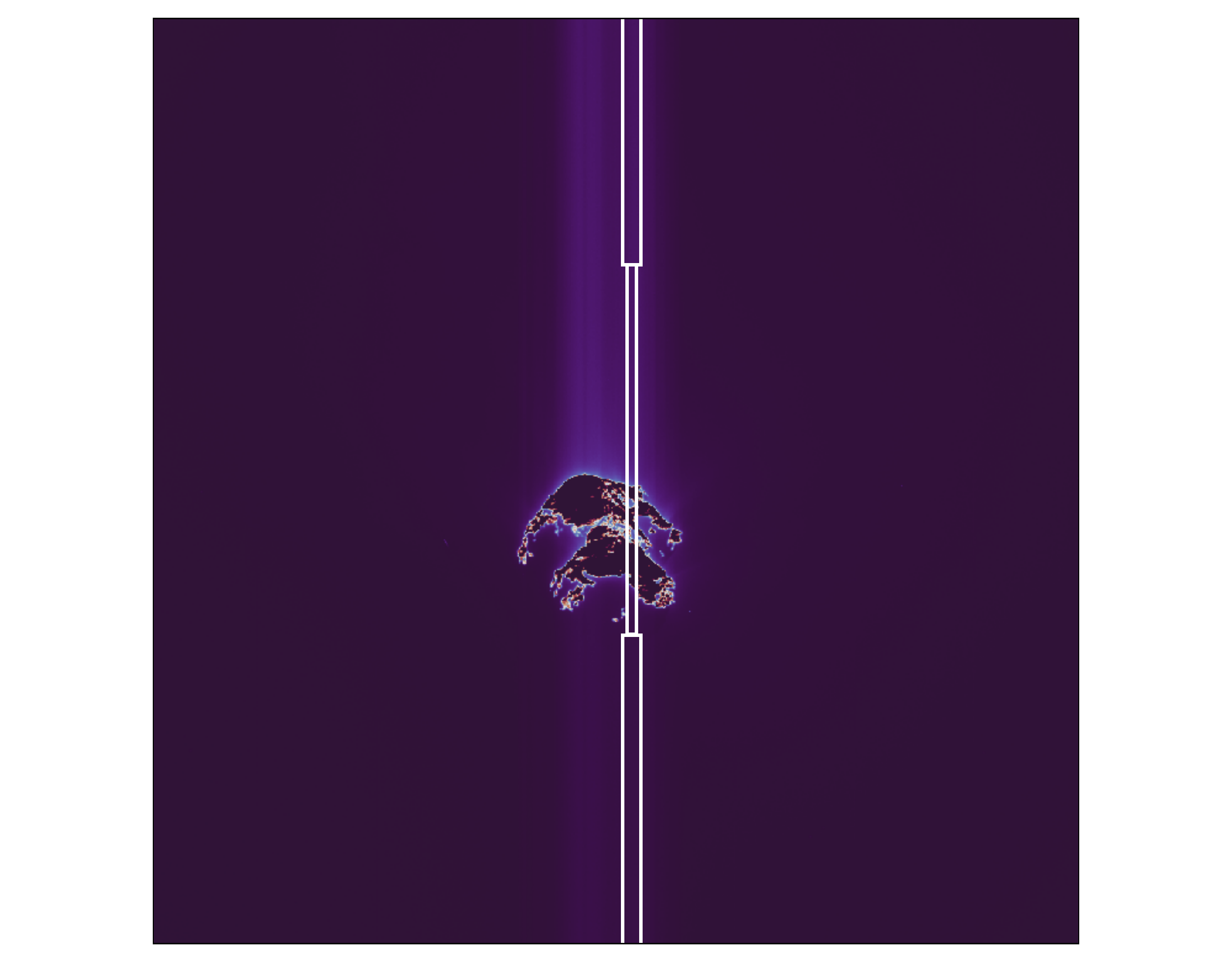}{0.33\linewidth}{a) 15:11 UTC}
\rotatefig{-90}{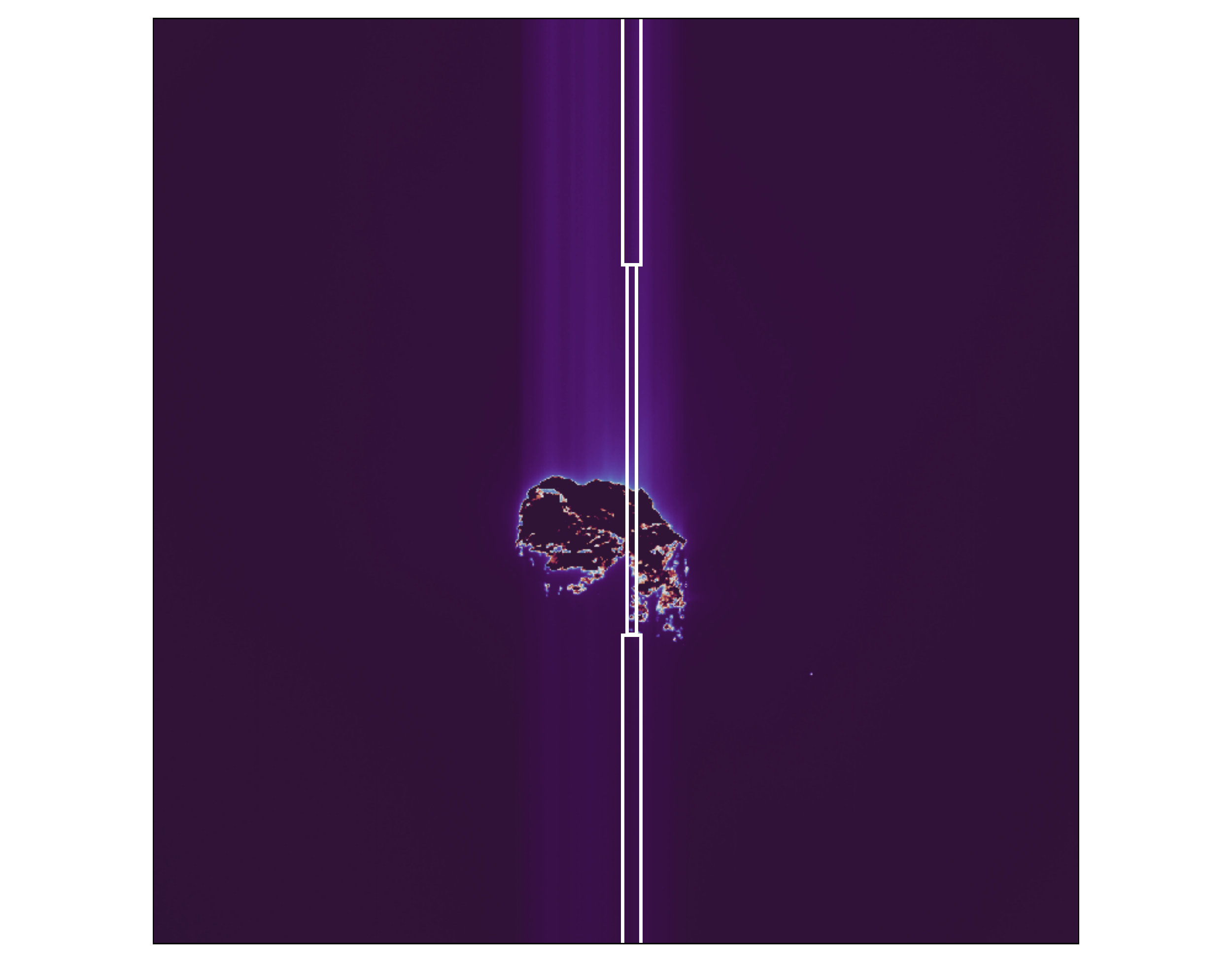}{0.33\linewidth}{b) 17:15 UTC}
\rotatefig{-90}{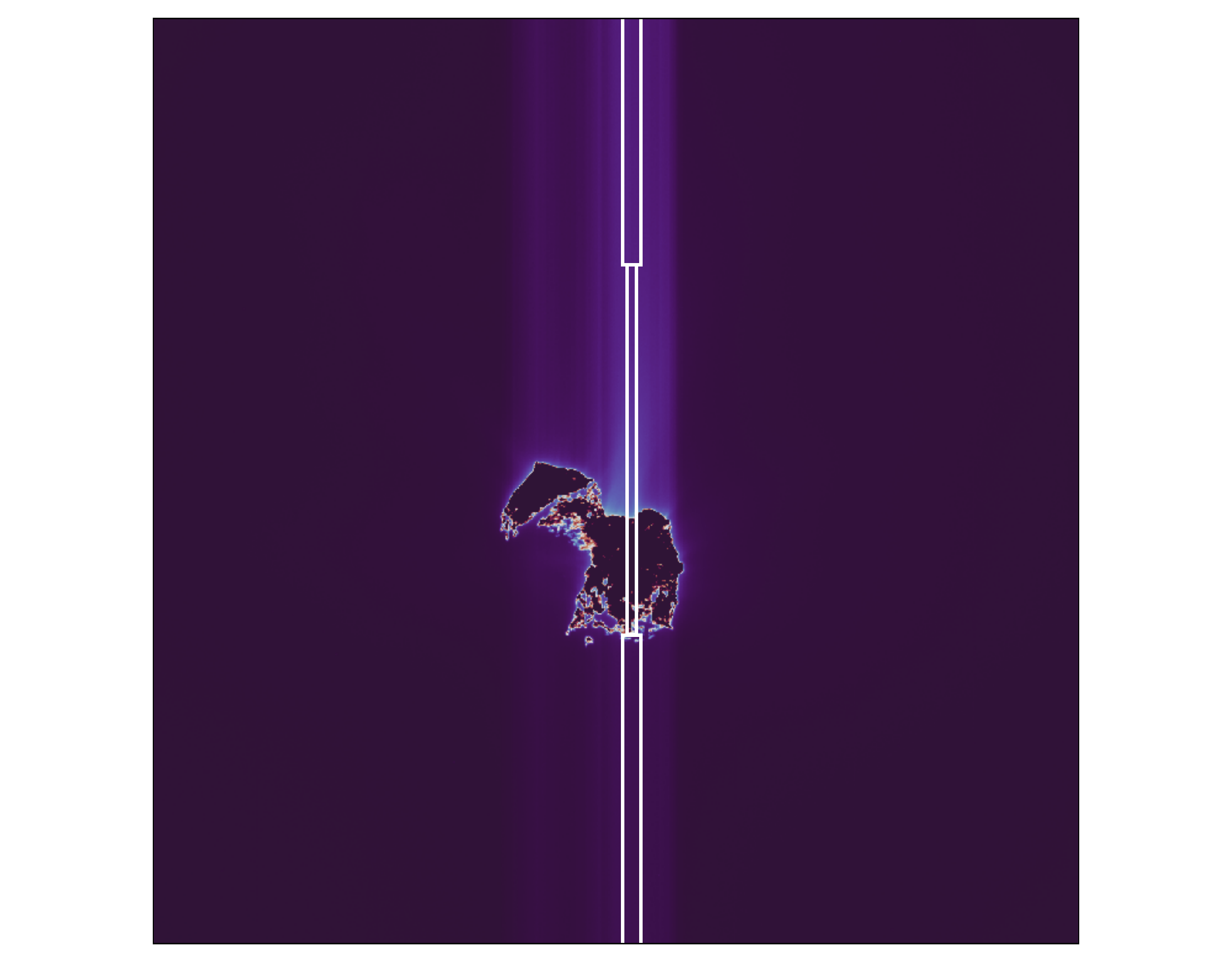}{0.33\linewidth}{c) 19:18 UTC}}
\caption{NAVCAM images from November 7 at 15:11, 17:15, and 19:18 UTC with nucleus pixels masked to highlight faint activity. The Alice slit is overlaid in white. The Sun is to the right in all three images. With the exception of faint jets emanating from the neck in image c) there is no evidence of significant background activity. }
\label{fig:NAVCAM_Jet}
\label{fig:11_07_3dT}
\end{figure}

\subsection{Alice Dataset}\label{Observations:Alice}
From 12:00 UTC until 19:18 UTC on November 7 the Alice instrument was in a stable pointing scheme, where there was minimal motion to the instrument's line of sight. At this time the Alice slit was centered on the nucleus of 67P, with the upper rows in the sunward direction and the lower rows in the anti-sunward direction (Figure \ref{fig:11_07_3dT}). A total of 38 exposures were taken in this period. In this period of time the distance from the spacecraft to the comet decreased from 237.0 to 228.8 km, the subspacecraft latitude was between -3.20$^{\circ}$ and -0.99$^{\circ}$, and the sub-spacecraft longitude ranged from 103.6$^{\circ}$ to -109.0$^{\circ}$. The end of the ``stare" scheme observations occurred at 19:18 UTC and Alice observations ceased until the ride-along observations described in detail by \citet{Noonan2021spatial}. Additional information on Alice observing schemes and planning can be found in \cite{pineau2018flight}.

\begin{figure}
\centering
 \includegraphics[width=0.6\columnwidth,clip=true]{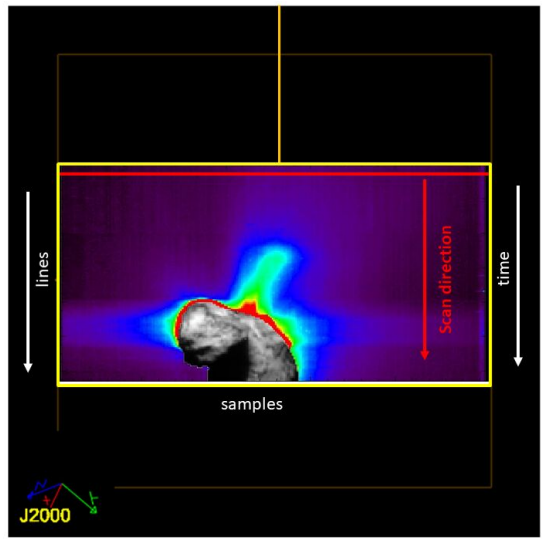}
    \caption{The figure shows the configuration of the nucleus of the comet and outburst with respect to the VIRTIS-M slit (horizontal red line) and its scan direction (red arrow). The VIRTIS-M data cubes are acquired with a scan in which each line corresponds to a given time (white arrows). The orange thick line shows the Sun direction. The spacecraft is approximately in a terminator orbit with a phase angle of 90$^\circ$, so that one side of the comet is illuminated by the Sun and the other side is in darkness.
    \label{fig:VM_fig_scan}}
\end{figure}

%\begin{figure*}
%\centering{
%\includegraphics[width=0.75\textwidth ,trim=1.0cm 2.0cm 0.0cm
%	1.0cm,clip=true]{figure_outburst_VM.pdf}
 %   \caption{Radiance value at 0.55 $\mu$m for the images V1\_00405532100 (A) and V1\_00405537500 (B),  acquired on November 7, showing the nucleus, dust coma, and outburst ejecta (collimated ejecta). A VM image of the nucleus at 0.55 $\mu$m is superimposed for better visualization. The cube details are listed in Table \ref{table:VM_observations}. The radiance has units of  W m$^{-2}$ sr$^{-1}$ $\mu$m$^{-1}$.}
 %   \label{fig:VM_figure_outburst}}
%\end{figure*}

\begin{figure*}
\centering
\includegraphics[width=0.85\textwidth,trim=0.0cm 0.0cm 0.0cm
	0.0cm]{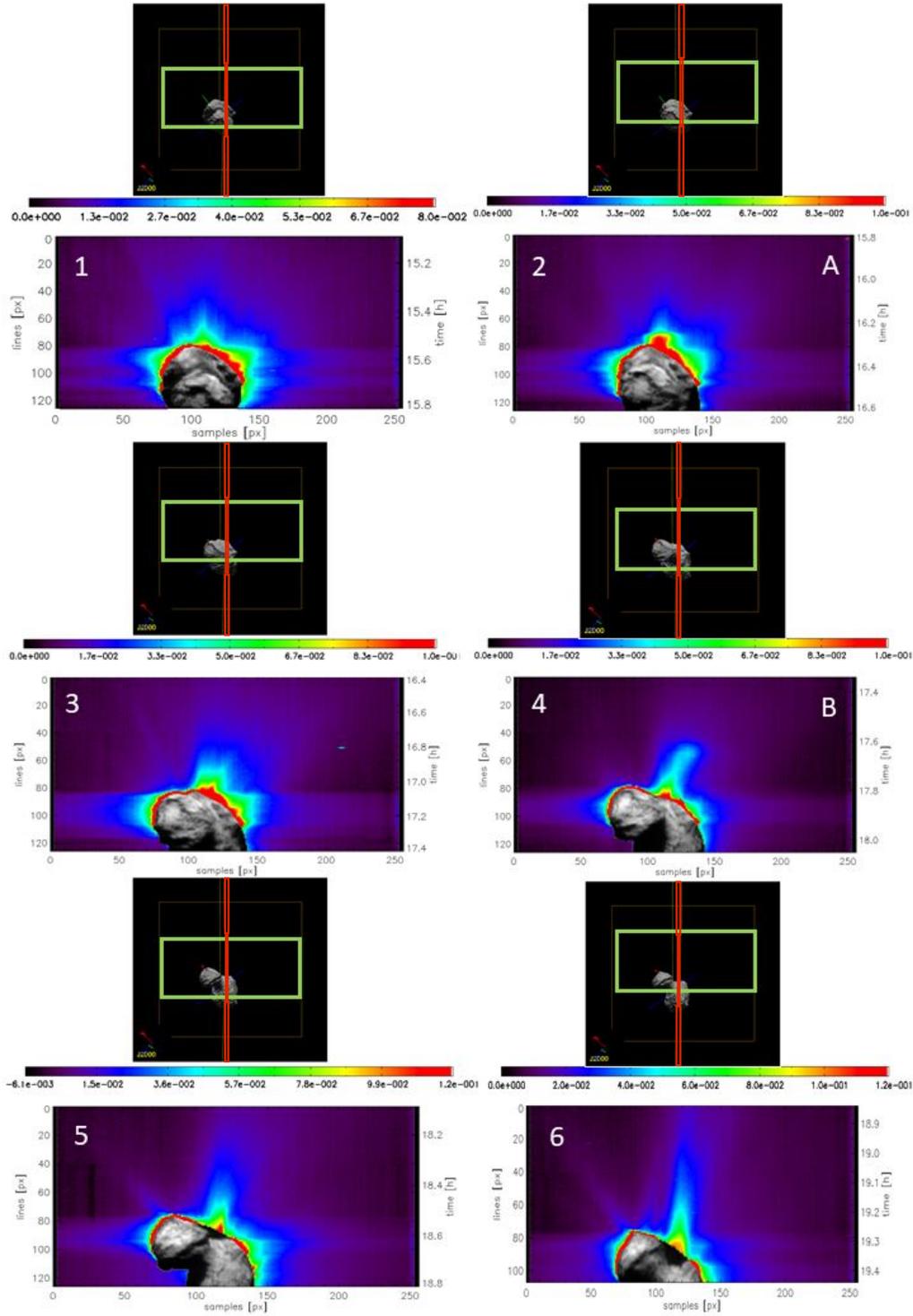}
\caption{VIRTIS-M images taken between 15:04 and 18:49 UTC on November 7. The maps are composite images where the comet nucleus image (taken as an average in the wavelength range between 0.45 and 0.55 $\mu$m) is superimposed on the maps of the dust continuum at 0.55 $\mu$m to highlight activity and the Sun is always at the top of each image. The images on the top of each VIRTIS-M image show the configuration of the nucleus with respect to the VIRTIS-M frames (green rectangles) and the Alice slit in red from the \textit{Rosetta} 3D tool. Frames 2 and 4 correspond to outbursts A and B,  respectively, with both notable extensions and intensities of dust relative to the other frames. The radiance has units of  W m$^{-2}$ sr$^{-1}$ $\mu$m$^{-1}$. The cube details are listed in Table \ref{tab:observations}. }
\label{fig:VM_figure_outburst}
\end{figure*} 

\subsection{VIRTIS-M Instrument Description}\label{Instrument:VIRTIS}
The Visual Infrared and Thermal Imaging Spectrometer (VIRTIS; \citet{Coradini2007}) was composed of two spectral instruments: VIRTIS-M and VIRTIS-H. 
VIRTIS-M was the visible (230 -- 1000 nm, 432 bands) and infrared (1000 -- 5000 nm, 432 bands) imaging spectrometer with a field of view of 3.6\degr\ (along the slit axis) and an instantaneous field of view (IFOV) of 250 $\mu$rad. 
The instrument acquired hyperspectral cubes by scanning in time the target scene line by line. 
The duration of the acquisition ($\Delta t$ in Table \ref{tab:VM_observations}) is given by the number of lines (including periodic dark current frames) times the internal repetition time, where the repetition time is the time between two consecutive steps necessary to move the internal scan mirror by one IFOV (Fig. \ref{fig:VM_fig_scan}). 
The integration time ($t_{exp}$  in Table \ref{tab:VM_observations}) is lower than the internal repetition time. 
The maximum 3.6\degr\ $\times$ 3.6\degr\ FOV was imaged by repeating acquisition on successive 256 scan mirror steps (lines). 
From a distance of 100 km this corresponds to a 6.4 km $\times$ 6.4 km swath with a resolution of 25 m pix$^{-1}$. 
As an example, we show in Fig. \ref{fig:VM_figure_outburst} the VIRTIS-M hyperspectral cubes with the line and time axes. 

\subsubsection{VIRTIS-M dataset}
Because of the failure in early May 2015 of the cryocooler, which is necessary to operate the IR channel, our analysis is restricted to VIS hyperspectral images. The spectra and images used for the analysis were reduced using the VIRTIS calibration pipeline \citep{Ammannito2006,Filacchionethesis2006}, with additional corrections derived from in-flight data. We removed defective pixels and cosmic-ray strikes using a median filter despiking algorithm, which was employed only in the spatial dimensions of the data and therefore left the spectral data intact.\\

On November 7, from 6:29 to 19:26 UTC, VIRTIS-M (VM) acquired 17 images that lasted about 40 min each. Six of these observations are contemporary with Alice observations, between 15:03 and 16:53 UTC, and provide useful context. As shown in Fig. \ref{fig:VM_figure_outburst}, in this range of time VM observed two outbursts (image cubes A and B) and small and strong jets in the other observations. The hyperspectral cubes are obtained from a target distance of 230 km with the FOV covering an area of 14.5 $\times$ 7.2 km$^2$. The spacecraft was approximately on a terminator orbit so that the Sun illuminated one side and the other side was in darkness (Fig. \ref{fig:VM_fig_scan}). Figure \ref{fig:VM_figure_outburst} displays the intensity maps of the dust continuum in units of W m$^{-2}$ sr$^{-1}$ $\mu$m$^{-1}$. The maps are a composite image where the comet nucleus (as an average in the wavelength range 0.45 - 0.55 $\mu$m) is superimposed on the image of the dust continuum averaged in a bandpass of 0.10 $\mu$m centered on 0.55 $\mu$m. The Sun is at the top of the image, and on the day side, the dust activity shows a predictable behavior that is correlated with the illumination conditions. Table \ref{tab:VM_observations} provides the geometry information for the hyperspectral cube calculated by a routine \citep{Acton1996} that uses the spacecraft trajectory and orientation stored in SPICE kernels, and the 67P SHAP5 shape model for the comet nucleus \citep{Jorda2016}. The image was acquired line by line by means of a scanning mirror, taking 20 seconds per line with the final image composed of a sequence of consecutive lines in the vertical direction (Fig. \ref{fig:VM_fig_scan}). The analysis of the VM continuum can be limited by in field straylight when the instrument slit is partially filled by the bright nucleus, and a sizable portion of the incoming photons is spread into the adjacent coma pixels. For the outburst events studied here, there is no stray light because they are out of these regions. 

\section{ALICE Results: Gas properties}\label{results:ALICE}
Here we discuss Alice UV observations of the period between 12:00 and 19:18 UTC on November 7 and describe properties of several useful Alice data products; spectra, light curves, and spatial profiles. 
\subsection{Spectra}\label{Spectra}
\begin{figure*}
\centering
\includegraphics[width=1.0\linewidth]{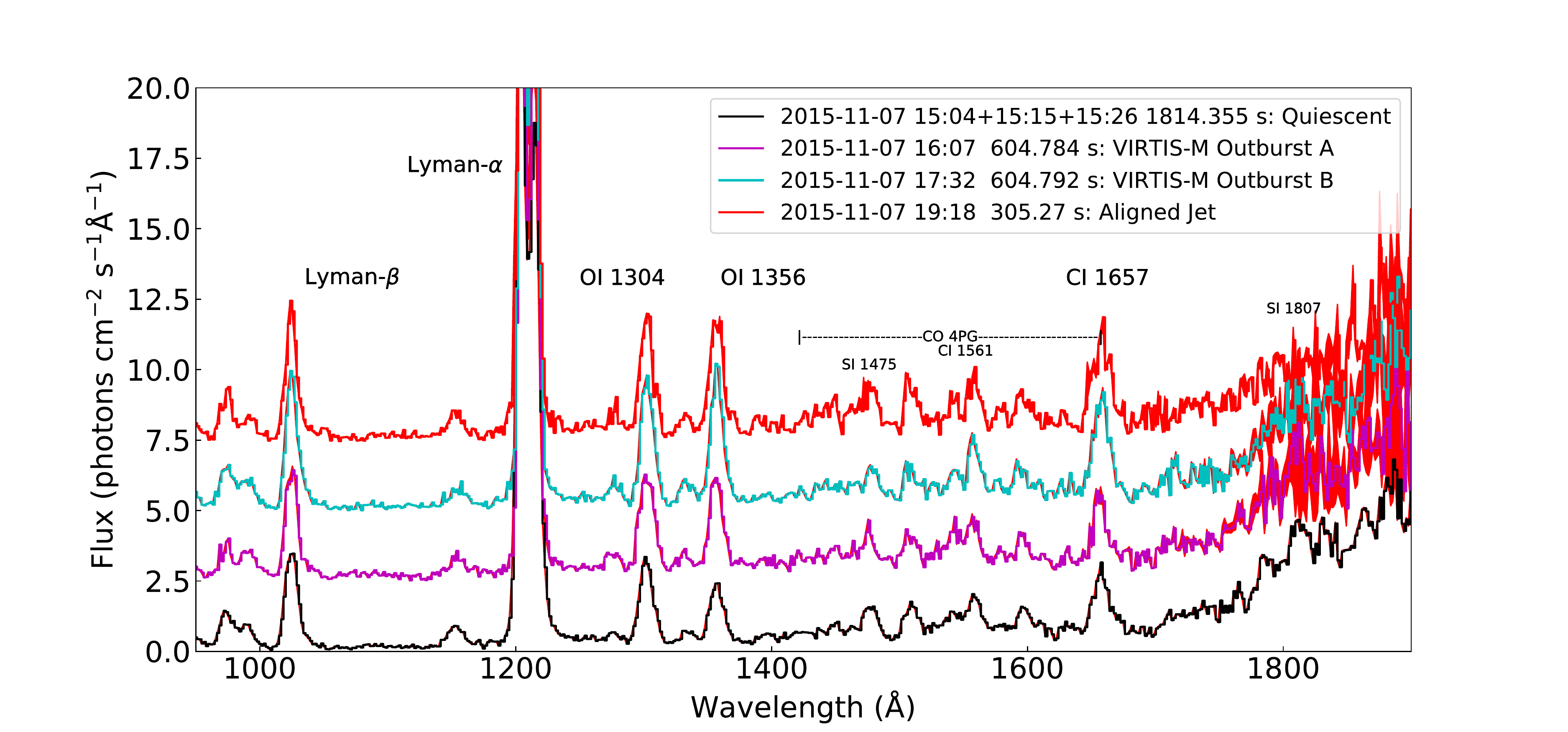} 
\caption{ Spectra derived from rows 15-18 taken during a quiescent period, the VIRTIS-M outburst detections, and near-alignment of a cometary jet with the Alice slit on November 7. Of particular interest is the \ion{O}{1}] 1356 and \ion{O}{1} 1304 ratio, indicative of dissociative electron impact. Spectra are offset by 2.5 photons cm$^{-2}$ s$^{-1}$ \AA$^{-1}$. There is also faint CO Fourth Positive emission in the 1400 \AA\ to 1600 \AA\ region mixed with emission from dissociative electron impact excitation of \ce{CO2}, discussed further in Section \ref{Discussion}. Errorbars are plotted but are largely smaller than the line-width at wavelengths less than 1800 \AA.}
\label{fig:stable_spectra}
\end{figure*}

\begin{figure*}
\centering
\includegraphics[width=1.0\linewidth]{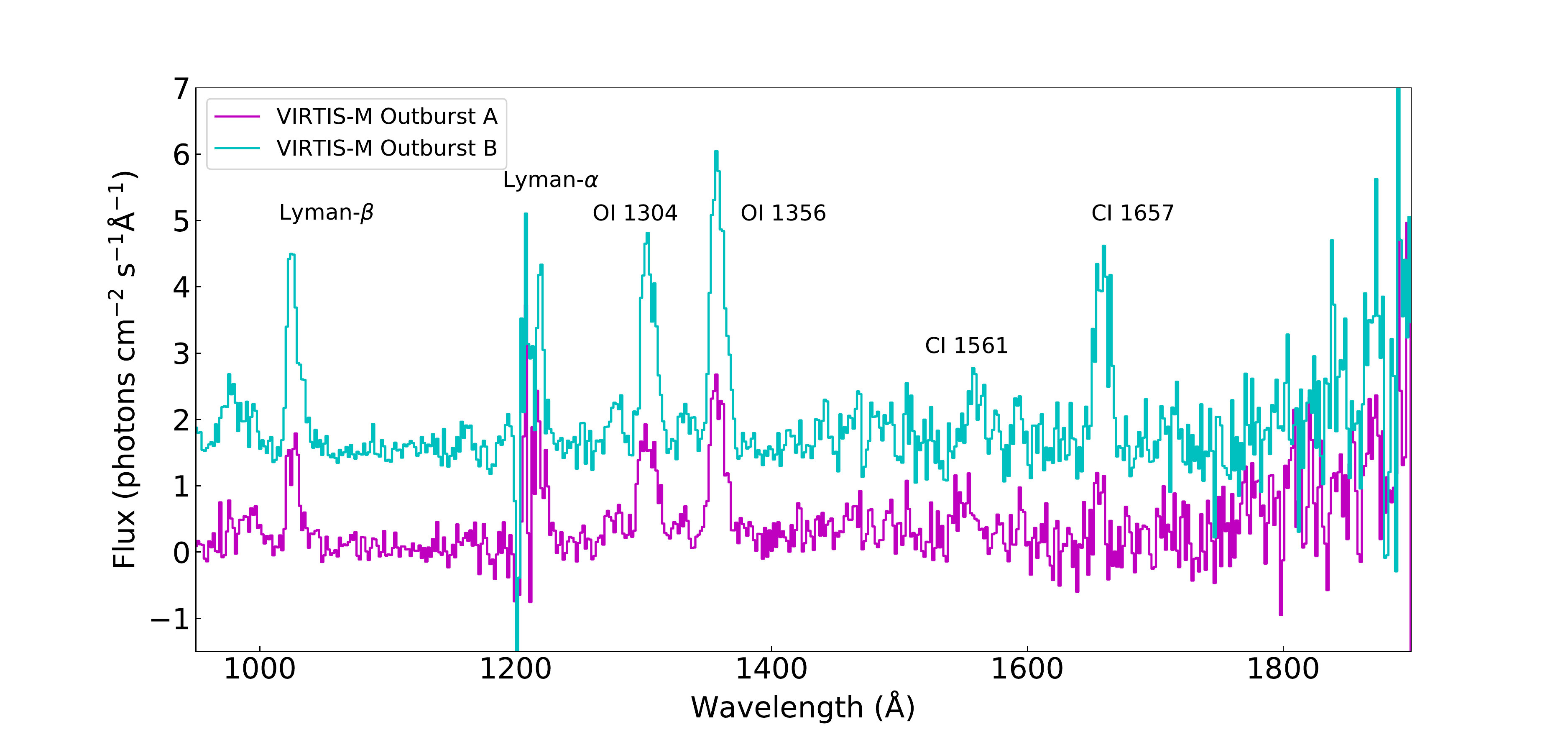} 
\caption{  Difference spectra resulting from the subtraction of the 15:04 UTC spectrum from the 16:07 and 17:32 UTC spectra on November 7. Spectra are offset by 2.5 photons cm$^{-2}$ s$^{-1}$ \AA$^{-1}$. Notice the substantial difference in \ion{O}{1}] 1356/\ion{O}{1} 1304 \AA\ emission between VIRTIS-M outbursts A and B. }
\label{fig:diff_spec}
\end{figure*}

Alice observations from 12:00 to 19:18 UTC were taken during a stable pointing scheme with little to no motion in the Alice slit relative to the nucleus for the duration. The exposure taken at 15:04 UTC on November 7, which is the quiescent background spectrum, contains the hydrogen emission features Lyman-$\alpha$ and -$\beta$, though Lyman-$\alpha$ has a non-standard line shape due to detector gain sag in that area (Fig. \ref{fig:stable_spectra}, blue spectrum).  The same spectrum shows the \ion{O}{1}] 1356/1304 \AA\  ratio is \textless 1, evidence that both dissociative electron impact and resonance scattering emission are within the Alice slit for relatively quiet periods of cometary activity \citep{kanik2003electron,feldman2016nature}, though resonance fluorescence along the Alice line of sight is dominant. \ion{O}{1}] 1356 \AA\ is the result of a spin-forbidden transition and is only present as as a result of dissociative electron impact, and its strength relative to the \ion{O}{1} 1304 \AA\ triplet can be used to infer compositions.  In cases where dissociative electron impact excitation on \ce{O2} or \ce{CO2} is dominant we would expect the \ion{O}{1}] 1356/\ion{O}{1} 1304 \AA\  ratio to approach 2 while if e+\ce{H2O} dominates the value is closer to 0.3 \citep{hall1998far,feldman2016nature,feldman2018fuv, noonan2018ultraviolet}. A \ion{O}{1}] 1356/\ion{O}{1} 1304 \AA\ ratio greater than 1 can be seen in the spectra taken at 16:07, 17:32, and 19:18 UTC on November 7, both in Figure \ref{fig:stable_spectra} and after quiescent subtraction to remove background coma emissions and interplanetary medium contribution to the Lyman series in Figure \ref{fig:diff_spec}. The carbon emission features for this period are dominated by \ion{C}{1} 1561 and 1657 \AA, both capable of being produced by photodissociation and dissociative electron impact of \ce{CO2} and \ce{CO}, though the line ratios suggest a mixture of resonance fluorescence and e+\ce{CO2} as the main contributor \citep{ajello1971dissociative,ajello1971emission,ajello2019uv}. \ce{CO} Fourth Positive emission appears blended with atomic carbon and sulfur features between 1400 and 1600 \AA\, specifically the 4-0, 5-1, 3-0, 4-1, 2-0, 1-0 and 0-0 bands, but far below the levels seen in spectra presented in \citet{feldman2018fuv} and are not characterized in this work. 


To highlight spectral characteristics we have chosen three spectra that correspond to two VIRTIS-M outbursts and one jet; the first corresponds to  outburst A at 16:07 UTC on November 7, the second to outburst B  at 17:32 UTC, and the third to a jet at 19:18 UTC (Fig. \ref{fig:stable_spectra}). These three spectra allow spectral comparison of two prominent types of cometary activity: outbursts and jets. Quiescent-subtracted spectra are displayed in Figure \ref{fig:diff_spec} to properly convey the spectral signature of each activity. Lyman-$\beta$ and the \ion{C}{1} 1561 and 1657 \AA\ features have no strong changes for the first VIRTIS-M outburst, but show much larger increases for both the second VIRTIS-M outburst and jet (Fig. \ref{fig:diff_spec}). Additionally, the \ion{O}{1} 1304 and 1356 \AA\ emissions for each spectrum show different characteristics; the first VIRTIS-M outburst shows an \ion{O}{1}] 1356/\ion{O}{1} 1304 ratio of $<$1, the second a ratio $>$1, and the jet a ratio of $\sim$1. 

The UV spectra contained in this analysis all share some weaker features that we will discuss here. First, the Ly-$\beta$ emission feature is blended with \ion{O}{1} 1025.72 \AA\ emission at the resolution of the Alice instrument. However, the \ion{O}{1} contribution to the Ly-$\beta$ + \ion{O}{1} 1025.72 \AA\ blend can be determined from e+\ce{O2} modelling done at 200 eV by \cite{ajello1985study} if the \ce{O2} column density is known, though we note that this energy is significantly above the expected mean energy for electrons in the near-nucleus environment \citep{clark2015suprathermal}. Because the \ce{O2} column density is typically not known a priori, we instead implement the line ratio for the \ion{O}{1} 1356/1025.72 \AA\ features from dissociative electron impact of \ce{O2}, which is approximately 35, and is therefore negligible compared to the typical strength of Lyman-$\beta$. Given the low $g$-factor for fluorescence of the \ion{O}{1} 1025.72 \AA\ transition we assume this ratio is accurate for our data. Second, there is evidence of \ion{S}{1} emission in Figure \ref{fig:stable_spectra} as a weak triplet at 1807, 1820, and 1826 \AA\, but these contributions are no longer clear in the quiescent-subtracted spectra in Figure \ref{fig:diff_spec} except in the aligned jet spectrum. This implies that there is weak \ion{S}{1} emission at 1425 and 1473 \AA\, blending with \ce{CO} Fourth Positive group emissions. However, given that the \ion{S}{1} and CO emissions are negligible in the quiescent subtracted spectra and are therefore not large components of either the outbursts or jet we will not focus on them. 

\begin{figure}
\centering
\includegraphics[width=1.0\linewidth]{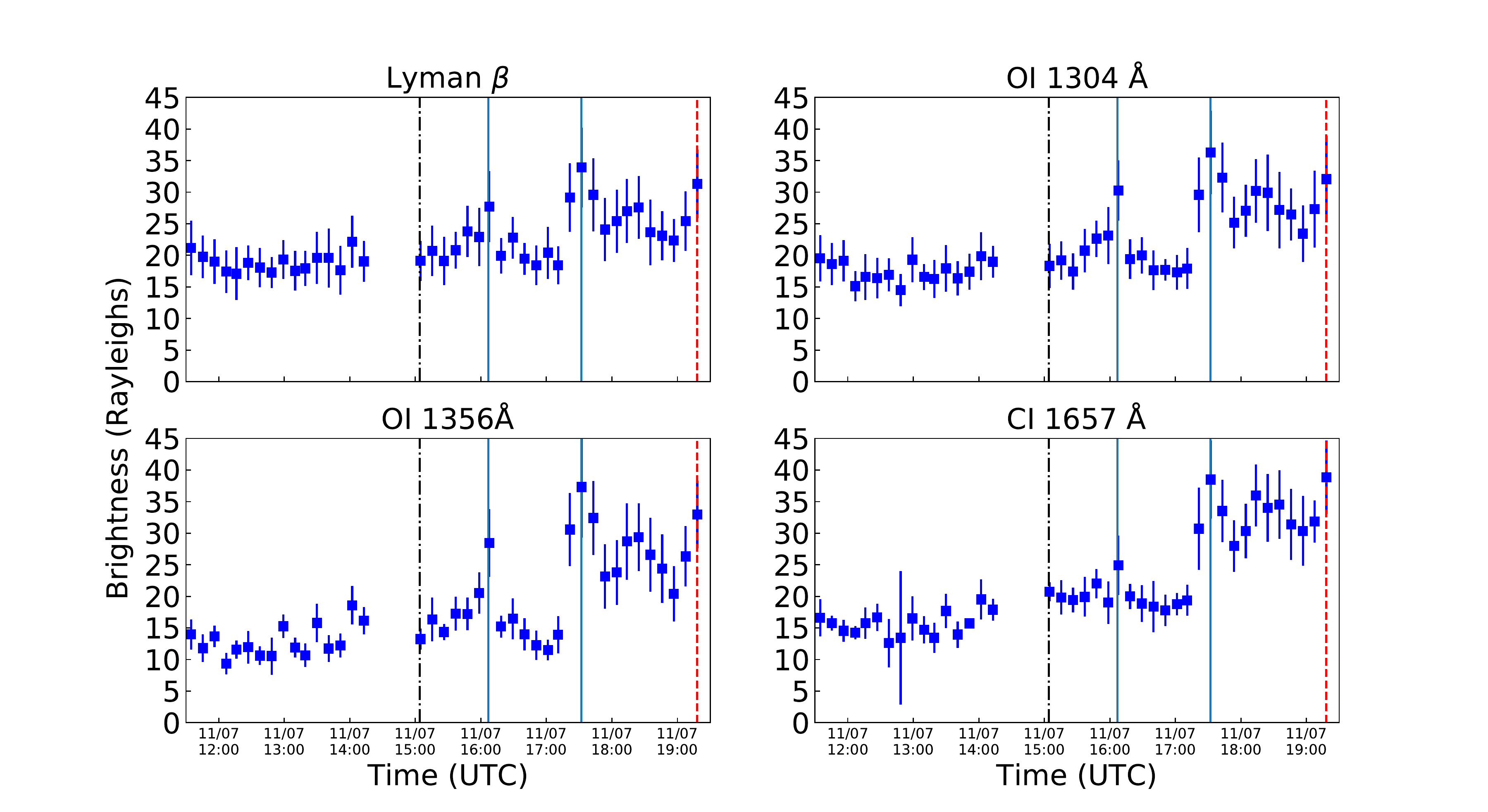} 
\caption{Light curves for dominant atomic emission features between 12:00 and 20:00 UTC on November 7, integrated over rows 15-18 on the Alice detector. Errorbars on each point are driven by variation between rows, which is amplified by the odd-even detector effect. The observation used for quiescent subtraction is marked with a black dot-and-dashed line, observations coinciding with VIRTIS-M outburst detections are marked with blue lines for outbursts A and B at 16:07 and 17:32 UTC, respectively. The observation of the nearly in-slit jet is marked with a red dashed line.}
\label{fig:time_vs_brightness}
\end{figure}

%\begin{figure}
%\centering
%\includegraphics[width=1.0\linewidth]{outburst_oi_ratio.pdf} 
%\caption{Ratio between the semi-forbidden \ion{O}{1}] 1356 and \ion{O}{1} 1304 \AA\ emission between 12:00 and 20:00 UTC on November 7, integrated over rows 17-22 on the Alice detector and plotted in Figure \ref{fig:time_vs_brightness}. The 1$\sigma$ error for each value is less than the size of the plotted point. The observation used for quiescent subtraction is marked with a black dot-and-dashed line, observations coinciding with VIRTIS-M outburst detections are marked with blue lines for outbursts A and B at 16:07 and 17:32 UTC, respectively. The observation of the nearly in-slit jet is marked with a red dashed line.}
%\label{fig:time_vs_oi_ratio}
%\end{figure}

\subsection{Light Curves}\label{lightcurves}
By integrating Ly-$\beta$, \ion{O}{1} 1304, \ion{O}{1}] 1356, and \ion{C}{1} 1657 \AA\ emission features in rows 17-22 of the Alice slit for observations taken during the period between 12:00 and 19:18 UTC on November 7 light curves detailing changes to cometary activity can be made, allowing confirmation of outbursts and other transient events. In Figure \ref{fig:time_vs_brightness} we clearly see stable atomic emissions from 12:00 until $\sim$16:00 UTC. At 16:07 UTC there is a sharp increase in \ion{O}{1} emissions that quickly subsides back to the quiescent levels. A much stronger increase in emissions then occurred at 17:32 UTC, had a 5-20 Rayleigh decrease by 18:00 UTC, which was then sustained until 19:18 UTC.  

From the Alice light curves several key pieces of information become clear. The first outburst identified by VIRTIS-M has a small \ce{CO2} component, indicated by the small change in \ion{C}{1} 1657 \AA\ emission relative to the normal comet activity between 12:00 and 16:00 UTC. Second, the first outburst is smaller not just in emission strength but also in duration compared to the second outburst. For the first outburst quiescent emission levels were reached in the Alice spectrum at 16:13 UTC, while for the second it appears substantial cometary activity followed the outburst and persisted until the end of Alice observations at 19:18 UTC. This may be an indication that the second outburst may have increased more typical cometary activity like jets, which are sustained longer than outbursts. Third, the \ion{O}{1}] 1356/\ion{O}{1} 1304 ratio for the period shows that outburst A has a substantially lower value than either outburst B or the aligned jet (Fig. \ref{fig:time_vs_brightness}). \ion{O}{1}] 1356/\ion{O}{1} 1304 ratios at or above 1 have previously been indicative of significant dissociative electron impact of \ce{CO2} and \ce{O2} in the inner coma \citep{feldman2018fuv}. The high \ion{O}{1}] 1356/\ion{O}{1} 1304 ratio persists well past 19:18 UTC into the period covered by \citet{Noonan2021spatial}, until 12 November 2015. This consistent and elevated ratio indicates that the dissociative electron impact emissions were elevated for days, regardless of comet rotation, which could be tied to increased cometary activity, plasma density, plasma energy, or a combination of all three. This paper will focus only on the outbursts that appear to initiate this extend period of elevated electron impact emissions. 

\subsection{Spatial Profiles}\label{spatialprof}

\begin{figure}
\centering
\includegraphics[width=0.5\linewidth]{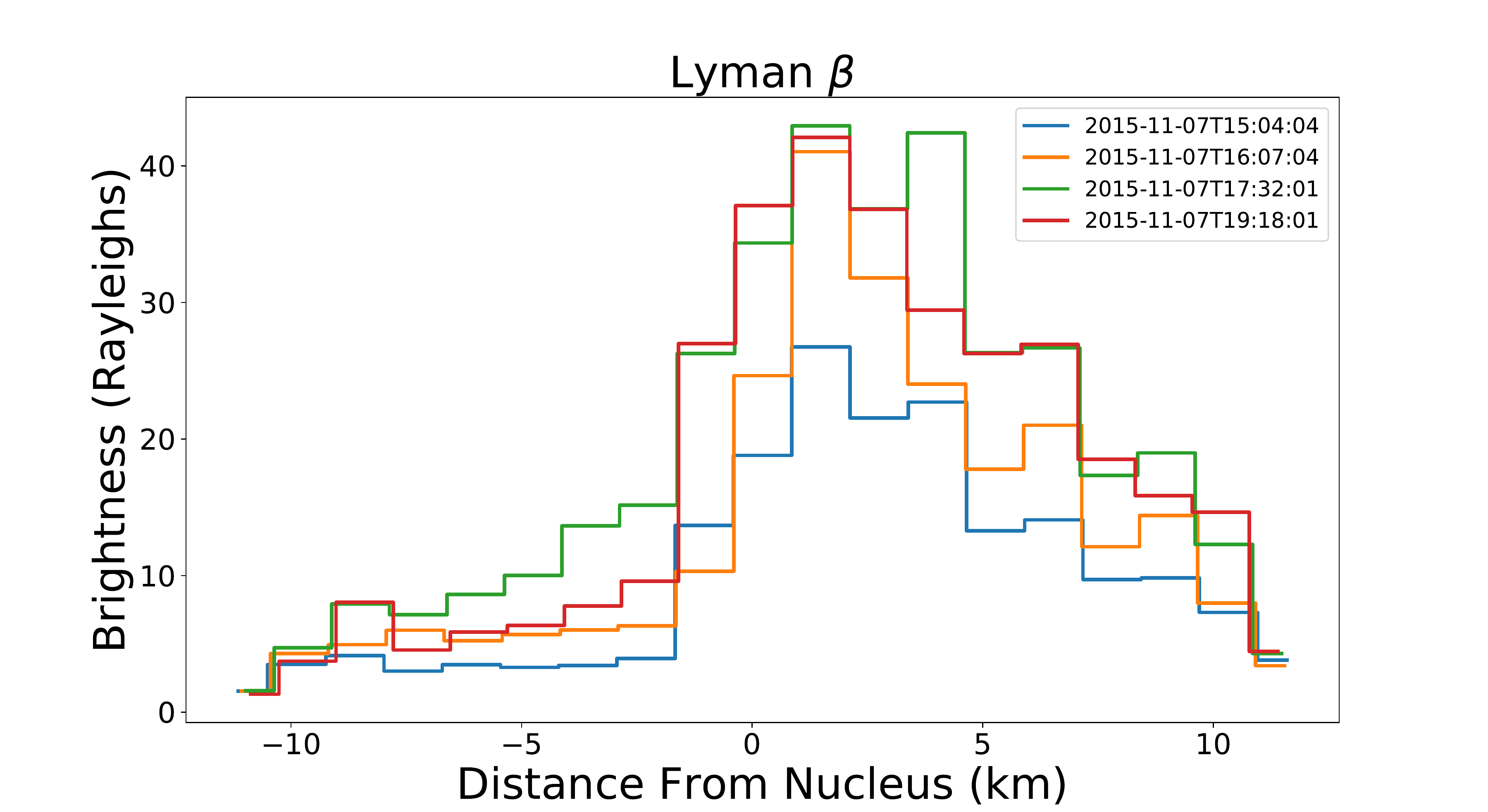} 
\includegraphics[width=0.5\linewidth]{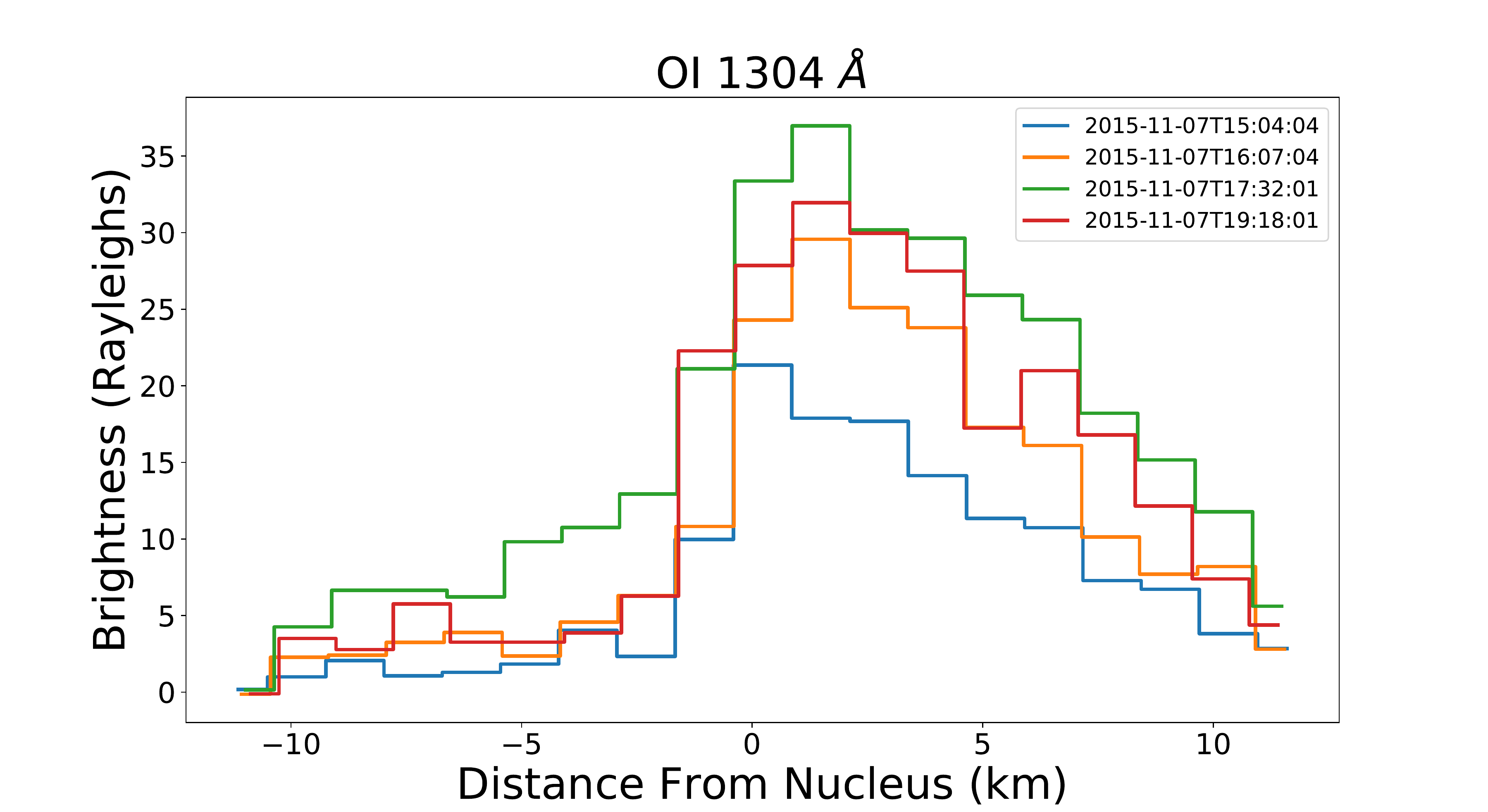} 
\includegraphics[width=0.5\linewidth]{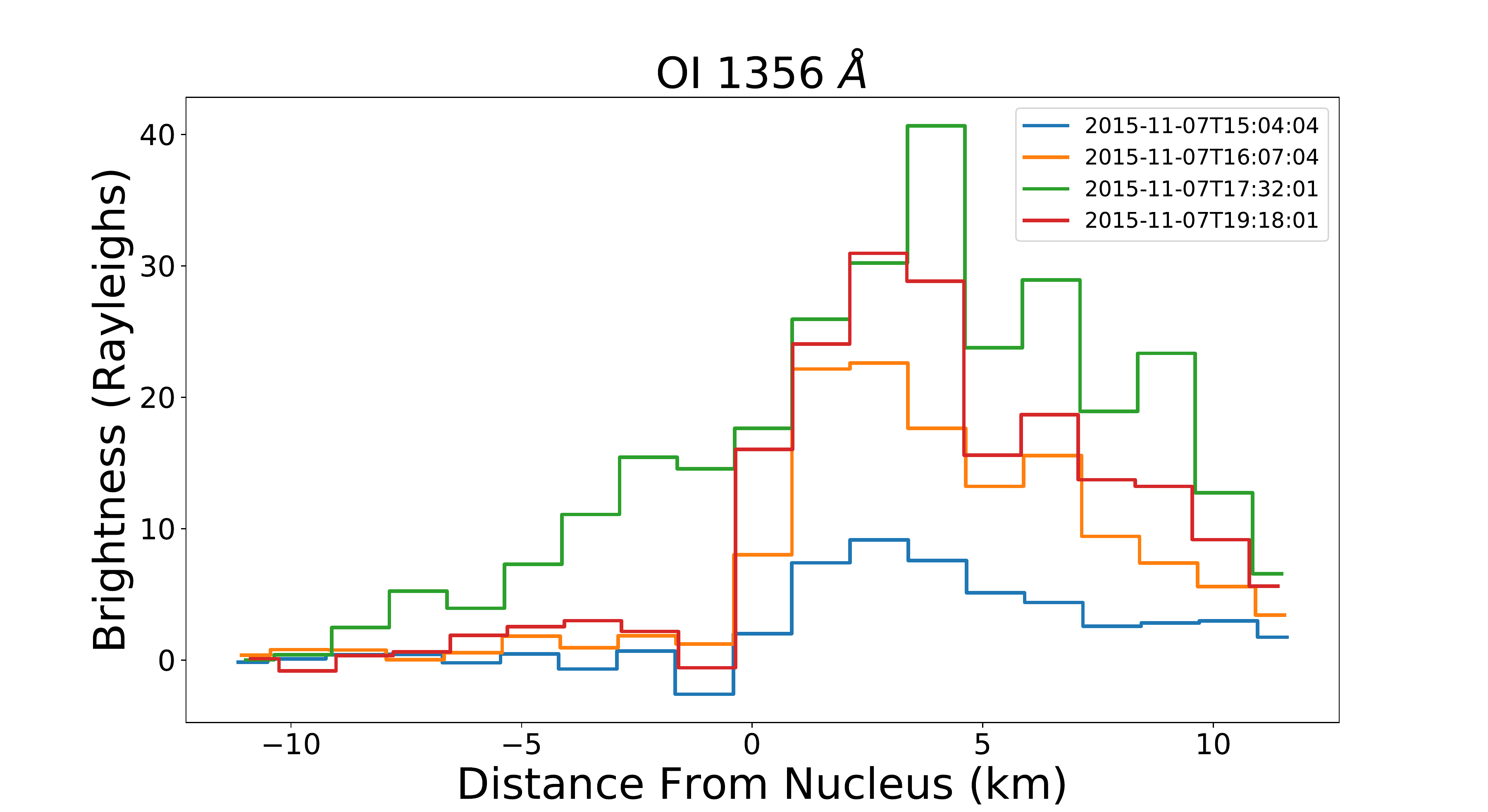} 
\includegraphics[width=0.5\linewidth]{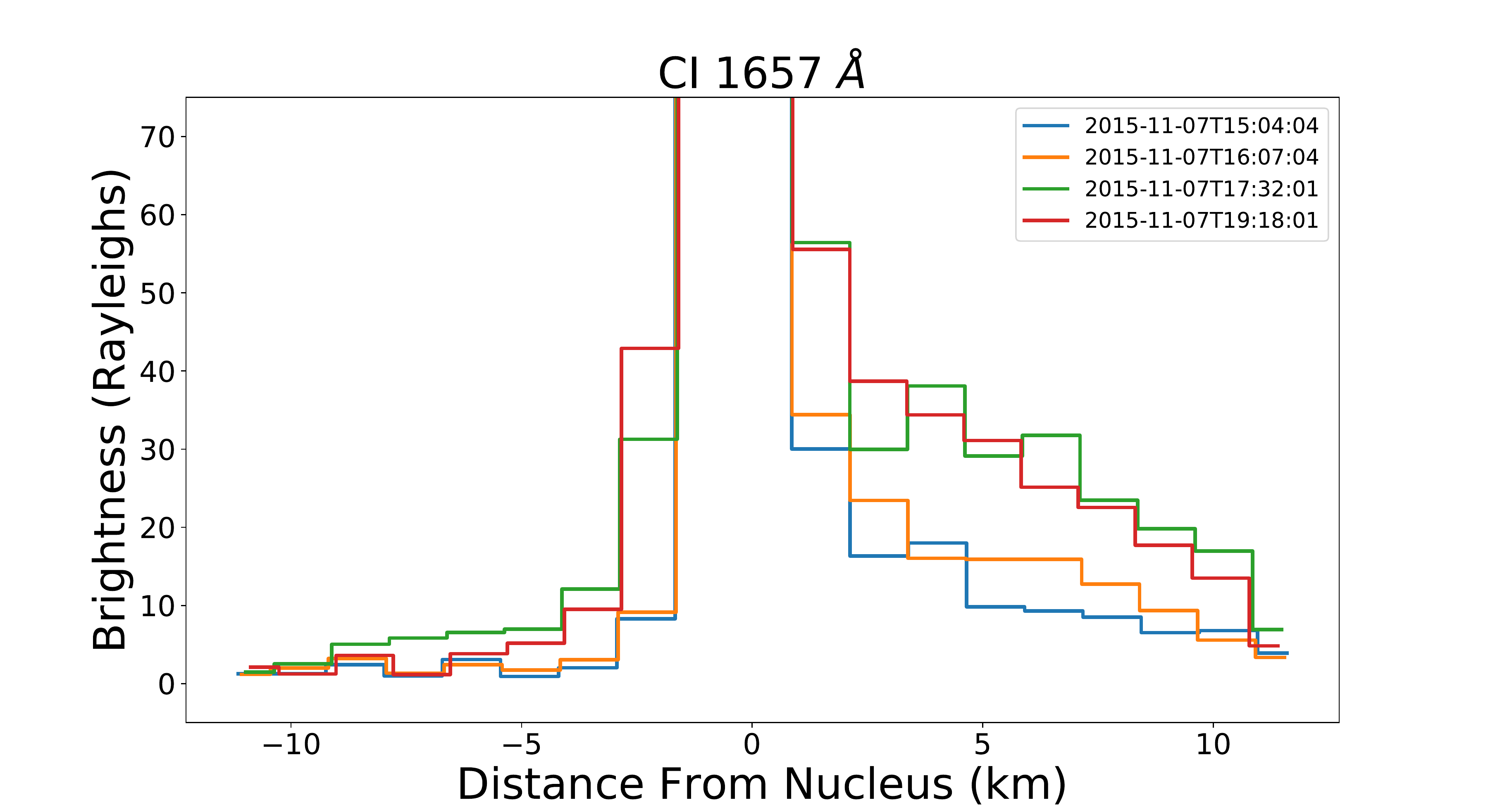} 
\caption{Spatial profiles of dominant emission features in Alice spectra at 15:04, 16:07, 17:32, and 19:18 UTC. The sunward direction is in the $+X$ direction.  Each detector row subtends approximately 1.2 km. Of particular note is the change in slope for \ion{O}{1}] 1356 \AA\ between quiescent, initial activity at 17:32 UTC, and the parallel jet observation at 19:18 UTC between 0 and 12 km.}
\label{fig:spatial_profiles}
\end{figure}

 During the stable pointing scheme on November 7 both the sunward and anti-sunward portions of the near-nucleus coma were observed in the same observations, an ideal geometry for developing one-dimensional spatial profiles of the near-nucleus coma. Several of these profiles are shown in Figure \ref{fig:spatial_profiles}. 

The earliest spatial profile is taken from the quiescent observation at 15:04 UTC and shows the inner coma as typically observed by Alice; strong emission on the sunward side, with weak emission on the anti-sunward side \citep[]{feldman2015measurements,feldman2018fuv}. This dichotomy is exhibited by all four emission features shown in the profiles in Figure \ref{fig:spatial_profiles}.  Emission strengths gathered from spectra taken at 16:07 and 17:32 UTC show the impact of outbursts A and B, while the emissions from 19:18 UTC are of the activity resulting from outburst B. All three emission features without substantial solar continuum contribution from the nucleus see an increase at the 0 km mark; Lyman $\beta$ increases from $\sim$22 to 30 Rayleighs at 17:32 UTC, \ion{O}{1} 1304 \AA\ rises from 19 to 25 Rayleighs, \ion{O}{1}] 1356 \AA\ from 0 to 10 Rayleighs. Additionally, \ion{O}{1}] 1356 \AA\  shows a factor of 5 increase in emission near the nucleus, from a maximum of 8 Rayleighs at 15:04 UTC to 40 Rayleighs at 17:32 UTC. The slope for the \ion{O}{1}] 1356 \AA\ spatial profile at 17:32 UTC is also steeper than in the quiescent period, approximately -1.86 $\pm$ 0.90 Rayleighs/km compared with the -0.45 $\pm$ 0.03 Rayleighs/km at 15:04 UTC. This slope increases between the 17:32 UTC and 19:18 UTC datasets to -2.07 $\pm$ 0.31 Rayleighs/km, possibly indicating a relaxation of the near-nucleus coma to the quiescent state at the end of the stable Alice observations on November 7 and possibly a weakening of the jet. 

The profiles for both Ly-$\beta$ and \ion{O}{1} 1304 \AA\ appear flat between 3 and 7 km for the observations taken at 17:32 and 19:18 UTC, something that is not captured in the linear fit to the slope. This feature is missing in both the \ion{O}{1}] 1356 and \ion{C}{1} 1657 \AA\ spatial profiles. There are two factors that make determining the significance of these changes difficult with the limited number of observations: the low spatial resolution and the odd-even effect experienced by Alice. The 1.2 km/pixel size combined with the tendency of the Alice detector to push counts to odd rows over even rows complicates peak determination, so all positions stated above have a pixel size uncertainty associated with them of $\pm$1.2 km for these stable pointing spatial profiles.

\section{VIRTIS Results: Dust properties}\label{results:VIRTIS}
In this section, we analyze the physical properties of the dusty outbursts observed by VIRTIS-M in terms of lightcurve, color, filling factor and dust mass loss.
In Table \ref{tab:out_char}, we list all the relevant information obtained from the analysis of our data set as  time, duration, longitude, and latitude of the estimated source region of the outburst, radiance level, and color at the maximum of the light curves. The duration is computed using the time at which the radiance returned to the pre-outburst value or when the coma observation starts and ends.

\begin{table*}
%\begin{sidewaystable}
	\centering   %\rotatebox{90}
	\caption{Dust outburst properties in the VIRTIS-M VIS channel}
	\label{tab:out_char}%[htbp]
  %  \begin{adjustbox}{max width=1\textwidth}
	\begin{tabular}{ccccccccccc} % four columns, alignment for each		
        \hline \hline 
        \\
	ID&	VIS 	  & Detection& Duration  &Local &Long & Lat &Max  & color &  \\
& filename&	& time & time&range  &range  &radiance & &    \\
 &	 & (h) & [min] &(h) & ($\circ$) & ($\circ$) &  &  $\%$/ 100 nm  \\
		\hline
		\\
	A&	  V1\_00405532100& [16:13, 16:18]& 4.8 & [17:00, 17:06] & [70.37, 73.40] & [-58.73, -38.56] & 0.08 & 13.1 $\pm$ 1.3 \\ 
	B&	  V1\_00405537500&[17:33, 17:48] &15 & [13:16, 17:01] & [25.18, 354.98] & [ -57.09,     -18.73] & 0.06 & 12.4 $\pm$ 2.4 \\
		\hline
	\end{tabular}
%    \end{adjustbox}

	\flushleft
     {\bf{Note}}:
{\it{Column 1}}: Assigned letter for each image cube.     
{\it{Column 2}}: Observation file name. 
%{\it{Column 2}}: Size of outburst defined in terms of life time of the event (fourth column) and radiance level at the maximum (eighth column).
{\it{Column 3}}: Start and stop detection time for the dust plume.
{\it{Column 4}}: Observed outburst life time.   
{\it{Column 5}}: Local time range of outburst source on surface (see Section \ref{sec:location} ).  
{\it{Column 6}}: Longitude range of dust plume source on surface (see Section \ref{sec:location}).
{\it{Column 7}}: Latitude range of outburst source on surface (see Section \ref{sec:location}).
{\it{Column 8}}: Radiance at 0.55 $\mu$m at the maximum of the outburst emission in W m$^{-2}$ sr$^{-1}$ $ \mu$m$^{-1}$.                                           
{\it{Column 9}}: color at the maximum of the outburst radiance  (see Section \ref{sec:color}). 
\end{table*}

\begin{figure*}
\centering{
	\includegraphics[width=0.9\textwidth,trim=.0cm 4cm 0.0cm
	3cm,clip=true]{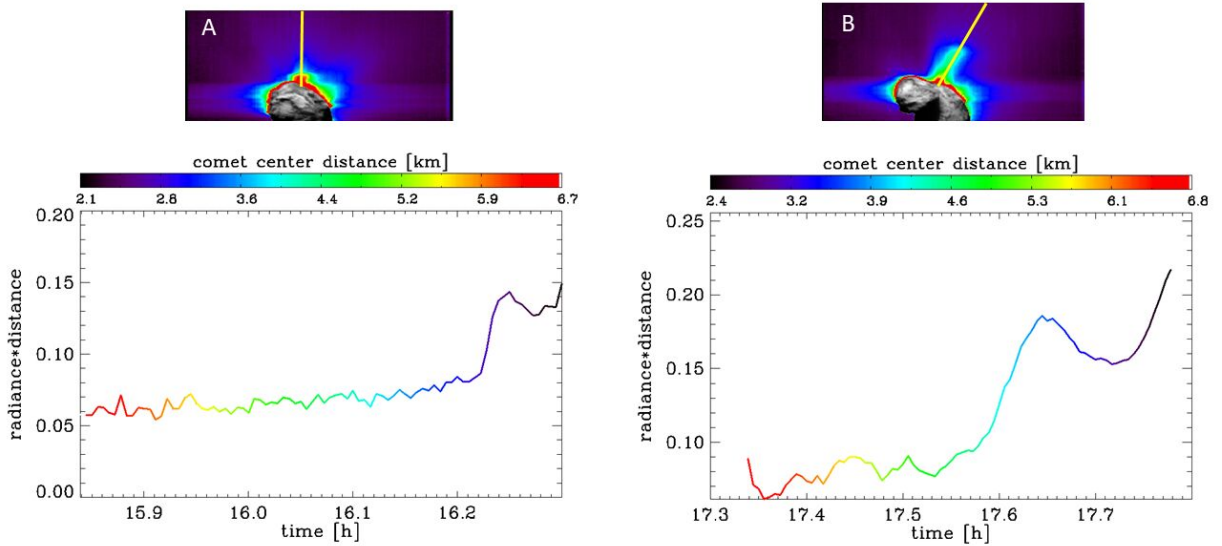}
    \caption{
    Light curves of the outbursts A (left) and B (right) acquired on November 7. The multicolor curves are the VM radiance at 0.55 $\mu$m multiplied by the distance from the comet center. The images on the top show the VM image frame for the observations A and B, and the yellow lines show where the lightcurve profiles have been extracted. The colorbar used for the VM light curves refer to the distance from the comet center and not the intensity displayed in the upper plots. For reference, a flatter curve represents a more linear relationship between radiance and distance from comet center. The deviations present at 16:13 UTC and 17:33 UTC are interpreted as the start of the outbursts.}
    \label{fig:VM_lightcurves}}
\end{figure*}

\subsection{Outburst morphology and light curves}

In this section, we analyze the evolution of the given outbursts and we characterize the spatial distribution of both the dust background and the ejected dust.
As shown in \citet{Rinaldi2018} to distinguish, in the VM data, between transient events and long-lasting features such as jets, which are stable for more than one comet rotation as shown by \citep{Vincent2016}, we adopted the following criteria: the transient events observed by VM were identified by their light curve (radiance at a given wavelength versus time) characterized by a sudden brightness increase in the coma that is associated with a release of gas and dust over a very short timescale, that is 5-30 min \citep{Farnham2007,Miles2016, Bockelee2017, Rinaldi2018}. The light curve of a transient event can only be derived when the scan occurs along the radial direction of the plume. The VM data are acquired with a temporal scan, as shown in Fig. \ref{fig:VM_fig_scan}, in which each line corresponds to a given time. This allows us to reconstruct the temporal evolution of the event. The two outbursts were captured in two consecutive VM data cubes, acquired 42 min apart. In Fig. \ref{fig:VM_figure_outburst} A, the dust distribution shows a wide structure whose behaviour is correlated with the illumination conditions, with an outburst of dust blobs in the direction of the Sun. The wide shape can be the result of a complex event, with more ejecta sources on different active regions of the surface. The wide and well-defined internal structures have a maximum intensity of 0.08 W m$^{-2}$ sr$^{-1}$. In Fig. \ref{fig:VM_figure_outburst} B, we see the onset of another less-intense event, manifested as a collimated structure with a maximum intensity of 0.06 W m$^{-2}$ sr$^{-1}$ (Table. \ref{tab:out_char}). 

In Fig. \ref{fig:VM_figure_outburst}, the temporal profiles (multicolor curves) in the bottom plots have been extracted along the yellow lines sown in the upper images. The colors used for the VM data points correspond to the distance from the comet center. According to the description of the shape model by \citet{Preusker2017}, this center is defined as the center of mass. Fig. \ref{fig:VM_lightcurves} shows outburst light curves where the radiance has been multiplied by the distance as a means for better visualising changes in the coma. Before the outburst event both profiles have an approximately constant value in time between 0.05 and 0.09. This is consistent with a cometary coma in a steady state, with constant dust production and outflow speed together with the conservation of dust grains. Each outburst in Fig. \ref{fig:VM_lightcurves} is identifiable by a sharp increase in the light curve that deviates from the radiance and distance product, which should be linear in nature for a steady state dust environment. Outburst A started at 16:13 h UTC and ended at 16:18 UTC. Outburst B is a less intense outburst starting at 17:33 UTC and ending at 17:48 UTC. Unfortunately, both curves do not cover the complete evolution of the outburst because the data do not show when the radiance returned to the pre-outburst value after the maximun. Both light curves show the typical behaviour of outburst evolution; a sudden increase of the dust radiance, reaching maximum intensity a few minutes later followed by a return to a typical dust environment by the next scan \citep{Belton2008, Knollenberg2016, Bockelee2019, Rinaldi2018}. 

 \begin{figure*}
\centering{
	\includegraphics[width=0.9\textwidth,trim=0cm 4.5cm 0.0cm
	3cm,clip=true]{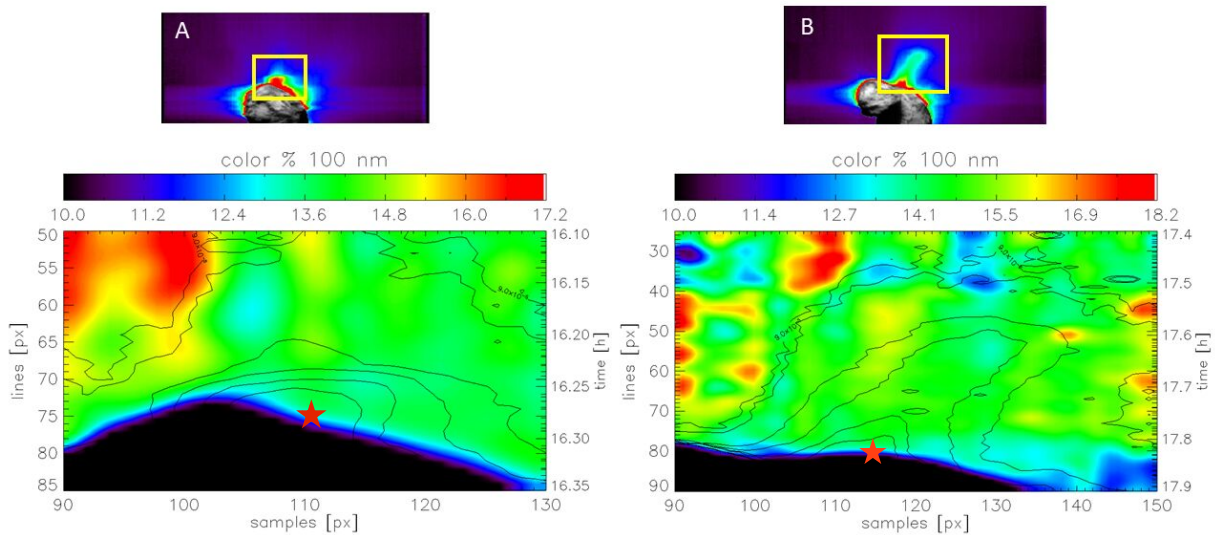}
    \caption{
    Images of the outburst of November 7 in the VIS at 0.55 $\mu$m (upper plots) and the VIS spatial distribution of the color (lower plots) calculated in the yellow square of the dust image. The colorbar used for the VM color maps refer to \% (100 nm)$^{-1}$ for the cometary dust in the lower plots and not the intensity displayed in the upper plots. The location of maximum radiance, assumed to be the outburst source, has been marked with a red star. Ejecta from outburst A or B do not display any evident color gradient, unlike other outbursts observed by VM \citep{Rinaldi2018}.}
    \label{fig:VM_color}}
\end{figure*}
 
\subsection{Color}  
\label{sec:color}
Previously in the literature it has been shown by \cite{Miles2016},  \cite{Bockelee2017}, and \cite{Rinaldi2018} that cometary outbursts have been associated with possible compositional and particle size changes. These changes would be evidenced by spatial and temporal evolution of the color, normalized reflectivity gradient, or reddening, measured in $\%$ (100 nm)$^{-1}$ \citep{JewittMeech1986}. The color can be calculated using the values of the reflectance at two or more wavelengths. The reflectance, $R$, is a dimensionless quantity calculated by dividing the measured scattered light intensity $I$ by the solar incident flux. Taking the wavelength dependent solar flux from \citet{Kurucz1994}, the mean reflectance gradient \% (100 nm)$^{-1}$, $r$, for a particular wavelength interval becomes:

\begin{equation}
    r=\frac{R_{\lambda_2} - R_{\lambda_1} }{\lambda_2 - \lambda_1} \times \frac{200}{R_{\lambda_2} + R_{\lambda_1}}.
	\label{eq:color}
\end{equation}

The reflectances used are averages over a narrow bandpass of 10 nm in width centered on 550 nm ($\lambda_1$) and 750 nm ($\lambda_2$), chosen to optimise the Signal-to-Noise Ratio (SNR) and so minimise the internal error in the color determination. We obtained a two-dimensional color map using the spectrum for each pixel in the image (Fig. \ref{fig:VM_figure_outburst}). The color uncertainties are evaluated with the method used by \citet{Rinaldi2018}, by propagating the formal error that is inversely proportional to the SNR. Outside the outburst the uncertainty is higher because the radiance and the SNR are both very low (Fig. \ref{fig:VM_figure_outburst}). For this reason, the color fluctuations outside the dust \emph{plume} region delimited by the contour lines in Fig. \ref{fig:VM_figure_outburst} are not realistic. Inside the outburst the radiance and the SNR give us an uncertainty of about 15-20$\%$.
In Fig. \ref{fig:VM_figure_outburst}, the two-dimensional color maps, for outbursts A and B, do not show evidence of different reddening values in the outburst dust continuum  with respect to the surrounding coma, which implies that we do not observe dust with different physical characteristics \citep{Bockelee2017,Rinaldi2018}. The color maps of both outbursts show a VIS color gradient at the maximum of the outburst ejecta with a value of 13.1 $\pm$ 1.3 \% (100 nm)$^{-1}$ for outburst A and a value of 12.4  $\pm$ 2.4 \% (100 nm)$^{-1}$ for outburst B.

\section{Additional Datasets}\label{results:Other}

\subsection{OSIRIS Images}\label{Instrument:OSRIS}\label{data:OSIRIS}
During the relevant time period there are 24 images taken by the Optical, Spectroscopic, and Infrared Remote Imaging System (OSIRIS) that are useful for comparing the observed outbursts to those described in \cite{vincent2016summer}.  The Narrow Angle Camera (NAC) has a field of view of 2.20$\times$2.22 degrees, which subtends approximately 9.0$\times$9.0 km at 67P's distance at 15:03 UTC and 8.8$\times$8.8 km for 19:18 UTC \citep{keller2007osiris}. Between 15:03 and 16:53 UTC on November 7 the OSIRIS NAC took 12 observations with exposure times of 0.675 seconds, long enough to capture cometary activity off of the comet's limb. There are no OSIRIS observations taken until November 9 after the 16:53 UTC observation on November 7. 

\begin{table}[]
    \centering
    \begin{tabular}{c|c|c}
        Observation Time  & Exposure Time (s) & Phase Angle ($^{\circ}$)\\
        \hline
         15:04:56 & 0.675 & 63.3 \\
         16:04:56 & 0.675 & 63.2 \\
         16:14:56 & 0.675 & 63.2 \\
         
    \end{tabular}
    \caption{OSIRIS Narrow Angle Camera observations used in this work.}
    \label{tab:osiris_obs}
\end{table}

\begin{figure}
    \begin{center}
        \includegraphics[trim = {5cm 0  5cm 0},width = 0.5\linewidth]{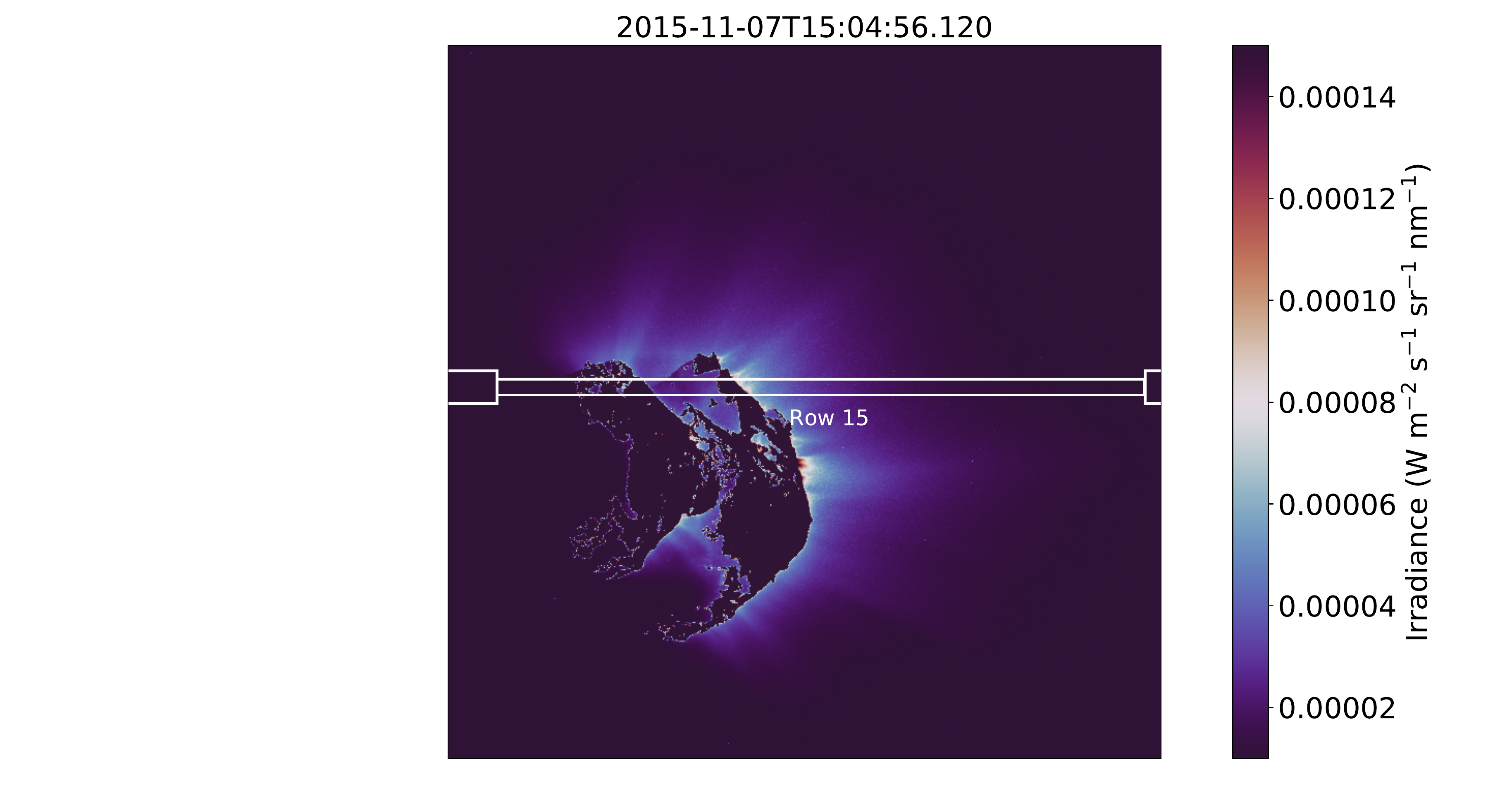}
        \includegraphics[trim = {5cm 0  5cm 0},width = 0.5\linewidth]{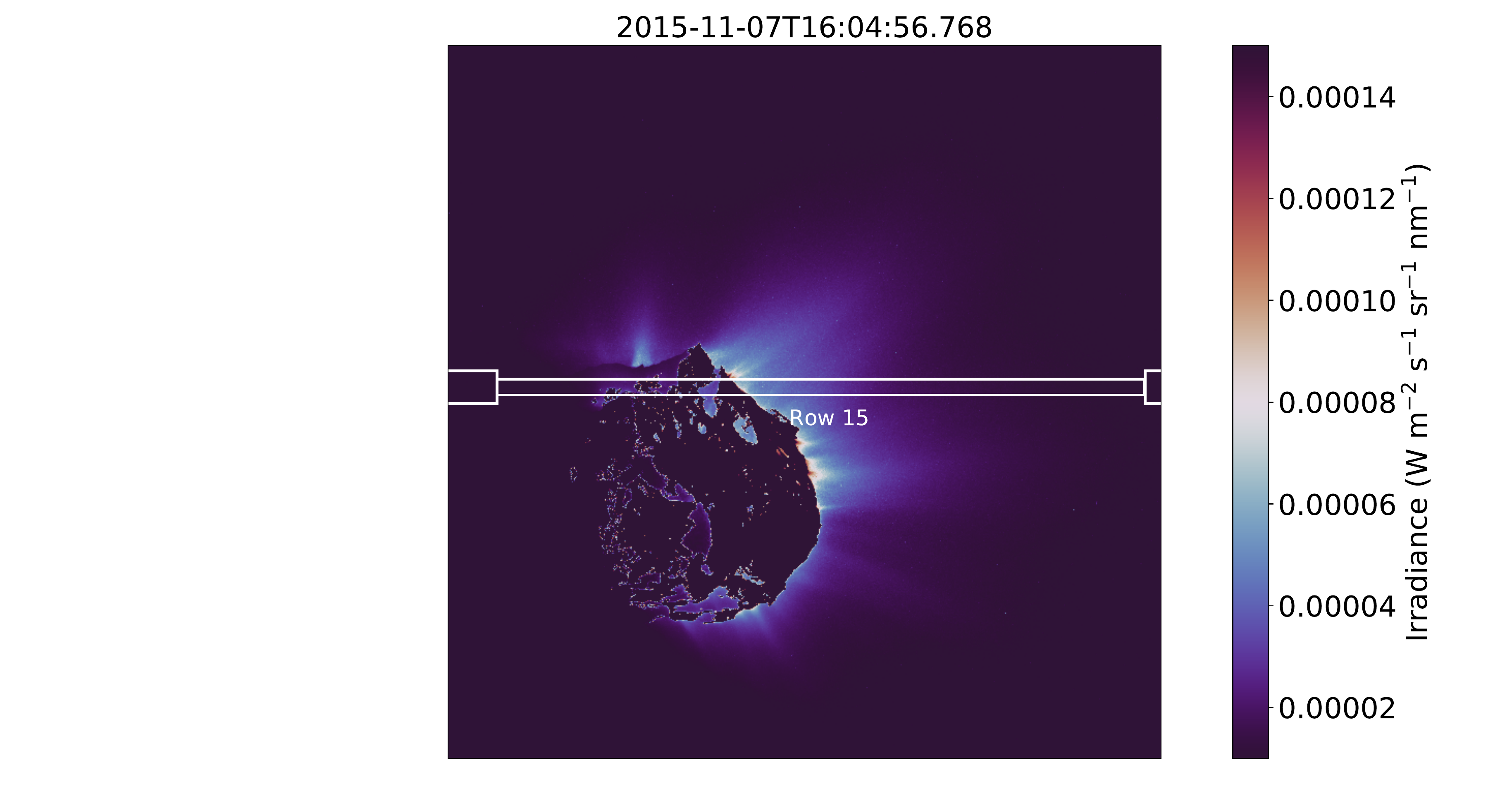}
        \includegraphics[trim = {5cm 0  5cm 0},width = 0.5\linewidth]{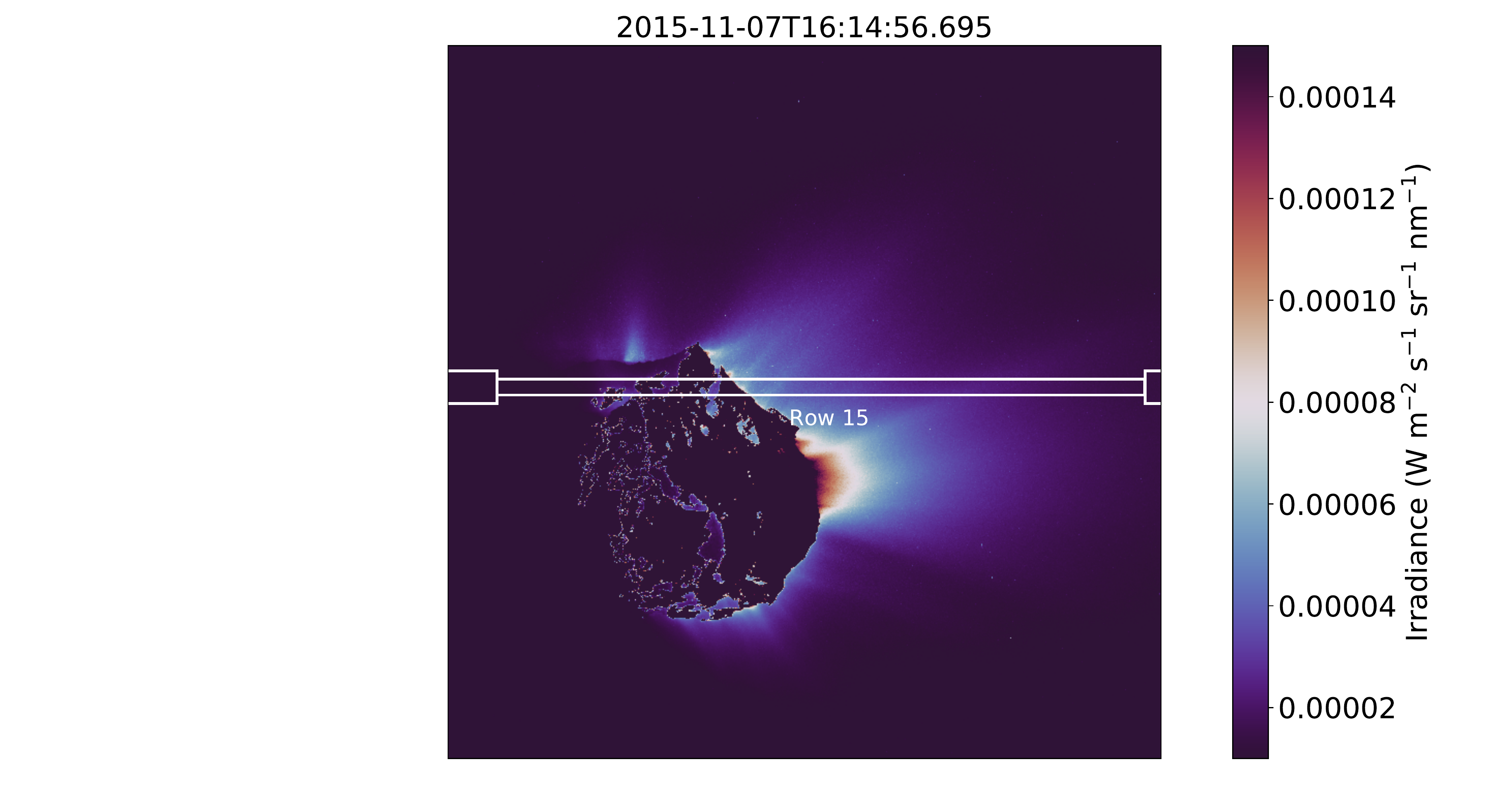}
    \end{center}
    \caption{Three OSIRIS NAC images with 0.675 s exposure times taken between 15:05 and 16:15 UTC on November 7. The overexposed nucleus has been masked in the images to highlight activity and the Sun is to the right in these images. Both the top and middle image are taken during periods of typical cometary activity. The top corresponds with the Alice spectrum taken at 15:04 UTC in Figure \ref{fig:stable_spectra}, which is used for Alice quiescent subtraction. The image in the middle was taken 15 minutes prior to the Alice outburst detected in the 16:07 UTC Alice spectrum and the outburst imaged at 16:15 by OSIRIS, which is displayed in the bottom image. The Alice slit is overlaid in white with the approximate location of row 15 identified.}
    \label{fig:Osiris_NAC_quiescent}
\end{figure}

Three of these observations are contemporary with key observations by Alice and VIRTIS-M and provide useful context (Fig. \ref{fig:Osiris_NAC_quiescent}). The 15:05 UTC  OSIRIS image taken within one minute of the Alice exposure at 15:05 UTC confirms that cometary activity is indeed quiescent at that period. This relatively quiet activity continued until at least 16:05 UTC (Fig. \ref{fig:Osiris_NAC_quiescent}b). The first outburst detected in the Alice and VIRTIS-M data occurred at some point in the next 3 minutes, because in the 16:15 UTC OSIRIS exposure a substantial amount of activity is detected (Fig. \ref{fig:Osiris_NAC_quiescent}c). This is in agreement with Alice data that shows a sharp increase in atomic emission at 16:07 UTC (Fig. \ref{fig:time_vs_brightness}). 

\begin{figure}
\begin{center}
\includegraphics[trim = {5cm 0  5cm 0},width = 0.75\linewidth]{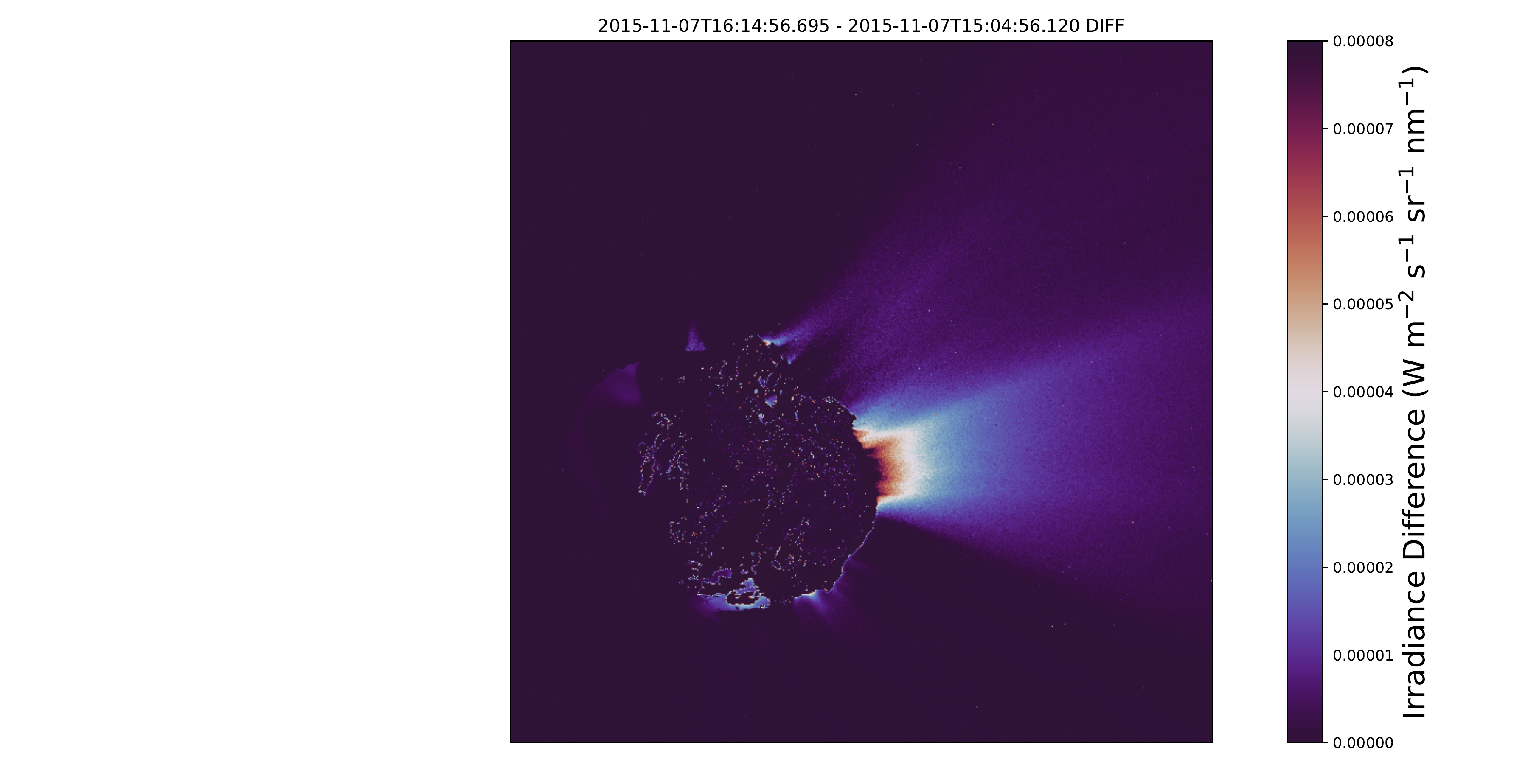}
\end{center}
\caption{Image displaying the change in irradiance between OSIRIS NAC observations at 16:04 and 16:13 UTC on November 7. The Sun is to the right in the image. The exposure taken at 16:04 UTC has been subtracted from the 16:14 UTC exposure to show the change in activity levels. Between the two images the comet nucleus rotated approximately 5\degr, leading to some artifacts in the difference image on the masked surface.  }
\label{fig:Osiris_NAC_diff}
\end{figure}

\begin{figure}
\begin{center}
        \includegraphics[width = 0.5\linewidth]{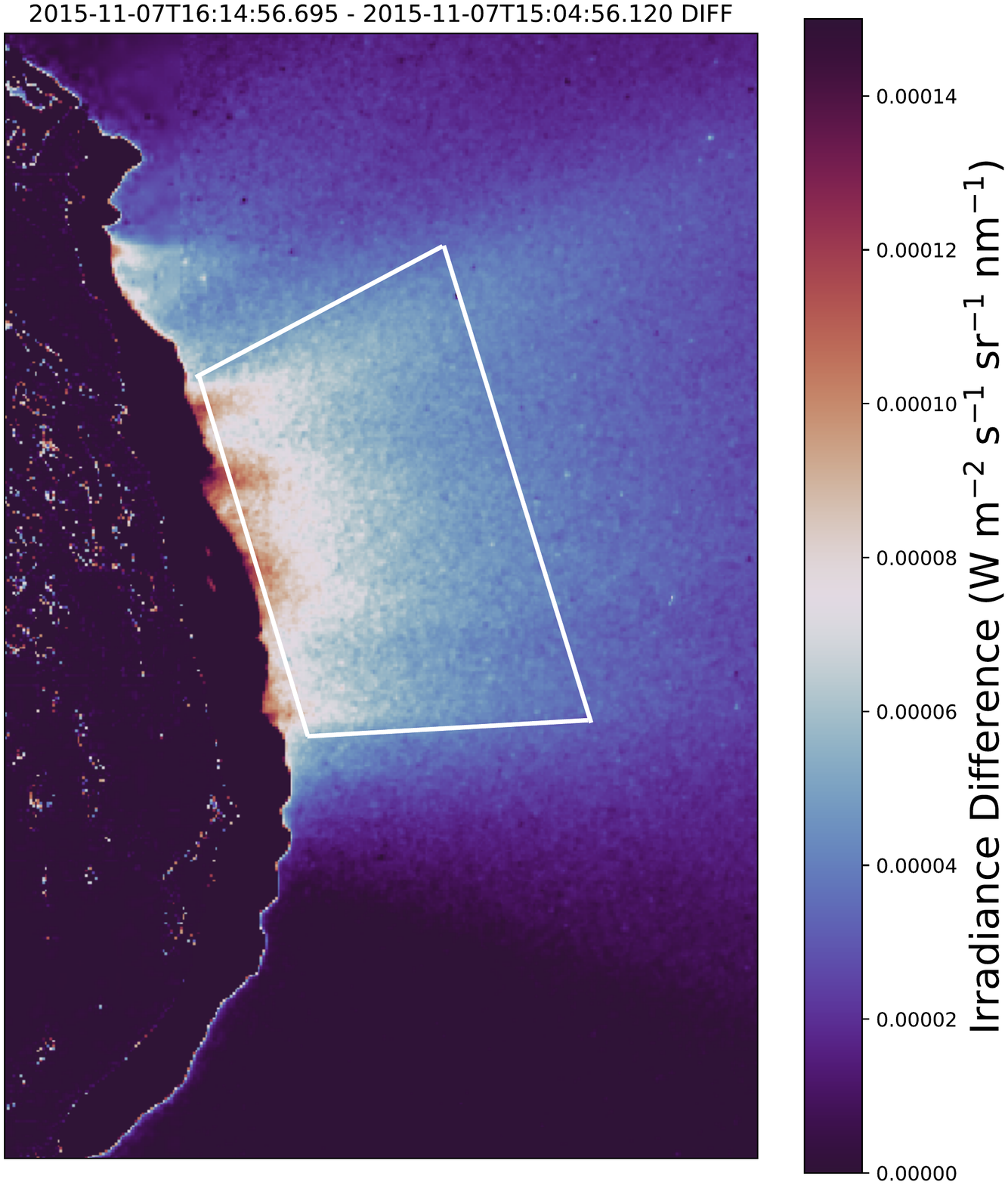}
\end{center}
    \caption{Image showing a close-up of the change in limb activity in Figure \ref{fig:Osiris_NAC_diff} with a shifted color scheme to highlight the outburst's fine structure. Note that four distinct areas are all active at 16:14 that were not at 16:04 UTC. The area integrated for outburst brightness as described in \cite{vincent2016summer} is shown by a white trapezoid.}
    \label{fig:Osiris_NAC_diff_zoom}
\end{figure}

The high resolution of the OSIRIS images allows us to examine the limb of the comet and the fine structure of the outburst. By first subtracting the 16:05 UTC image to create an irradiance difference image (Fig. \ref{fig:Osiris_NAC_diff}) and then enlarging the sunward limb of the nucleus (Fig. \ref{fig:Osiris_NAC_diff_zoom}) we can identify four new active sites on the limb of the nucleus that were not active ten minutes prior, at most. This suggests that outburst A is new activity and not the extension of existing active sites.   

For comparison of outburst A to outbursts previously studied by OSIRIS we completed the same calculations for relative intensity as described in \cite{vincent2016summer}. The integrated irradiance (W m$^{-2}$ sr$^{-1}$ nm$^{-1}$) was found by summing the individual pixel irradiances in a trapezoidal area extending between 50-500 meters from the comet limb. The total irradiance is then multiplied by 4$\pi r_{c}^{2}$, where r$_{c}$ is the spacecraft-comet distance, to calcuate the luminosity of the outburst. For outburst A we find that the luminosity was 1.12$\times$10$^{12}$ W,  an order of magnitude weaker than the strongest outburst recorded in \cite{vincent2016summer} but consistent with the other 33 outbursts described in Table 3 of that same paper. Without OSIRIS data for outburst B we can't calculate the relative intensity, but a simple scaling of the maximum radiance observed by VIRTIS-M for each outburst provides a reasonable estimate of the relative luminosity. In effect,
\begin{equation}
  \frac{R_{B,VM}}{R_{A,VM}} = \frac{L_{B,O}}{L_{A,O}} \longrightarrow \frac{R_{B,VM}}{R_{A,VM}}\times(L_{A,O}) = L_{B,O} 
\end{equation}
, where $R$ indicates the max radiance of outbursts A and B from VIRTIS in W m$^{-2}$ sr$^{-1}$ $ \mu$m$^{-1}$ and $L$ indicates the luminosity of the outbursts in the OSIRIS images in W. From this we find  (0.06/0.08$)\times$1.12$\times$10$^{12}$ W = 8.4$\times$10$^{11}$ W
, or 8\% of the maximum radiance of 1.18$\times$10$^{13}$ W. This is still greater than almost half of the outbursts reported in \cite{vincent2016summer}.  These OSIRIS context observations show that the outbursts observed by Alice, VIRTIS-M, and OSIRIS on November 7 are not unique compared to previously observed outbursts except in two ways: the instrument datasets available over the relevant time period and the distinct differences between the two outbursts themselves despite their close temporal proximity.

\subsection{NAVCAM Images}
\begin{figure}
\centering
\gridline{\rotatefig{-90}{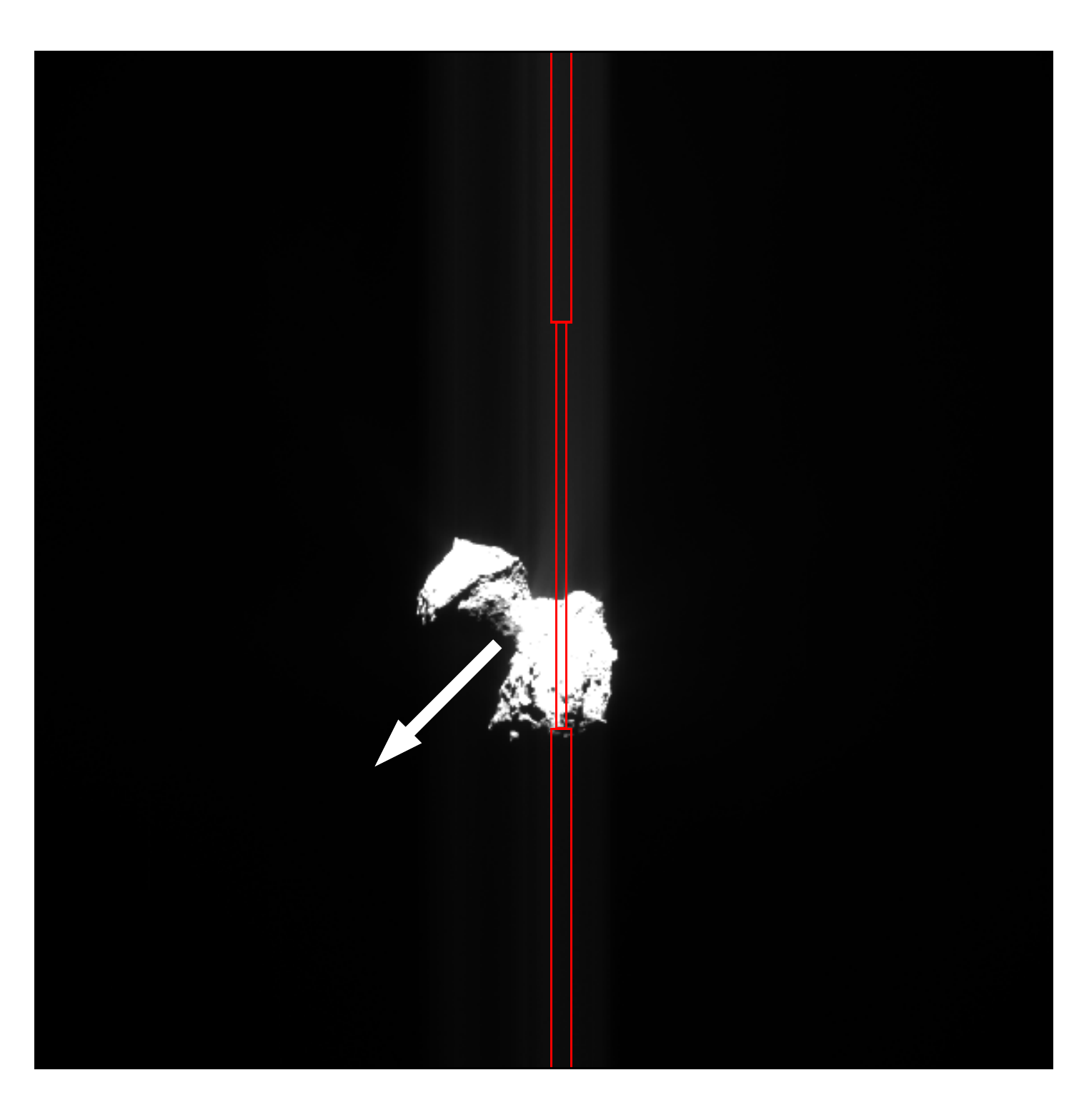}{0.33\linewidth}{a) NAVCAM}
\rotatefig{-90}{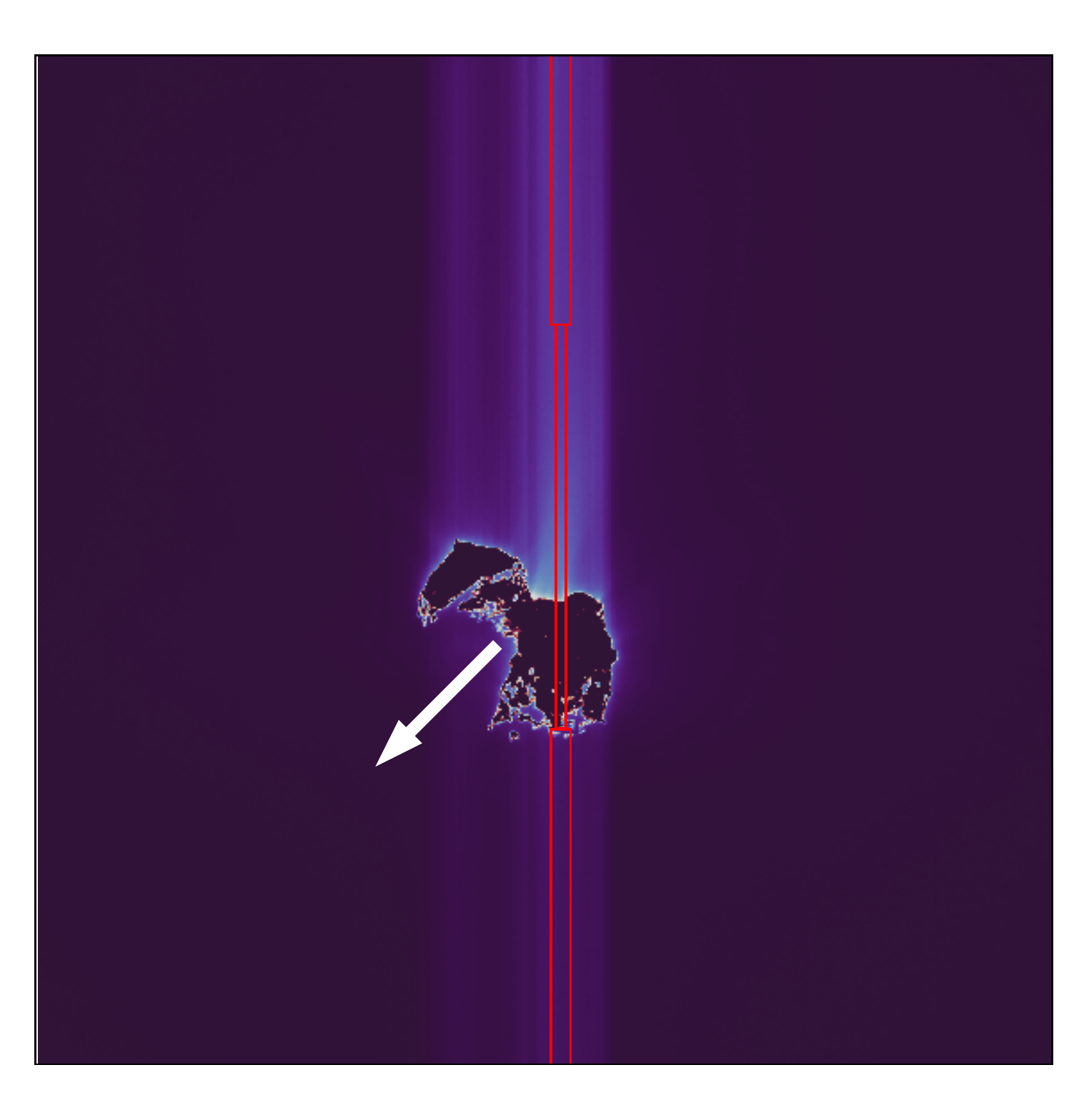}{0.33\linewidth}{b) NAVCAM, Nucleus Masked}\rotatefig{-90}{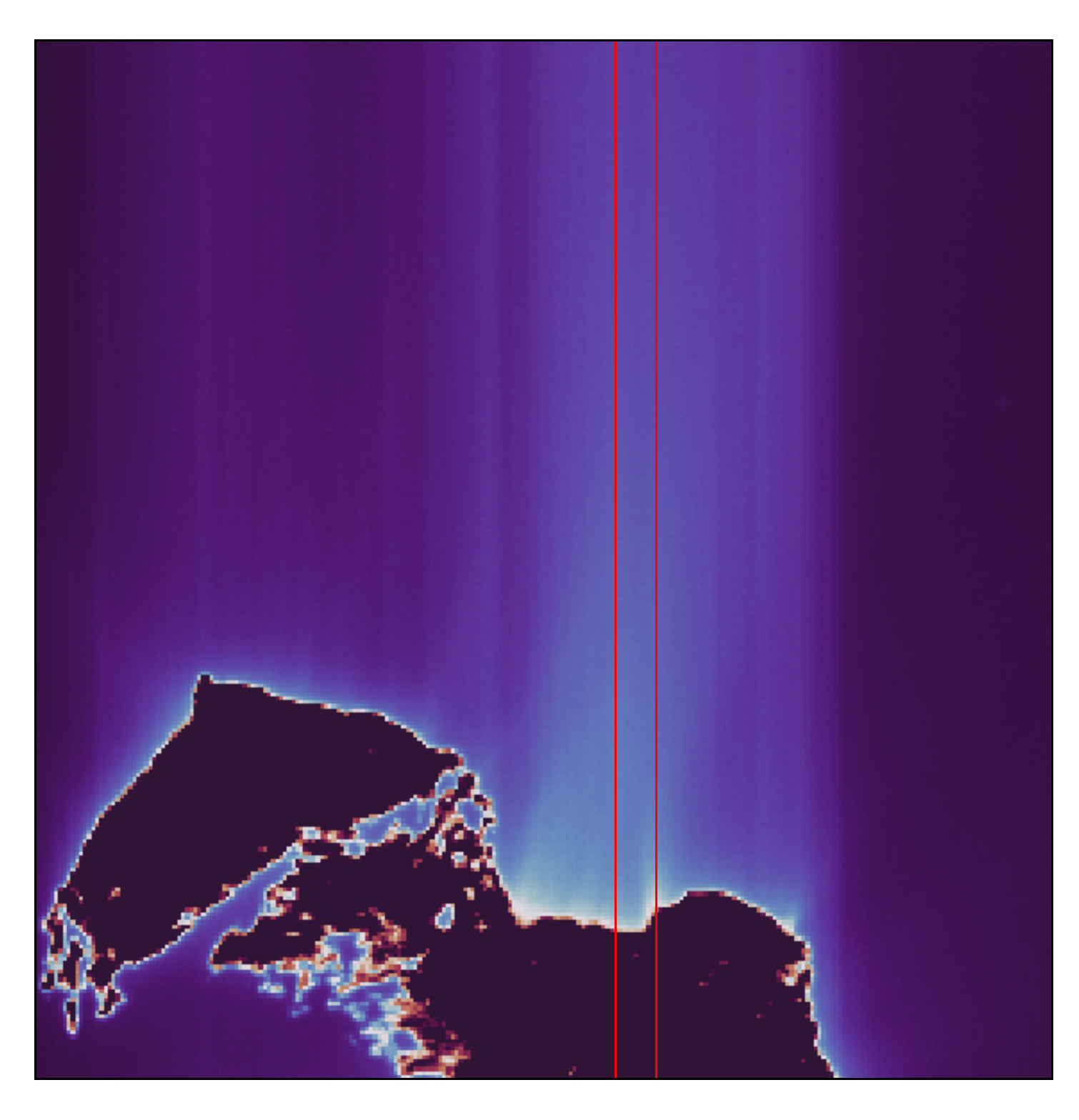}{0.33\linewidth}{c) Enlarged version of (b) to show detail}}
\caption{NAVCAM image from November 7 at 19:18 UTC, with Alice slit overlay in red. Subfigures b) and c) have been stretched to show cometary activity in the southern hemisphere. The slit was aligned with the activity at that time. The spectrum is shown in Figure \ref{fig:stable_spectra}.  
This geometry is representative of the stable pointing scheme. For this period the Alice slit subtends 22 km at the nucleus distance, approximately 1.2 km per pixel. The white vector in the image denotes the rotational axis of 67P. The Sun is to the right in all three images. }
\label{fig:NAVCAM_Jet_zoom}
\end{figure}

Of the three available NAVCAM images taken during the period in question only one shows increased cometary activity coincident with Alice observations of an emissions increase. These three NAVCAM images are shown in Figure \ref{fig:NAVCAM_Jet}. In the NAVCAM image at 19:18 UTC, a jet can be seen extending along the Alice slit (Figure \ref{fig:NAVCAM_Jet_zoom}). This particular NAVCAM image was taken at the same time as an Alice exposure, displayed in red in Figure \ref{fig:stable_spectra}, and informs us that the spectrum is not taken during a period of strong outbursting but jet activity. The morphology of the 19:18 NAVCAM image is unlike the outbursts detected by VIRTIS-M and observed in the OSIRIS images, and appears to be more collimated than the other activities. This is particularly useful for identifying differences in composition and excitation processes between outbursts and more common cometary jets.  

In addition, the alignment of the first jet with the Alice slit means that the spatial profiles shown in Figure \ref{fig:stable_spectra} can provide insight into dominant emission mechanisms within the jet itself and how they compare to small and large outbursts, as discussed in the next section.

\begin{figure}
\centering
\gridline{\fig{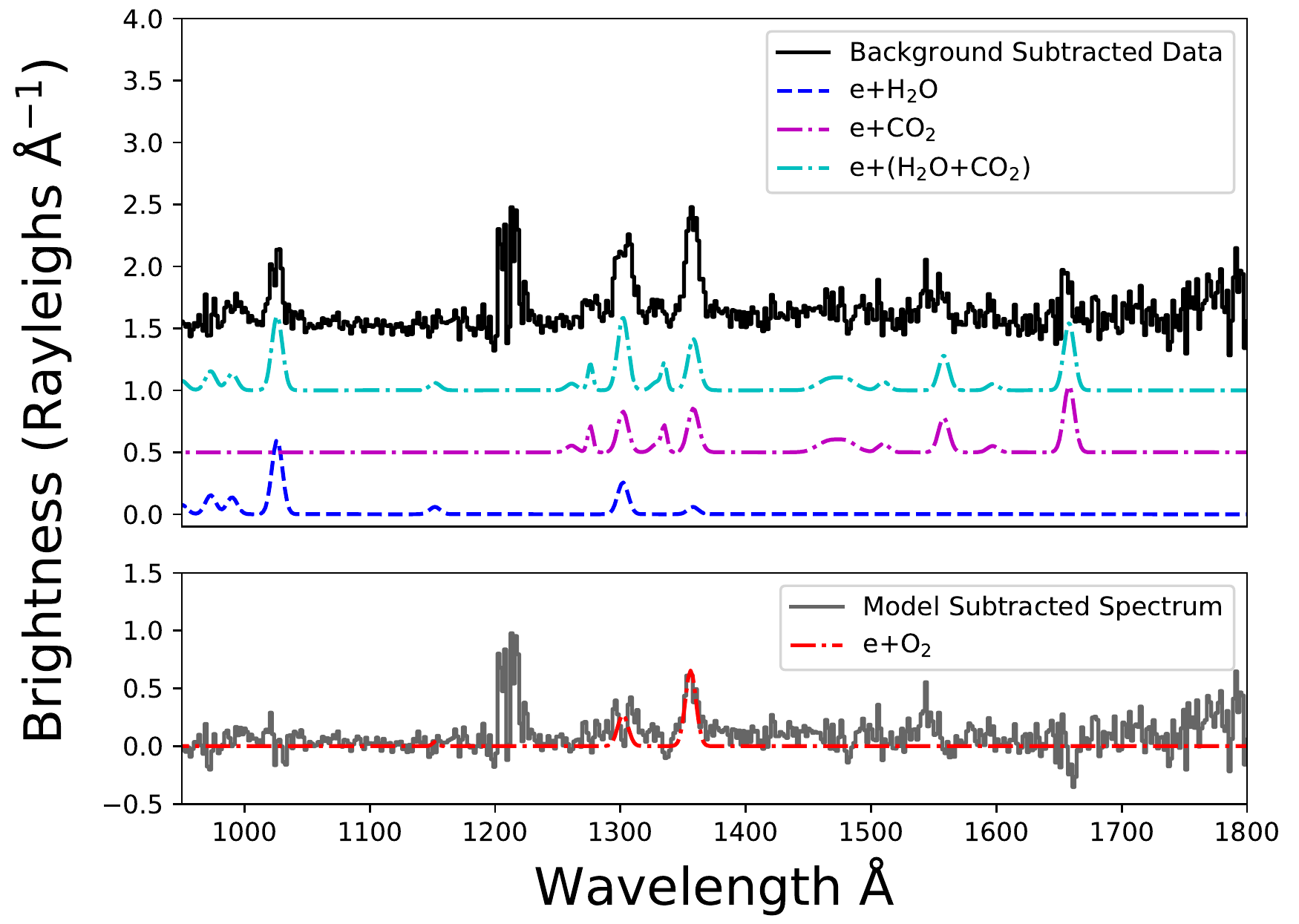}{0.5\linewidth}{a) 2015/11/07 16:07 UTC}}
\gridline{\fig{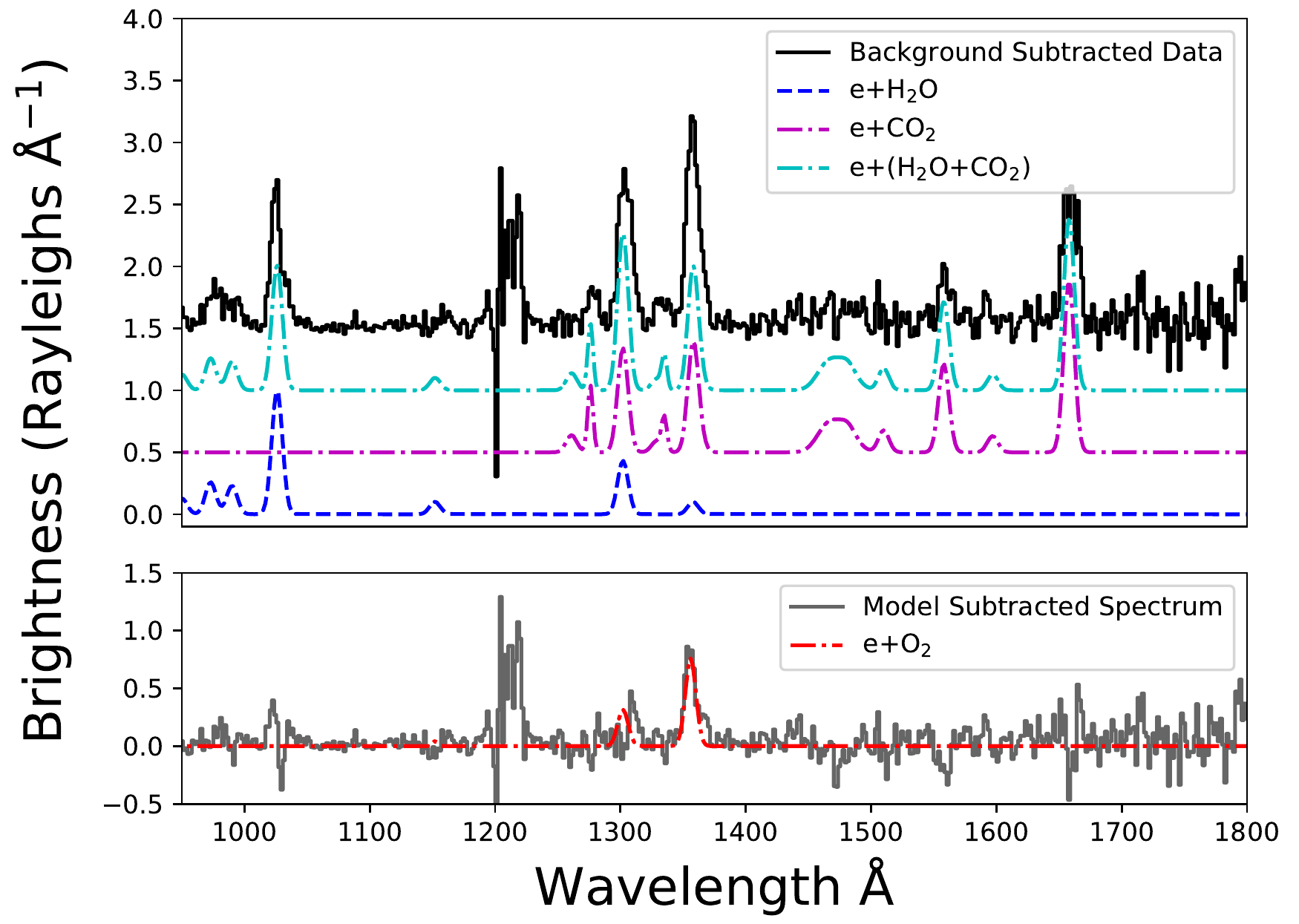}{0.5\linewidth}{b) 2015/11/07 17:32 UTC}}
\gridline{\fig{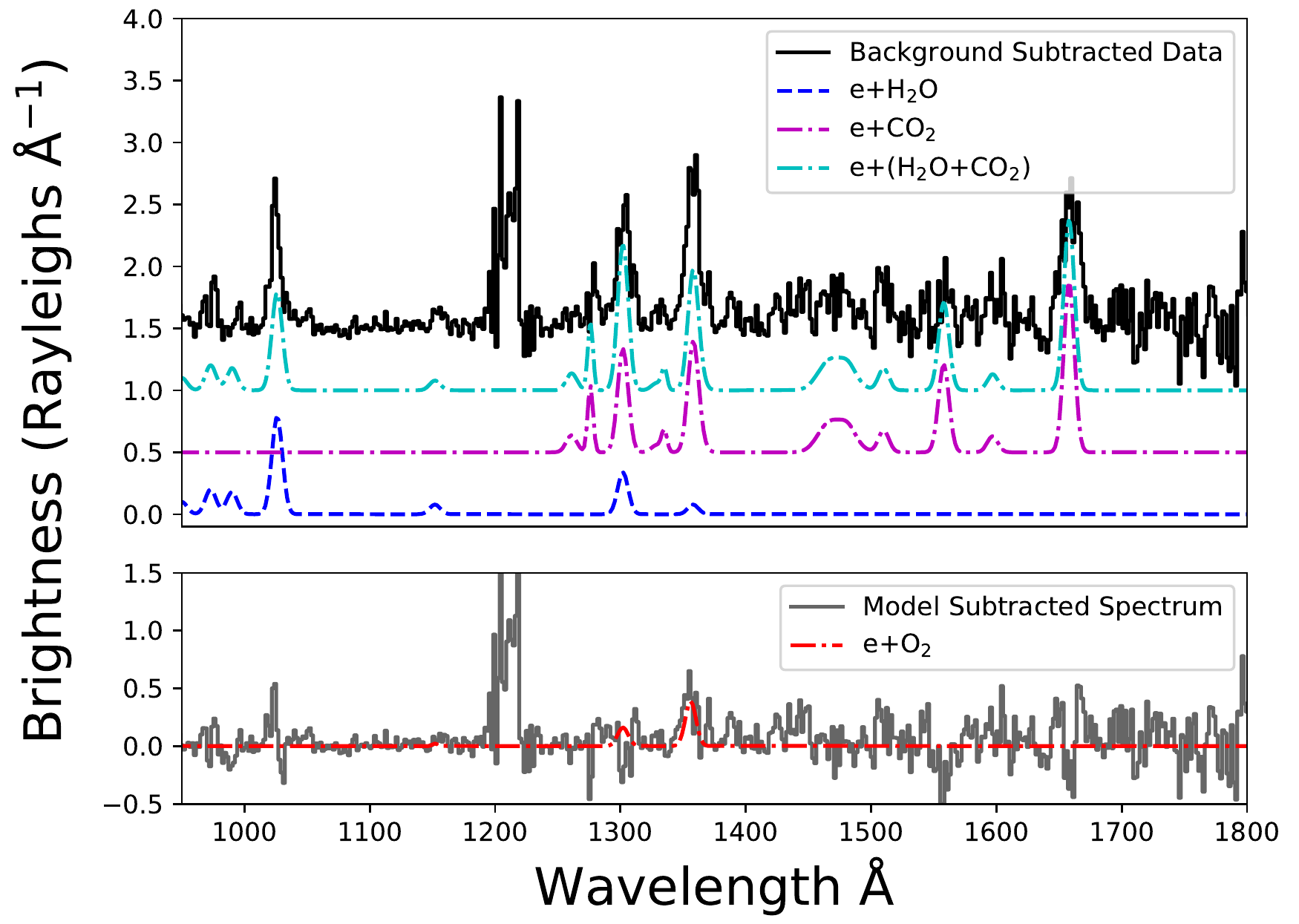}{0.5\linewidth}{c) 2015/11/07 19:18 UTC}}
\caption{Alice quiescent-subtracted spectra showing the spectrum near outburst A (a), outburst B (b), and the outburst and/or jet (c). The spectrum from 15:04 UTC shown in Figure \ref{fig:stable_spectra} is used as the quiescent spectrum. The modeled relative molecular abundances are reported in Table \ref{table:compositions}.}
\label{fig:e+mol_maps}
\end{figure}

%\begin{figure}
%\centering
%\gridline{\fig{e_impact_O2_fit_1607UTC7NOV.pdf}{0.5\linewidth}{2015/11/07 16:07 UTC}}
%\gridline{\fig{e_impact_O2_fit_1732UTC7NOV.pdf}{0.5\linewidth}{2015/11/07 17:32 UTC}}
%\gridline{\fig{e_impact_O2_fit_1918UTC7NOV.pdf}{0.5\linewidth}{2015/11/07 19:18 UTC}}
%\caption{Fits of e+\ce{O2} derived from \citet{kanik2003electron} to the residual spectrum from Figure \ref{fig:e+mol_maps} following the subtraction of the e+(\ce{CO2}+\ce{H2O}) synthetic spectrum.  Column densities are reported in Table \ref{table:compositions}. respectively. We note that the 19:18 UTC spectrum did not produce a detection and that the displayed e+\ce{O2} spectrum is to illustrate the upper limit on \ce{O2} column density.  }
%\label{fig:e+O2_maps}
%\end{figure}

\section{Discussion}\label{Discussion}
The observations from Alice and VIRTIS-M data show two outbursts of different magnitudes, both in gas and dust emissions, occurring within a short period followed by approximately 1.5 hours of elevated activity before Alice observations cease. Such an overlap presents the opportunity for dust and gas analysis and a discussion about the implications for cometary outburst mechanisms. 

\subsection{Excitation Mechanisms}\label{subsec:ex_pro}
The substantial presence of \ion{O}{1}] 1356 \AA\ emission and the spectral fit of dissociative electron impact emission of \ce{H2O} and \ce{CO2} makes it clear that the dominant atomic emission mechanism relevant for near-nucleus cometary activity is driven by electron-neutral interactions, whether jets or outbursts. The substantial increase of the extent of the spatial profile for the semi-forbidden emission of \ion{O}{1}] 1356 \AA\ in Figure \ref{fig:spatial_profiles} for the outbursts and jet shows that the near-nucleus plasma environment is critically important for understanding the UV emissions. Without simultaneous plasma measurements taken within the outbursts it is difficult to know how much of this change in dissociative electron impact emission is due to increases in the neutral density, electron density, or electron energy. Improved modeling of the plasma effects on the UV emissions from 67P has been executed for periods where \textit{Rosetta} was much closer to the nucleus, and electron distribution information from RPC-IES could be combined with ROSINA molecular abundances and VIRTIS or MIRO column densities to calculate emission rates \citep{galand2020far,stephenson2021multi}. For this set of observations we are without those datasets, and must fall back to models that can only fit the relative abundances of \ce{CO2} and \ce{O2} to \ce{H2O}. Such models rely on the emission line ratios derived from laboratory work with dissociative electron impact and are insensitive to the plasma properties near the nucleus.

Given the concurrent observations from VIRTIS-M indicating the substantial presence of dust it is useful to compare the activity in question to the dusty outburst of 2016 February 19 discussed in \citet{grun20162016} and \citet{hajra2017}. \cite{hajra2017} showed that the near-nucleus plasma environment experienced a 3-fold increase in total electron density while simultaneously experiencing a 2-9 fold decrease in electrons greater than 10 eV. If a dusty outburst released little new volatile material and neutral density remained constant while the average electron energy decreased and the ''cold" electron density increased, this would result in the calculation of a lower limit for column density assuming the 100 eV cross sections from the literature are implemented. Such a decrease to two of the three components for determining dissociative electron impact excitation rates would produce a significant decrease in dissociative electron impact excitation in Alice data. Put simply, comet dust outbursts should produce less dissociative electron impact excitation emissions than gas outbursts, most easily identifiable by the \ion{O}{1}] 1356 \AA\ emission feature. 

If dusty outburst A possessed similar plasma properties as that discussed in \citet{hajra2017}, Alice spectra appear consistent with the decrease in electron energy, and thus lower emissions than would be typically expected from a outburst. In contrast, the presence of high-threshold energy \ion{C}{2} 1335 \AA\ emission at 17:32 UTC in outburst B (Figure \ref{fig:e+mol_maps}) is interesting given the VIRTIS-M result that outburst B was a weaker dust outburst than outburst A, despite Alice measurements showing increased gas emissions. The weaker dust outburst B may not have damped the electron energies to nearly the same level as the stronger dust outburst A, made evident in the appearance of atomic emission features with larger threshold energies like \ion{C}{1} 1279 and \ion{C}{2} 1335 \AA, at 26 and 40 eV respectively \citep{ajello1971dissociative,ajello2019uv}. This suggests that gas and dust outbursts do experience different levels of dissociative electron impact emission and present unique spectral signatures in the UV, specifically of high-threshold energy atomic emission features. However, to be sure of the correlation more outbursts with similar gas/dust properties observed by both Alice and the plasma instruments on board \emph{Rosetta} must be analyzed.

\subsection{Gas Composition}\label{subsec:Composition}

Given the progression from weak outgassing at 12:00 UTC to elevated activity levels with a substantial jet at 19:18 UTC here we analyze the composition of three key spectra. To identify the unique composition of the activity these spectra have been quiescent subtracted using the 15:04 UTC spectrum, leaving emissions produced by the newly introduced neutrals from the respective outbursts or jet. These emissions can be fit with a relative abundance model for dissociative electron impact emission of \ce{H2O}, \ce{CO2}, \ce{CO}, and \ce{O2} at 100 eV \citep{makarov2004kinetic,ajello2019uv,mumma1972dissociative, ajello1971dissociative, ajello1971emission,kanik2003electron}. The model first determines the relative abundance of \ce{CO2}/\ce{H2O} by taking the line ratios of the Lyman-$\beta$ and \ion{C}{1} emission features. The relative abundances were then used to model the contribution from e+\ce{H2O} and e+\ce{CO2} to \ion{O}{1} emission features, using published line ratios from the aforementioned literature. These model spectra were then subtracted from the Alice data and the residual \ion{O}{1} emission features were fit with an \ce{O2} dissociative electron impact model. The calculated abundances for \ce{CO2} and \ce{O2} relative to \ce{H2O} for 16:07, 17:32, and 19:18 UTC are shown in Table \ref{table:compositions}.

This model is not without caveat. We make the assumption that atomic cross section ratios measured at 100 eV are accurate for the range of electron energies above each transition's threshold energies. This holds true for the main atomic emissions (Ly-$\beta$, \ion{O}{1} 1304, \ion{O}{1}] 1356, and \ion{C}{1} 1657), but not all observed Alice features. An extension of this assumption requires that atomic transitions that have higher threshold energies (e.g. \ion{C}{2} 1335 \AA) need to be properly addressed.  This is done via a scaling factor implemented in the fitting algorithm. Physically this is an attempt to capture the plasma environment's difference from the modeled Maxwellian distribution and the depletion of electrons with energies higher than threshold energy of certain atomic emissions. This model parameter is unitless and allowed to vary between 0 and 1, and in the fits shown in Figure \ref{fig:e+mol_maps} this parameter ranges between 0.1 and 0.25. 

\begin{table*}
\begin{center}
\begin{tabular}{c c c c c}
\hline
Observation ID & UTC Time  & \ce{CO2}/\ce{H2O} & \ce{O2}/\ce{H2O} \\
\hline
ra\_151107160746\_hisa\_lin  & 16:07 & 0.6 & 0.21 \\
ra\_151107173215\_hisa\_lin  & 17:32 & 1.0 & 0.14 \\
ra\_151107191814\_hisa\_lin  & 19:18 & 1.2 & $\lesssim$0.10 \\

\end{tabular}
\caption{Modelled relative abundances for \ce{H2O}, \ce{CO2}, and \ce{O2} for the quiescent-subtracted spectra shown in Figure \ref{fig:e+mol_maps}. The errorbars on these modelled relative abundances are between 30 and 35\%, the result of uncertainties in emission brightness, emission cross section ratio measurements in the literature, and to a lesser degree, the model fitting error. Errors are higher for lower Ly-$\beta$ brightnesses, as this increases the brightness measurement uncertainty for the determination of the baseline \ce{H2O}, by definition set to 1 to find the relative abundances of \ce{CO2} and \ce{O2}. }
\label{table:compositions}
\end{center}    
\end{table*}

 The calculated relative abundances for outburst A, outburst B, and the jet yield unique insight into the progression of cometary activity. At the peak of outburst A the \ce{CO2}/\ce{H2O} ratio was 0.6$\pm$0.2, while outburst B exhibits a different composition of \ce{CO2}/\ce{H2O}=1.0$\pm$0.3. These relative abundances are particularly interesting because the values, when paired with the brightness of the emission features, suggest that outburst B had a much stronger gas component than outburst A, despite the larger dust irradiances measured by VIRTIS-M for outburst A, and that there was a much stronger \ce{CO2} component for outburst B, indicating more volatile rich material. The elevated emission from \ce{CO2} continues after the outburst appears to be over; the jet has a \ce{CO2}/\ce{H2O} ratio of approximately 1.2$\pm$0.4 before Alice observations cease. This implies that the sublimation of \ce{H2O} is largely responsible for lifting the dust in the outbursts, not \ce{CO2}. This result requires confirmation through separate events, which in turn will require analysis of more outburst events in the Alice and VIRTIS-M catalog, as the implications for the interpretation of cometary activity are significant.
 
 We point out that there is excess \ion{C}{1} 1657~\AA\ in the 19:18 UTC spectrum that is being fit with the e+\ce{CO2} spectra with a strength of about 2-3 Rayleighs, contributing to an over-subtraction of the \ion{O}{1} 1304 \AA\ emission feature when subtracting the model from the observations. This 2-3 Rayleigh emission can partially be explained as the result of photodissociation and excitation of \ion{C}{1} from the newly introduced \ce{CO2} column, which has an excitation rate of $\sim$1.1$\times$ 10$ ^{-8}$ photons s$^{-1}$ molecule$^{-1}$ at 1 au \citep{wu1978atomic}. This produces a predicted integrated brightness of $\sim$3 Rayleighs for a column density of $\sim$3$\pm$0.7$\times10^{15}$. Without concurrent water column measurements from VIRTIS or MIRO to compare to, we must compare to the closest date available, when the MIRO instrument measured a total water column of 2$\times$10$^{16}$ cm$^{-2}$ just a few days later on 11 November 2015 \citep{Biver2019}. Using a direct comparison of the derived \ce{CO2} column density from the excess \ion{C}{1} 1657 \AA\ emission and the later water column measurement would lead to a \ce{CO2}/\ce{H2O} ratio of 0.15, but this would be comparing a quiescent-subtracted derived value to a total coma value, and is therefore misleading. To order of magnitude the required \ce{CO2} column density to explain the excess \ion{C}{1} 1657 \AA\ emission is near the measured water column density, and therefore any weak additional \ion{C}{1} 1657~\AA\ is likely from photodissociation of \ce{CO2}. All other \ion{C}{1} and \ion{C}{2} emission features experience a negligible amount of solar resonance scattering and are fit well with the e+\ce{CO2} synthetic spectrum for those two spectra. The final jet observation at 19:18 UTC shows residual emission of \ion{C}{1} 1657, 1561, and what may be weak CO Fourth Positive Group emission, specifically the (0-1), (0-0), and (1-0) bands between 1400 and 1600~\AA. We also note that these features may also represent low-energy electron impact on \ce{CO2}, described in \citet{ajello2019uv} but these  are not implemented in our models owing to their low contribution to the overall emission. The calculated \ce{CO2}/\ce{H2O} ratios of 0.57-1.5 fall within the range expected for the regional composition of the southern hemisphere of 67P from both Alice and ROSINA observations \citep{feldman2018fuv, mall2016high}.
 
 The \ce{O2} column density is more difficult to calculate. Both outburst A and outburst B show excess \ion{O}{1}] 1356 \AA\ emission in the e+(\ce{H2O} + \ce{CO2}) subtracted spectra that could be an indication of e+\ce{O2}, but the residuals are poorly fit by the electron impact model owing to lack of corresponding \ion{O}{1} 1304 \AA\ emission. However, given the lack of \ce{CO} Fourth Positive Group emissions there is not another likely neutral molecule that could be the source of the emissions, so we can cautiously treat this excess as an indicator of e+\ce{O2}. We can constrain the \ce{O2}/\ce{H2O} to be approximately 0.21$\pm$0.07 and 0.14$\pm$0.05 for outbursts A and B. By contrast there is little excess \ion{O}{1}] 1356 \AA\ emission detected in the model subtracted jet spectrum (Figure \ref{fig:e+mol_maps}). These \ce{O2}/\ce{H2O} abundances are consistent with previous Alice observations of outbursts and transient events \citep{feldman2016nature,noonan2018ultraviolet}, but substantially lower than values taken later on November 7 2015 \citep{Noonan2021spatial}. There are no atomic or molecular emissions present in the spectra that would suggest dissociative excitation of additional neutrals plays a significant role at this time, and therefore we can treat these particular outbursts as representative of a typical small scale cometary outburst. 

The post-outburst period for outbursts A and B are very different, offering insight into the different areas exposed during the course of each. Quiescent atomic emission levels are reached quickly after outburst A but following outburst B there is an increased brightness of atomic C and O features in Fig. \ref{fig:time_vs_brightness} that lasts at least until 19:18 UTC. While the compositions are relatively similar, this suggests that outburst B exposed fresh material with substantial \ce{H2O} and \ce{CO2}, which began to sublimate immediately. This may explain the lack of excess \ion{O}{1}] 1356 \AA\ in the 19:18 UTC jet observation; the more volatile \ce{O2} has been depleted, leaving a spectrum well fit by e+(\ce{H2O}+\ce{CO2}) (Fig. \ref{fig:e+mol_maps}). Therefore outburst A, which was substantially weaker in atomic emission, must have occurred in an area depleted in volatiles and was unable to sustain activity. Outburst B on the other hand, saw a sharp appearance of \ion{O}{1}] 1356 \AA\, possibly indicating \ce{O2}, followed by sustained sublimation of \ce{CO2} and \ce{H2O}. This would suggest \ce{O2} as the initiating volatile for a rapid outflow outburst model like that of \cite{skorov2016model}.

\begin{figure*}
\centering
 \includegraphics[width=0.9\textwidth,trim=0cm 3cm 0.0cm
	2cm,clip=true]{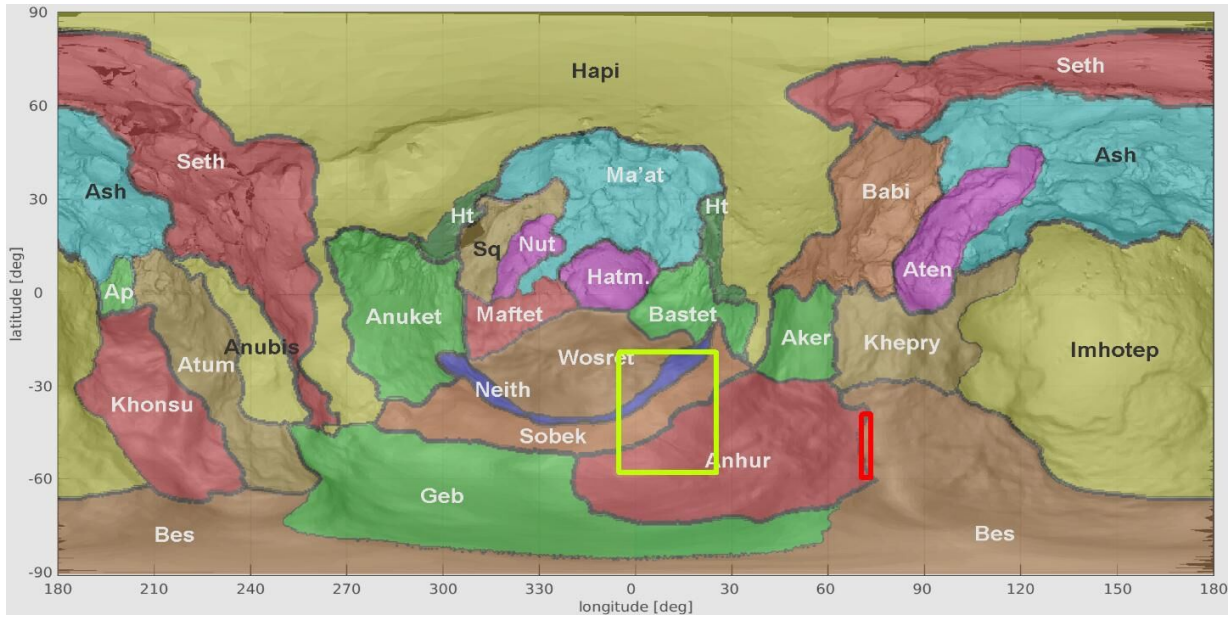}
    \caption{Source regions of the outbursts detected by VIRTIS-M. The red  and yellow rectangles correspond to the outbursts A and B, respectively. The map is centered on the small lobe, the big lobe covers the left-hand and right-hand side of the map, and the
contact area between the two lobes covers mainly the top of the map (regions Hapi and Seth). The boundary regions, shown in the map, have been defined by \citet{ElMaarry2015, ElMaarry2016}}
    \label{fig:map}
\end{figure*}

\subsection{Outburst Location}
\label{sec:location}

Fig. \ref{fig:map} shows the outburst source regions projected on a morphological map of 67P/CG displaying the region boundaries defined by  \citet{ElMaarry2015,ElMaarry2016}. 
In the VIRTIS-M images analyzed in this work, the dust ejecta are seen only in one image, and the source location of the outbursts is not visible. Hence, we can only roughly identify the source location. We selected the pixels inside the outburst ejecta and determined the projected latitude and longitude on the surface at the intersection with the vector passing through the center of the target (tangent point). 
Both outburst sources are located in the Southern hemisphere, approximately at the latitude range from -20$^\circ$ to -80$^\circ$. 
The location of outburst A is in the boundary region, between Anhur and Bes. Such boundary areas are characterised by discontinuities in the local terrain, either textural or topographic as observed in \citet{Vincent2016} and \citet{Fornasier2018}. The location of outburst B is in the boundary region, between Anhur, Neith, Wosret and Sobek. These regions are located in the  southern part of the neck. This region is considerably more complex texturally than in the north.  Neith is bounded by Wosret on one side and Sobek on the other. It forms the major steep cliff from an edge (the Neith-Wosret boundary) down into the neck itself. The surface is very rough on intermediate scales. There do not appear to be
any large scale structures \citep{Thomas2018}. Neither of these projected source regions has a previously detected outburst from \cite{vincent2016summer} within their bounds, but the Anhur region has been classified as very active and volatile-rich \citep{Fornasier2019a,Fornasier2019b}.

\subsection{Comparing Alice and VIRTIS-M}
\begin{figure}
\centering
 \includegraphics[width=0.9\textwidth]{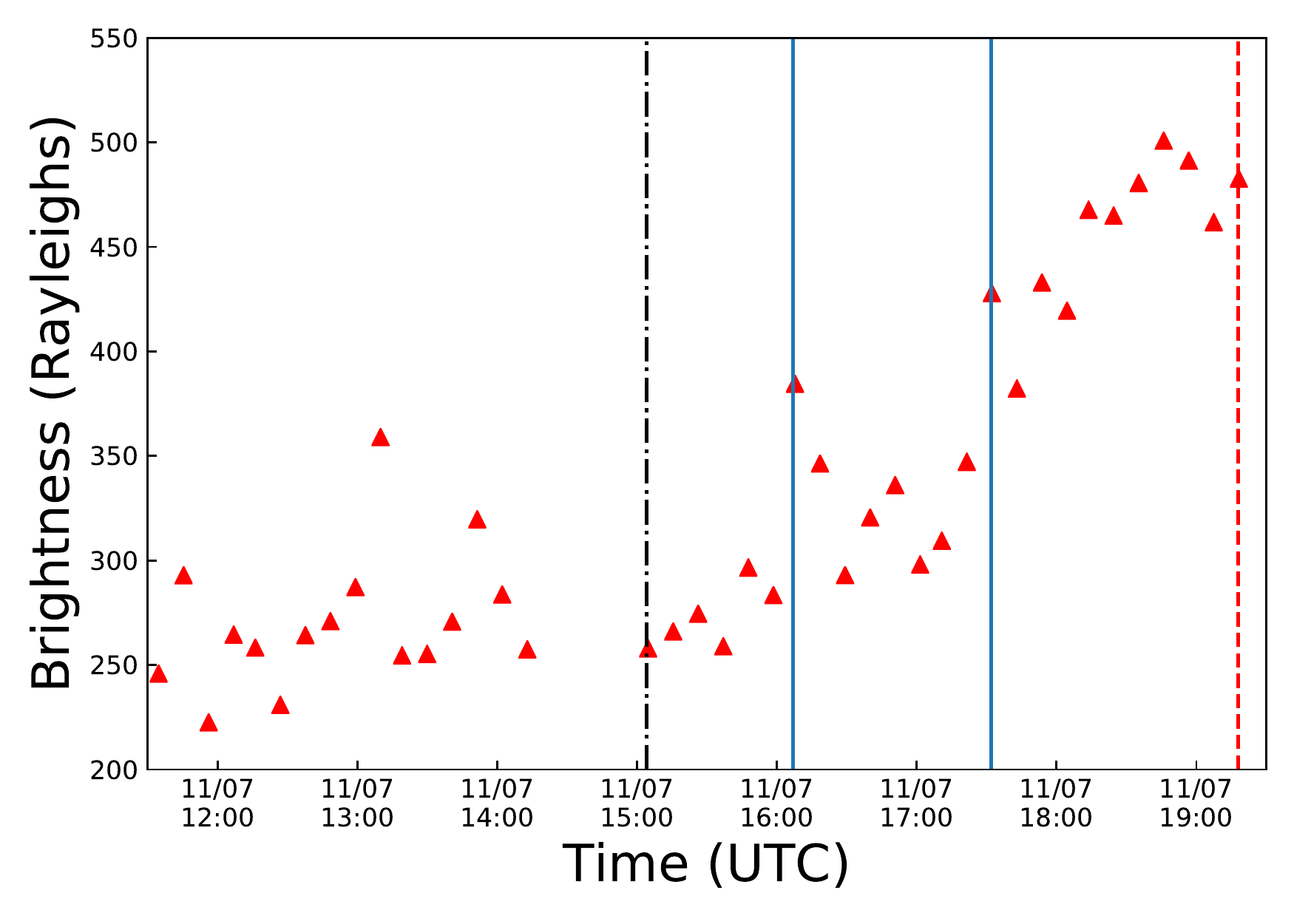}
    \caption{Brightness in Rayleighs for row 16 of the Alice detector, integrated between 1850 and 1950 \AA . Vertical lines represent quiescent (black dot-dash), outburst (blue solid), and jet (red dashed) observations. For this period of stable pointing row 16 was just off of the nucleus limb and only contains solar continuum emission reflected off dust. Of particular note are the $\sim$36\% and  $\sim$22\% increases in brightnesses over the previous continuum measurements for outbursts A and B, respectively. Error bars on each measurement are smaller than the size of the points. }
    \label{fig:cont_lightcurve}
\end{figure}

Alice and VIRTIS-M observations are largely divided into gas and dust characterization, but close inspection of the overlapping characteristics of the dust components are warranted with this dataset. Here we discuss three different methods to compare dust reflected solar continuum between the UV and visible observations. If we assume that the reflected UV light is directly proportional to the reflected visible light (VIRTIS-M 5500 \AA) (Table \ref{tab:VM_observations}), referencing the SORCE solar continuum for November 7 from the the LISIRD database\footnote{http://lasp.colorado.edu/lisird/data/sorce\_ssi\_l3/}, and that the visible and UV albedos are similar, we find that the total dust contribution to the continuum between 1850-1950 \AA\ should be approximately $\sim$93 and 70 Rayleighs for outbursts A and B, respectively. These are consistent with the changes in brightness observed in the nearest row to the nucleus limb of 100 and 80 Rayleighs for outbursts A and B (Figure \ref{fig:cont_lightcurve}). These zeroth order estimates could be further refined by addressing albedo and phase function variations as the UV albedo of 67P has been measured to have a strong blue slope, leading to a lower albedo at wavelengths between 1850 and 1950 \AA\ \citep{feaga2015far}. The UV albedo of the comet surface red-ward of 1830 \AA\ is not well characterized, so we can try the other edge case and assume that the albedo between 1800 and 1830 \AA, 0.03 at zero-degree phase, is a good proxy for the UV dust reflectance. The ratio of UV to visible albedo then needs to be accounted for in the expected dust contribution to the continuum. Using the VIRTIS-M albedo at 5500 \AA, 0.06 \citep{Capaccioni2015}, we get a ratio of 0.5 and an expected Alice continuum brightnesses of 47 and 35 Rayleighs for outbursts A and B, respectively. These values both underpredict the observed brightness, so a more representative albedo must be found via a different method.

As an alternative we can derive the albedo from the measured Alice and VIRTIS-M brightnesses and the VIRTIS-M albedo to find UV/visible albedos ratios of 1.07 and 1.14, corresponding to UV albedos of 0.065 and 0.069 when scaling the albedo of 0.06 at 5500 \AA\ from \cite{Capaccioni2015}.  These slightly increased FUV albedos for outbursts A and B may represent the increased presence of water ice grains, which have higher UV albedos than carbonaceous material redward of 1700 \AA\  \citep{hendrix2010ultraviolet,Steffl2015}. However, a strong decrease in color slope was not observed in the VIRTIS-M dust data, which would be expected if a large number of large water ice grains ($\sim$mm-size) were incorporated into the outburst plume \citep{calvin1995spectra}. For smaller grains this effect is much less pronounced, as the albedo is nearly constant between 5000 and 8000 \AA\ for 20 $\mu$m size ice grains.  Outbursts characterized from OSIRIS data by \citet{Fornasier2019b} exhibited a blue slope, albeit in the visible wavelength range, and were inferred to contain icy grains. 

\subsubsection{Hybrid Gas-Dust Outbursts}
The early division of Alice outbursts into gas and dust types was observationally driven, with gas outbursts lacking dust components and dust outbursts lacking a gas component \citep{feldman2016nature,Steffl2015}. Hybrid outbursts have been identified with Alice observations \citep{Steffl2018}, and here we build on the concept. What the combined VIRTIS-M and Alice observations here portray is a range of outburst types bracketed by gas outbursts at one end and dust outbursts at the other. Previously characterized Alice gas outbursts may have had a weak FUV solar continuum reflectance that was not initially recognized owing to its limited brightness and distribution from the nucleus limb, which would then lead to a new classification in this combined gas/dust outburst type. 

\cite{feldman2016nature} interpreted the gas outbursts observed by Alice as evidence of an outburst mechanism relying on a deepening fracture exposing volatile gases, likely driven by \ce{O2}, rather than CO as described by \cite{skorov2016model}. Conversely, other outbursts were linked to cliff collapses and geological failure \citep{steckloff2016rotationally,vincent2016summer,pajola2017pristine}. The outbursts studied in this work do not fit neatly into either classification with the lack of direct observations of the landscape during the outburst due to the large cometocentric distance, making it difficult to directly identify the mechanical process that caused either outburst. However, there are several key components of the outbursts that provide clues. 
\begin{enumerate}
    \item Alice measurements of atomic emissions indicative of dissociating neutrals and VIRTIS-M VIS maps of reflected solar continuum from dust, are inversely correlated for outbursts A and B.
    \item UV spectra for outburst A are depleted in high-threshold energy atomic emissions from e+\ce{CO2}, a sign that the local plasma environment is depleted in electrons above 25 eV.
    \item Following outburst B there is sustained sublimation observed by Alice, indicating exposure of fresh volatiles.
    \item Analysis of the FUV continuum shows that there is good agreement with the expected brightness from VIRTIS-M measurements of the max radiance for outburst A with the measured UV comet albedo of 0.041, but outburst B is unexpectedly bright in the FUV and is better matched with an albedo of 0.069, possibly indicative of the presence of water ice grains in the outburst plume. However, substantial color differences were not detected between the plumes and surrounding coma in the VIRTIS data, 
    \item The source regions for each outburst lie in the Anhur region, a known outburst region with consolidated material and large discontinuities. 
\end{enumerate}
These results point to possible mechanisms for each outburst. Outburst A requires a smaller gas contribution, a large dust contribution, and cannot spur further sublimation from the exposed region. This suggests that structural failure of a largely volatile-depleted cliff is a likely mechanism for outburst A; the crumbling cliff face produces a large quantity of dust but has already been largely depleted of volatiles and did not expose substantial new material. The lack of high-threshold energy emissions also corroborates previous plasma observations of a cliff collapse outburst \citep{hajra2017}, which show an overall increase in total electron density but a depletion of electrons greater than 40 eV. The Anhur region has had substantial activity, both jets and outbursts, and has substantial volatile content \citep{Fornasier2019a,Fornasier2019b}.

Outburst B requires a smaller dust contribution that is higher in UV albedo, implying the presence of water ice grains, and a larger gas contribution without damping the near-nucleus plasma environment to preserve high-threshold energy atomic emissions from e+\ce{CO2}. These conditions are more closely aligned with the outbursts described in \cite{feldman2016nature} and by the fracture propogation model of \cite{skorov2016model} than outburst A. However, we have two other conditions that need to be addressed: the sustained activity following the outburst and the presence of water ice grains. Sustained activity, especially with the substantial amount of \ce{CO2} observed, requires a volatile-rich region to be exposed. The water ice grains could be the result of \ce{CO2} gas sublimating deep in the fracture dragging icy grains from the interior out with the flow before the grains begin to sublimate as well. This mechanism could also explain the relatively flat extension of the Ly-$\beta$ and \ion{O}{1} 1304 \AA\ spatial profile in Figure \ref{fig:spatial_profiles} for 17:32 and 19:38 UTC; an extended source of \ce{H2O} has become available, one that was not present in outburst A at 16:07 UTC. 

Without direct observations of the source regions during the outbursts a multi-instrument approach contains a wealth of information that can be used to constrain the outburst mechanisms. Further cross-analysis of as of yet unidentified outbursts in the Alice and VIRTIS datasets should yield improved spectral criteria to identify outburst mechanisms remotely.  

\section{Summary}\label{Summary}
In this paper we analyzed three distinct activity types observed within five hours on 2015 November 7: a small outburst, a large outburst, and a cometary jet. The following results were shown from the Alice, OSIRIS, and VIRTIS-M datasets regarding the outbursts on November 7:
\begin{enumerate}
    \item Alice, OSIRIS, and VIRTIS-M observations of outbursts indicated that these are not uniquely strong or compositionally distinct and are therefore likely representative of other outbursts at 67P. 
    \item Outbursts A and B display both gas and dust components and are therefore not entirely ``gas" nor ``dust" outbursts as previously described in Alice literature \citep{feldman2016nature,Steffl2015} but are better classified as ``hybrid" outbursts \citep{Steffl2018}.
    \item The dust color as measured by VIRTIS-M is 13.1\% (100 nm)$^{-1}$ and 12.4\% (100 nm)$^{-1}$ for outbursts A and B respectively. 
    \item Outburst A likely originated in the far east Anhur region, outburst B likely originated in the boundary regions of the southern neck, between Worset, Neith, Sobek and Anhur.
    \item Alice observations of the gas components show that outburst B was approximately twice as strong based on atomic emissions, which is inverse of the VIRTIS-M measured dust irradiances. VIRTIS-M observations of the dust component show that outburst A had a maximum radiance approximately 2.5$\times$ that of outburst B. 
    \item Alice spectra taken during outburst A show a lower \ce{CO2}/\ce{H2O} ratio (0.6) than in outburst B (1.0). Outburst A also has a higher \ce{O2}/\ce{H2O}, 0.21$\pm$0.07 compared to 0.14$\pm$0.05 for outburst B. 
    \item Comparison of the Alice FUV continuum between 1850 and 1950 \AA\ to VIRTIS-M dust irradiances shows that each outburst produced enough FUV reflectance to be detected in the Alice data, once investigated closely. 
    \item Outburst A is likely the result of structural surface feature failure (i.e. mass wasting). Outburst B contains elevated \ce{CO2}, indicating a more pristine surface origin (i.e. fracture deepening) and sustains increased activity for over two hours.  
    \item The jet resulting from outburst B has a moderate \ce{CO2}/\ce{H2O} ratio (1.2) and appears depleted of \ce{O2} compared to earlier activity, evidence that \ce{O2} may have initiated the outburst and exposed new volatile-rich material that sublimated \ce{CO2} and \ce{H2O} at least until 19:18 UTC.
\end{enumerate}
Analysis of these outbursts shows that mixed gas and dust outbursts have features within their FUV and visible spectra that can help constrain the initiating outburst mechanisms. Obtaining spectra of the relevant near-nucleus coma in the future will be difficult without improved space-based UV-capable observatories or spacecraft, to say nothing of the temporal resolution required to properly identify outbursts. Even identifying outbursts in the optical wavelength ranges with imaging, where sensitivities are significantly higher, is difficult and requires both high-cadence and high-sensitivity observations \citep[Kelley et al. in press] {Boehnhardt2016wendelstein,knight2017gemini,Farnham2019tess,farnham2021narrowband}. However, the \textit{Rosetta} mission dataset contains many more events to analyze. The outbursts discussed in this article are just two out of the dozens still requiring analysis in the \textit{Rosetta} datasets awaiting further interrogation. The large dataset from the Alice spectrograph on \textit{Rosetta} is unique, with nearly constant observation of a near-nucleus coma in the UV for two and a half years, providing a cometary UV dataset unlikely to be equaled for years to come. Multi-instrument analysis of cometary outbursts has already proven to be of critical importance for understanding outburst mechanics and there is much more to be done with the data available. Cometary outbursts require further investigations to verify spectroscopic characteristics, both within the \textit{Rosetta} datasets and in theoretical work.

\begin{acknowledgements}

\textit{Acknowledgements.} This work was made possible thanks to the ESA/NASA \textit{Rosetta} mission with contributions from ESA member states and NASA. The Alice team would like to acknowledge the support of NASA's Jet Propulsion Laboratory, specifically through contract 1336850 to the Southwest Research Institute. JWN and JWP would also like to acknowledge funding from NASA's \textit{Rosetta} Data Analysis Program through grant number 80NSSC19K1304. JWN would also like to acknowledge Peter Stephenson for insightful discussion regarding dissociative electron impact environments and modeling. The team also acknowledges the useful comments from the reviewer, which improved this manuscript.

We thank the following institutions and agencies for support of this work: Italian Space Agency (ASI, Italy) contract number I/024/12/1, center National d'\'Etudes Spatiales (CNES, France), DLR (Germany), NASA (USA) Rosetta Program, and Science and Technology Facilities Council (UK). VIRTIS was built by a consortium, which includes Italy, France, and Germany, under the scientific responsibility of the Istituto di Astrofisica e Planetologia Spaziali of INAF, Italy, which also guides the scientific operations. The VIRTIS instrument development, led by the prime contractor Leonardo-Finmeccanica (Florence, Italy), has been funded and managed by ASI, with contributions from Observatoire de Meudon financed by CNES, and from DLR. We thank the Rosetta Science Ground Segment and the Rosetta Mission Operations center for their support throughout all the phases of the mission. The VIRTIS calibrated data will be available through the ESA's Planetary Science Archive Website (www.rssd.esa.int) and is available upon request until posted to the archive.\\
     
\end{acknowledgements} 

\bibliographystyle{aasjournal}

\bibliography{uv_mapping,biblio_VIRTIS}

\begin{thebibliography}{}
\expandafter\ifx\csname natexlab\endcsname\relax\def\natexlab#1{#1}\fi
\providecommand{\url}[1]{\href{#1}{#1}}

\bibitem[{{Acton}(1996)}]{Acton1996}
{Acton}, C.~H. 1996, \planss, 44, 65

\bibitem[{Agarwal {et~al.}(2017)Agarwal, Della~Corte, Feldman, Geiger,
  Merouane, Bertini, Bodewits, Fornasier, Gr{\"u}n, Hasselmann,
  {et~al.}}]{agarwal2017evidence}
Agarwal, J., Della~Corte, V., Feldman, P., {et~al.} 2017, \mnras, 469, s606

\bibitem[{Ajello \& Franklin(1985)}]{ajello1985study}
Ajello, J., \& Franklin, B. 1985, \jcp, 82, 2519

\bibitem[{Ajello {et~al.}(2019)Ajello, Malone, Evans, Holsclaw, Hoskins, Jain,
  McClintock, Liu, Veibell, Deighan, {et~al.}}]{ajello2019uv}
Ajello, J., Malone, C., Evans, J., {et~al.} 2019, \jgr: Space Physics

\bibitem[{Ajello(1971{\natexlab{a}})}]{ajello1971dissociative}
Ajello, J.~M. 1971{\natexlab{a}}, \jcp, 55, 3156

\bibitem[{Ajello(1971{\natexlab{b}})}]{ajello1971emission}
---. 1971{\natexlab{b}}, \jcp, 55, 3169

\bibitem[{{Ammannito} {et~al.}(2006){Ammannito}, {Filacchione}, {Coradini},
  {Capaccioni}, {Piccioni}, {de Sanctis}, {Dami}, \& {Barbis}}]{Ammannito2006}
{Ammannito}, E., {Filacchione}, G., {Coradini}, A., {et~al.} 2006, Review of
  Scientific Instruments, 77, 093109

\bibitem[{{Belton} {et~al.}(2008){Belton}, {Feldman}, {A'Hearn}, \&
  {Carcich}}]{Belton2008}
{Belton}, M.~J.~S., {Feldman}, P.~D., {A'Hearn}, M.~F., \& {Carcich}, B. 2008,
  \icarus, 198, 189

\bibitem[{Biver {et~al.}(2019)Biver, Bockel{\'e}e-Morvan, Hofstadter, Lellouch,
  Choukroun, Gulkis, Crovisier, Schloerb, Rezac, Von~Allmen,
  {et~al.}}]{Biver2019}
Biver, N., Bockel{\'e}e-Morvan, D., Hofstadter, M., {et~al.} 2019, \aap, 630,
  A19

\bibitem[{Bockel{\'e}e-Morvan {et~al.}(2017)Bockel{\'e}e-Morvan, Rinaldi,
  Erard, Leyrat, Capaccioni, Drossart, Filacchione, Migliorini, Quirico,
  Mottola, {et~al.}}]{bockelee2017comet}
Bockel{\'e}e-Morvan, D., Rinaldi, G., Erard, S., {et~al.} 2017, \mnras, 469,
  S443

\bibitem[{{Bockel{\'e}e-Morvan} {et~al.}(2017){Bockel{\'e}e-Morvan}, {Rinaldi},
  {Erard}, {Leyrat}, {Capaccioni}, {Drossart}, {Filacchione}, {Migliorini},
  {Quirico}, {Mottola}, {Tozzi}, {Arnold}, {Biver}, {Combes}, {Crovisier},
  {Longobardo}, {Blecka}, \& {Capria}}]{Bockelee2017}
{Bockel{\'e}e-Morvan}, D., {Rinaldi}, G., {Erard}, S., {et~al.} 2017, \mnras,
  469, S443

\bibitem[{{Bockel{\'e}e-Morvan} {et~al.}(2019){Bockel{\'e}e-Morvan}, {Leyrat},
  {Erard}, {Andrieu}, {Capaccioni}, {Filacchione}, {Hasselmann}, {Crovisier},
  {Drossart}, {Arnold}, {Ciarniello}, {Kappel}, {Longobardo}, {Capria}, {De
  Sanctis}, {Rinaldi}, \& {Taylor}}]{Bockelee2019}
{Bockel{\'e}e-Morvan}, D., {Leyrat}, C., {Erard}, S., {et~al.} 2019, \aap, 630,
  A22

\bibitem[{Bodewits {et~al.}(2016)Bodewits, Lara, A\textsc{\char13}Hearn,
  La~Forgia, Gicquel, Kovacs, Knollenberg, Lazzarin, Lin, Shi,
  {et~al.}}]{bodewits2016changes}
Bodewits, D., Lara, L.~M., A\textsc{\char13}Hearn, M.~F., {et~al.} 2016, \aj,
  152, 130

\bibitem[{{Boehnhardt} {et~al.}(2016){Boehnhardt}, {Riffeser}, {Kluge}, {Ries},
  {Schmidt}, \& {Hopp}}]{Boehnhardt2016wendelstein}
{Boehnhardt}, H., {Riffeser}, A., {Kluge}, M., {et~al.} 2016, \mnras, 462, S376

\bibitem[{Calvin {et~al.}(1995)Calvin, Clark, Brown, \&
  Spencer}]{calvin1995spectra}
Calvin, W.~M., Clark, R.~N., Brown, R.~H., \& Spencer, J.~R. 1995, Journal of
  Geophysical Research: Planets, 100, 19041

\bibitem[{{Capaccioni} {et~al.}(2015){Capaccioni}, {Coradini}, {Filacchione},
  {Erard}, {Arnold}, {Drossart}, {De Sanctis}, {Bockelee-Morvan}, {Capria},
  {Tosi}, {Leyrat}, {Schmitt}, {Quirico}, {Cerroni}, {Mennella}, {Raponi},
  {Ciarniello}, {McCord}, {Moroz}, {Palomba}, {Ammannito}, {Barucci},
  {Bellucci}, {Benkhoff}, {Bibring}, {Blanco}, {Blecka}, {Carlson}, {Carsenty},
  {Colangeli}, {Combes}, {Combi}, {Crovisier}, {Encrenaz}, {Federico}, {Fink},
  {Fonti}, {Ip}, {Irwin}, {Jaumann}, {Kuehrt}, {Langevin}, {Magni}, {Mottola},
  {Orofino}, {Palumbo}, {Piccioni}, {Schade}, {Taylor}, {Tiphene}, {Tozzi},
  {Beck}, {Biver}, {Bonal}, {Combe}, {Despan}, {Flamini}, {Fornasier},
  {Frigeri}, {Grassi}, {Gudipati}, {Longobardo}, {Markus}, {Merlin}, {Orosei},
  {Rinaldi}, {Stephan}, {Cartacci}, {Cicchetti}, {Giuppi}, {Hello}, {Henry},
  {Jacquinod}, {Noschese}, {Peter}, {Politi}, {Reess}, \&
  {Semery}}]{Capaccioni2015}
{Capaccioni}, F., {Coradini}, A., {Filacchione}, G., {et~al.} 2015, Science,
  347, aaa0628

\bibitem[{Chaufray {et~al.}(2017)Chaufray, Bockel{\'e}e-Morvan, Bertaux, Erard,
  Feldman, Capaccioni, Schindhelm, Leyrat, Parker, Filacchione,
  {et~al.}}]{chaufray2017rosetta}
Chaufray, J.-Y., Bockel{\'e}e-Morvan, D., Bertaux, J.-L., {et~al.} 2017,
  \mnras, 469, S416

\bibitem[{Clark {et~al.}(2015)Clark, Broiles, Burch, Collinson, Cravens, Frahm,
  Goldstein, Goldstein, Mandt, Mokashi, {et~al.}}]{clark2015suprathermal}
Clark, G., Broiles, T., Burch, J., {et~al.} 2015, \aap, 583, A24

\bibitem[{{Coradini} {et~al.}(2007){Coradini}, {Capaccioni}, {Drossart},
  {Arnold}, {Ammannito}, {Angrilli}, {Barucci}, {Bellucci}, {Benkhoff},
  {Bianchini}, {Bibring}, {Blecka}, {Bockelee-Morvan}, {Capria}, {Carlson},
  {Carsenty}, {Cerroni}, {Colangeli}, {Combes}, {Combi}, {Crovisier}, {De
  Sanctis}, {Encrenaz}, {Erard}, {Federico}, {Filacchione}, {Fink}, {Fonti},
  {Formisano}, {Ip}, {Jaumann}, {Kuehrt}, {Langevin}, {Magni}, {McCord},
  {Mennella}, {Mottola}, {Neukum}, {Palumbo}, {Piccioni}, {Rauer}, {Saggin},
  {Schmitt}, {Tiphene}, \& {Tozzi}}]{Coradini2007}
{Coradini}, A., {Capaccioni}, F., {Drossart}, P., {et~al.} 2007, \ssr, 128, 529

\bibitem[{{El-Maarry} {et~al.}(2015){El-Maarry}, {Thomas}, {Giacomini},
  {Massironi}, {Pajola}, {Marschall}, {Gracia-Bern{\'a}}, {Sierks}, {Barbieri},
  {Lamy}, {Rodrigo}, {Rickman}, {Koschny}, {Keller}, {Agarwal}, {A'Hearn},
  {Auger}, {Barucci}, {Bertaux}, {Bertini}, {Besse}, {Bodewits}, {Cremonese},
  {Da Deppo}, {Davidsson}, {De Cecco}, {Debei}, {G{\"u}ttler}, {Fornasier},
  {Fulle}, {Groussin}, {Gutierrez}, {Hviid}, {Ip}, {Jorda}, {Knollenberg},
  {Kovacs}, {Kramm}, {K{\"u}hrt}, {K{\"u}ppers}, {La Forgia}, {Lara},
  {Lazzarin}, {Lopez Moreno}, {Marchi}, {Marzari}, {Michalik}, {Naletto},
  {Oklay}, {Pommerol}, {Preusker}, {Scholten}, {Tubiana}, \&
  {Vincent}}]{ElMaarry2015}
{El-Maarry}, M.~R., {Thomas}, N., {Giacomini}, L., {et~al.} 2015, \aap, 583,
  A26

\bibitem[{{El-Maarry} {et~al.}(2016){El-Maarry}, {Thomas}, {Gracia-Bern{\'a}},
  {Pajola}, {Lee}, {Massironi}, {Davidsson}, {Marchi}, {Keller}, {Hviid},
  {Besse}, {Sierks}, {Barbieri}, {Lamy}, {Koschny}, {Rickman}, {Rodrigo},
  {A'Hearn}, {Auger}, {Barucci}, {Bertaux}, {Bertini}, {Bodewits}, {Cremonese},
  {Da Deppo}, {De Cecco}, {Debei}, {G{\"u}ttler}, {Fornasier}, {Fulle},
  {Giacomini}, {Groussin}, {Gutierrez}, {Ip}, {Jorda}, {Knollenberg}, {Kovacs},
  {Kramm}, {K{\"u}hrt}, {K{\"u}ppers}, {Lara}, {Lazzarin}, {Lopez Moreno},
  {Marschall}, {Marzari}, {Naletto}, {Oklay}, {Pommerol}, {Preusker},
  {Scholten}, {Tubiana}, \& {Vincent}}]{ElMaarry2016}
{El-Maarry}, M.~R., {Thomas}, N., {Gracia-Bern{\'a}}, A., {et~al.} 2016, \aap,
  593, A110

\bibitem[{{Farnham} {et~al.}(2019){Farnham}, {Kelley}, {Knight}, \&
  {Feaga}}]{Farnham2019tess}
{Farnham}, T.~L., {Kelley}, M. S.~P., {Knight}, M.~M., \& {Feaga}, L.~M. 2019,
  \apjl, 886, L24

\bibitem[{Farnham {et~al.}(2021)Farnham, Knight, Schleicher, Feaga, Bodewits,
  Skiff, \& Schindler}]{farnham2021narrowband}
Farnham, T.~L., Knight, M.~M., Schleicher, D.~G., {et~al.} 2021, The Planetary
  Science Journal, 2, 7

\bibitem[{{Farnham} {et~al.}(2007){Farnham}, {Wellnitz}, {Hampton}, {Li},
  {Sunshine}, {Groussin}, {McFadden}, {Crockett}, {A'Hearn}, {Belton},
  {Schultz}, \& {Lisse}}]{Farnham2007}
{Farnham}, T.~L., {Wellnitz}, D.~D., {Hampton}, D.~L., {et~al.} 2007, \icarus,
  187, 26

\bibitem[{Feaga {et~al.}(2015)Feaga, Protopapa, Schindhelm, Stern,
  A\textsc{\char13}Hearn, Bertaux, Feldman, Parker, Steffl, \&
  Weaver}]{feaga2015far}
Feaga, L.~M., Protopapa, S., Schindhelm, E., {et~al.} 2015, \aap, 583, A27

\bibitem[{Feldman {et~al.}(2011)Feldman, Steffl, Parker, A’Hearn, Bertaux,
  Stern, Weaver, Slater, Versteeg, Throop, {et~al.}}]{feldman2011rosetta}
Feldman, P.~D., Steffl, A.~J., Parker, J.~W., {et~al.} 2011, Icarus, 214, 394

\bibitem[{Feldman {et~al.}(2015)Feldman, A\textsc{\char13}Hearn, Bertaux,
  Feaga, Parker, Schindhelm, Steffl, Stern, Weaver, Sierks,
  {et~al.}}]{feldman2015measurements}
Feldman, P.~D., A\textsc{\char13}Hearn, M.~F., Bertaux, J.-L., {et~al.} 2015,
  \aap, 583, A8

\bibitem[{Feldman {et~al.}(2016)Feldman, A\textsc{\char13}Hearn, Feaga,
  Bertaux, Noonan, Parker, Schindhelm, Steffl, Stern, \&
  Weaver}]{feldman2016nature}
Feldman, P.~D., A\textsc{\char13}Hearn, M.~F., Feaga, L.~M., {et~al.} 2016,
  \apj Letters, 825, L8

\bibitem[{Feldman {et~al.}(2018)Feldman, A\textsc{\char13}Hearn, Bertaux,
  Feaga, Keeney, Knight, Noonan, Parker, Schindhelm, Steffl,
  {et~al.}}]{feldman2018fuv}
Feldman, P.~D., A\textsc{\char13}Hearn, M.~F., Bertaux, J.-L., {et~al.} 2018,
  \aj, 155, 9

\bibitem[{{Filacchione}(2006)}]{Filacchionethesis2006}
{Filacchione}, G. 2006, PhD thesis, (Universita degli Studi di Napoli Federico
  II)

\bibitem[{{Fornasier} {et~al.}(2018){Fornasier}, {Hoang}, {Hasselmann},
  {Feller}, {Barucci}, {Deshapriya}, {Sierks}, {Naletto}, {Lamy}, {Rodrigo},
  {Koschny}, {Davidsson}, {Agarwal}, {Barbieri}, {Bertaux}, {Bertini},
  {Bodewits}, {Cremonese}, {Da Deppo}, {Debei}, {De Cecco}, {Deller},
  {Ferrari}, {Fulle}, {Gutierrez}, {Guttler}, {Ip}, {Keller}, {K{\"u}ppers},
  {La Forgia}, {Lara}, {Lazzarin}, {Lin}, {Lopez Moreno}, {Marzari}, {Mottola},
  {Pajola}, {Shi}, {Toth}, \& {Tubiana}}]{Fornasier2018}
{Fornasier}, S., {Hoang}, V.~H., {Hasselmann}, P.~H., {et~al.} 2018, ArXiv
  e-prints, arXiv:1809.03997

\bibitem[{{Fornasier} {et~al.}(2019{\natexlab{a}}){Fornasier}, {Hoang},
  {Hasselmann}, {Feller}, {Barucci}, {Deshapriya}, {Sierks}, {Naletto}, {Lamy},
  {Rodrigo}, {Koschny}, {Davidsson}, {Agarwal}, {Barbieri}, {Bertaux},
  {Bertini}, {Bodewits}, {Cremonese}, {Da Deppo}, {Debei}, {De Cecco},
  {Deller}, {Ferrari}, {Fulle}, {Gutierrez}, {G{\"u}ttler}, {Ip}, {Keller},
  {K{\"u}ppers}, {La Forgia}, {Lara}, {Lazzarin}, {Lin}, {Lopez Moreno},
  {Marzari}, {Mottola}, {Pajola}, {Shi}, {Toth}, \& {Tubiana}}]{Fornasier2019a}
---. 2019{\natexlab{a}}, \aap, 630, A7

\bibitem[{{Fornasier} {et~al.}(2019{\natexlab{b}}){Fornasier}, {Feller},
  {Hasselmann}, {Barucci}, {Sunshine}, {Vincent}, {Shi}, {Sierks}, {Naletto},
  {Lamy}, {Rodrigo}, {Koschny}, {Davidsson}, {Bertaux}, {Bertini}, {Bodewits},
  {Cremonese}, {Da Deppo}, {Debei}, {De Cecco}, {Deller}, {Ferrari}, {Fulle},
  {Gutierrez}, {G{\"u}ttler}, {Ip}, {Jorda}, {Keller}, {Lara}, {Lazzarin},
  {Lopez Moreno}, {Lucchetti}, {Marzari}, {Mottola}, {Pajola}, {Toth}, \&
  {Tubiana}}]{Fornasier2019b}
{Fornasier}, S., {Feller}, C., {Hasselmann}, P.~H., {et~al.}
  2019{\natexlab{b}}, \aap, 630, A13

\bibitem[{{Galand} {et~al.}(2020){Galand}, {Feldman}, {Bockel{\'e}e-Morvan},
  {Biver}, {Cheng}, {Rinaldi}, {Rubin}, {Altwegg}, {Deca}, {Beth},
  {Stephenson}, {Heritier}, {Henri}, {Parker}, {Carr}, {Eriksson}, \&
  {Burch}}]{galand2020far}
{Galand}, M., {Feldman}, P.~D., {Bockel{\'e}e-Morvan}, D., {et~al.} 2020,
  Nature Astronomy, 4, 1084

\bibitem[{{Gr{\"u}n} {et~al.}(2016){Gr{\"u}n}, {Agarwal}, {Altobelli},
  {Altwegg}, {Bentley}, {Biver}, {Della Corte}, {Edberg}, {Feldman}, {Galand},
  {Geiger}, {G{\"o}tz}, {Grieger}, {G{\"u}ttler}, {Henri}, {Hofstadter},
  {Horanyi}, {Jehin}, {Kr{\"u}ger}, {Lee}, {Mannel}, {Morales}, {Mousis},
  {M{\"u}ller}, {Opitom}, {Rotundi}, {Schmied}, {Schmidt}, {Sierks},
  {Snodgrass}, {Soja}, {Sommer}, {Srama}, {Tzou}, {Vincent},
  {Yanamandra-Fisher}, {A'Hearn}, {Erikson}, {Barbieri}, {Barucci}, {Bertaux},
  {Bertini}, {Burch}, {Colangeli}, {Cremonese}, {Da Deppo}, {Davidsson},
  {Debei}, {De Cecco}, {Deller}, {Feaga}, {Ferrari}, {Fornasier}, {Fulle},
  {Gicquel}, {Gillon}, {Green}, {Groussin}, {Guti{\'e}rrez}, {Hofmann},
  {Hviid}, {Ip}, {Ivanovski}, {Jorda}, {Keller}, {Knight}, {Knollenberg},
  {Koschny}, {Kramm}, {K{\"u}hrt}, {K{\"u}ppers}, {Lamy}, {Lara}, {Lazzarin},
  {L{\`o}pez-Moreno}, {Manfroid}, {Epifani}, {Marzari}, {Naletto}, {Oklay},
  {Palumbo}, {Parker}, {Rickman}, {Rodrigo}, {Rodr{\`i}guez}, {Schindhelm},
  {Shi}, {Sordini}, {Steffl}, {Stern}, {Thomas}, {Tubiana}, {Weaver},
  {Weissman}, {Zakharov}, \& {Taylor}}]{Grun2016}
{Gr{\"u}n}, E., {Agarwal}, J., {Altobelli}, N., {et~al.} 2016, \mnras, 462,
  S220

\bibitem[{Gr{\"u}n {et~al.}(2016)Gr{\"u}n, Agarwal, Altobelli, Altwegg,
  Bentley, Biver, Della~Corte, Edberg, Feldman, Galand,
  {et~al.}}]{grun20162016}
Gr{\"u}n, E., Agarwal, J., Altobelli, N., {et~al.} 2016, \mnras, 462, S220

\bibitem[{{Hajra} {et~al.}(2017){Hajra}, {Henri}, {Valli{\`e}res}, {Galand},
  {H{\'e}ritier}, {Eriksson}, {Odelstad}, {Edberg}, {Burch}, {Broiles},
  {Goldstein}, {Glassmeier}, {Richter}, {Goetz}, {Tsurutani}, {Nilsson},
  {Altwegg}, \& {Rubin}}]{hajra2017}
{Hajra}, R., {Henri}, P., {Valli{\`e}res}, X., {et~al.} 2017, \aap, 607, A34

\bibitem[{Hall {et~al.}(1998)Hall, Feldman, McGrath, \& Strobel}]{hall1998far}
Hall, D., Feldman, P., McGrath, M.~A., \& Strobel, D. 1998, \apj, 499, 475

\bibitem[{Hendrix {et~al.}(2010)Hendrix, Hansen, \&
  Holsclaw}]{hendrix2010ultraviolet}
Hendrix, A.~R., Hansen, C.~J., \& Holsclaw, G.~M. 2010, Icarus, 206, 608

\bibitem[{{Jewitt} \& {Meech}(1986)}]{JewittMeech1986}
{Jewitt}, D., \& {Meech}, K.~J. 1986, \apj, 310, 937

\bibitem[{{Jorda} {et~al.}(2016){Jorda}, {Gaskell}, {Capanna}, {Hviid}, {Lamy},
  {{\v D}urech}, {Faury}, {Groussin}, {Guti{\'e}rrez}, {Jackman}, {Keihm},
  {Keller}, {Knollenberg}, {K{\"u}hrt}, {Marchi}, {Mottola}, {Palmer},
  {Schloerb}, {Sierks}, {Vincent}, {A'Hearn}, {Barbieri}, {Rodrigo}, {Koschny},
  {Rickman}, {Barucci}, {Bertaux}, {Bertini}, {Cremonese}, {Da Deppo},
  {Davidsson}, {Debei}, {De Cecco}, {Fornasier}, {Fulle}, {G{\"u}ttler}, {Ip},
  {Kramm}, {K{\"u}ppers}, {Lara}, {Lazzarin}, {Lopez Moreno}, {Marzari},
  {Naletto}, {Oklay}, {Thomas}, {Tubiana}, \& {Wenzel}}]{Jorda2016}
{Jorda}, L., {Gaskell}, R., {Capanna}, C., {et~al.} 2016, \icarus, 277, 257

\bibitem[{Kanik {et~al.}(2003)Kanik, Noren, Makarov, Vattipalle, Ajello, \&
  Shemansky}]{kanik2003electron}
Kanik, I., Noren, C., Makarov, O., {et~al.} 2003, \jgr: Planets, 108

\bibitem[{Keller {et~al.}(2007)Keller, Barbieri, Lamy, Rickman, Rodrigo,
  Wenzel, Sierks, A’Hearn, Angrilli, Angulo, {et~al.}}]{keller2007osiris}
Keller, H.~U., Barbieri, C., Lamy, P., {et~al.} 2007, Space science reviews,
  128, 433

\bibitem[{{Knight} {et~al.}(2017){Knight}, {Snodgrass}, {Vincent}, {Conn},
  {Skiff}, {Schleicher}, \& {Lister}}]{knight2017gemini}
{Knight}, M.~M., {Snodgrass}, C., {Vincent}, J.-B., {et~al.} 2017, \mnras, 469,
  S661

\bibitem[{{Knollenberg} {et~al.}(2016){Knollenberg}, {Lin}, {Hviid}, {Oklay},
  {Vincent}, {Bodewits}, {Mottola}, {Pajola}, {Sierks}, {Barbieri}, {Lamy},
  {Rodrigo}, {Koschny}, {Rickman}, {A'Hearn}, {Barucci}, {Bertaux}, {Bertini},
  {Cremonese}, {Davidsson}, {Da Deppo}, {Debei}, {De Cecco}, {Fornasier},
  {Fulle}, {Groussin}, {Guti{\'e}rrez}, {Ip}, {Jorda}, {Keller}, {K{\"u}hrt},
  {Kramm}, {K{\"u}ppers}, {Lara}, {Lazzarin}, {Lopez Moreno}, {Marzari},
  {Naletto}, {Thomas}, {G{\"u}ttler}, {Preusker}, {Scholten}, \&
  {Tubiana}}]{Knollenberg2016}
{Knollenberg}, J., {Lin}, Z.~Y., {Hviid}, S.~F., {et~al.} 2016, \aap, 596, A89

\bibitem[{{Kurucz}(1994)}]{Kurucz1994}
{Kurucz}, R.~L. 1994, in IAU Symposium, Vol. 154, Infrared Solar Physics, ed.
  D.~M. {Rabin}, J.~T. {Jefferies}, \& C.~{Lindsey}, 523

\bibitem[{Lin {et~al.}(2017)Lin, Knollenberg, Vincent, A’hearn, Ip, Sierks,
  Barbieri, Lamy, Rodrigo, Koschny, {et~al.}}]{lin2017investigating}
Lin, Z.-Y., Knollenberg, J., Vincent, J.-B., {et~al.} 2017, Monthly Notices of
  the Royal Astronomical Society, 469, S731

\bibitem[{Makarov {et~al.}(2004)Makarov, Ajello, Vattipalle, Kanik, Festou, \&
  Bhardwaj}]{makarov2004kinetic}
Makarov, O.~P., Ajello, J.~M., Vattipalle, P., {et~al.} 2004, \jgr: Space
  Physics, 109

\bibitem[{Mall {et~al.}(2016)Mall, Altwegg, Balsiger, Bar-Nun, Berthelier,
  Bieler, Bochsler, Briois, Calmonte, Combi, {et~al.}}]{mall2016high}
Mall, U., Altwegg, K., Balsiger, H., {et~al.} 2016, \apj, 819, 126

\bibitem[{{Miles} {et~al.}(2016){Miles}, {Faillace}, {Mottola}, {Raab},
  {Roche}, {Soulier}, \& {Watkins}}]{Miles2016}
{Miles}, R., {Faillace}, G.~A., {Mottola}, S., {et~al.} 2016, \icarus, 272, 327

\bibitem[{Mumma {et~al.}(1972)Mumma, Stone, Borst, \&
  Zipf}]{mumma1972dissociative}
Mumma, M., Stone, E., Borst, W., \& Zipf, E. 1972, \jcp, 57, 68

\bibitem[{Noonan {et~al.}(submitted)Noonan, {Bockel{\'e}e-Morvan}, Feldman,
  {et~al.}}]{Noonan2021spatial}
Noonan, J.~W., {Bockel{\'e}e-Morvan}, D., Feldman, P.~F., {et~al.} submitted,
  \aj

\bibitem[{Noonan {et~al.}(2018)Noonan, Stern, Feldman, Broiles, Wedlund,
  Edberg, Schindhelm, Parker, Keeney, Vervack~Jr,
  {et~al.}}]{noonan2018ultraviolet}
Noonan, J.~W., Stern, S.~A., Feldman, P.~D., {et~al.} 2018, \aj, 156, 16

\bibitem[{Pajola {et~al.}(2017)Pajola, H{\"o}fner, Vincent, Oklay, Scholten,
  Preusker, Mottola, Naletto, Fornasier, Lowry, {et~al.}}]{pajola2017pristine}
Pajola, M., H{\"o}fner, S., Vincent, J.-B., {et~al.} 2017, Nature Astronomy, 1,
  1

\bibitem[{Pineau {et~al.}(2018)Pineau, Parker, Steffl, Schindhelm, Medina,
  Stern, Birath, \& Versteeg}]{pineau2018flight}
Pineau, J.~P., Parker, J.~W., Steffl, A.~J., {et~al.} 2018, in 2018 SpaceOps
  Conference, 2589

\bibitem[{{Preusker} {et~al.}(2017){Preusker}, {Scholten, F.}, {Matz, K.-D.},
  {Roatsch, T.}, {Hviid, S. F.}, {Mottola, S.}, {Knollenberg, J.}, {K\"uhrt,
  E.}, {Pajola, M.}, {Oklay, N.}, {Vincent, J.-B.}, {Davidsson, B.},
  {A\'{}Hearn, M. F.}, {Agarwal, J.}, {Barbieri, C.}, {Barucci, M. A.},
  {Bertaux, J.-L.}, {Bertini, I.}, {Cremonese, G.}, {Da Deppo, V.}, {Debei,
  S.}, {De Cecco, M.}, {Fornasier, S.}, {Fulle, M.}, {Groussin, O.},
  {Guti\'errez, P. J.}, {G\"uttler, C.}, {Ip, W-H.}, {Jorda, L.}, {Keller, H.
  U.}, {Koschny, D.}, {Kramm, J. R.}, {K\"uppers, M.}, {Lamy, P.}, {Lara, L.
  M.}, {Lazzarin, M.}, {Lopez Moreno, J. J.}, {Marzari, F.}, {Massironi, M.},
  {Naletto, G.}, {Rickman, H.}, {Rodrigo, R.}, {Sierks, H.}, {Thomas, N.}, \&
  {Tubiana, C.}}]{Preusker2017}
{Preusker}, F., {Scholten, F.}, {Matz, K.-D.}, {et~al.} 2017, A\&A, 607, L1.
\newblock \url{https://doi.org/10.1051/0004-6361/201731798}

\bibitem[{{Rinaldi} {et~al.}(2018){Rinaldi}, {Bockel{\'e}e-Morvan},
  {Ciarniello}, {Tozzi}, {Capaccioni}, {Ivanovski}, {Filacchione}, {Fink},
  {Doose}, {Taylor}, {Kappel}, {Erard}, {Leyrat}, {Raponi}, {D'Aversa},
  {Capria}, {Longobardo}, {Palomba}, {Tosi}, {Migliorini}, {Rotundi}, {Della
  Corte}, \& {Salatti}}]{Rinaldi2018}
{Rinaldi}, G., {Bockel{\'e}e-Morvan}, D., {Ciarniello}, M., {et~al.} 2018,
  \mnras, 481, 1235

\bibitem[{Skorov {et~al.}(2016)Skorov, Rezac, Hartogh, Bazilevsky, \&
  Keller}]{skorov2016model}
Skorov, Y.~V., Rezac, L., Hartogh, P., Bazilevsky, A., \& Keller, H. 2016,
  \aap, 593, A76

\bibitem[{Steckloff {et~al.}(2016)Steckloff, Graves, Hirabayashi, Melosh, \&
  Richardson}]{steckloff2016rotationally}
Steckloff, J.~K., Graves, K., Hirabayashi, M., Melosh, H.~J., \& Richardson,
  J.~E. 2016, Icarus, 272, 60

\bibitem[{{Steffl} {et~al.}(2015){Steffl}, {Feldman}, {A'Hearn}, {Bertaux},
  {Feaga}, {Keeney}, {Knight}, {Noonan}, {Parker}, {Schindhelm}, {Stern},
  {Vervack}, \& {Weaver}}]{Steffl2015}
{Steffl}, A.~J., {Feldman}, P.~D., {A'Hearn}, M.~F., {et~al.} 2015, in
  AAS/Division for Planetary Sciences Meeting Abstracts \#47, AAS/Division for
  Planetary Sciences Meeting Abstracts, 413.08

\bibitem[{{Steffl} {et~al.}(2018){Steffl}, {A'Hearn}, {Bertaux}, {Feaga},
  {Feldman}, {Keeney}, {Noonan}, {Parker}, {Stern}, \& {Weaver}}]{Steffl2018}
{Steffl}, A.~J., {A'Hearn}, M.~F., {Bertaux}, J.-L., {et~al.} 2018, in
  AAS/Division for Planetary Sciences Meeting Abstracts \#50, AAS/Division for
  Planetary Sciences Meeting Abstracts, 110.03

\bibitem[{Stephenson {et~al.}(2021)Stephenson, Galand, Feldman, Beth, Rubin,
  Bockel{\'e}e-Morvan, Biver, Cheng, Parker, Burch,
  {et~al.}}]{stephenson2021multi}
Stephenson, P., Galand, M., Feldman, P., {et~al.} 2021, Astronomy \&
  Astrophysics, 647, A119

\bibitem[{Stern {et~al.}(2007)Stern, Slater, Scherrer, Stone, Versteeg,
  A\textsc{\char13}Hearn, Bertaux, Feldman, Festou, Parker,
  {et~al.}}]{stern2007alice}
Stern, S.~A., Slater, D., Scherrer, J., {et~al.} 2007, \ssr, 128, 507

\bibitem[{{Thomas} {et~al.}(2018){Thomas}, {El Maarry}, {Theologou},
  {Preusker}, {Scholten}, {Jorda}, {Hviid}, {Marschall}, {K{\"u}hrt},
  {Naletto}, {Sierks}, {Lamy}, {Rodrigo}, {Koschny}, {Davidsson}, {Barucci},
  {Bertaux}, {Bertini}, {Bodewits}, {Cremonese}, {Da Deppo}, {Debei}, {De
  Cecco}, {Fornasier}, {Fulle}, {Groussin}, {Guti{\`e}rrez}, {G{\"u}ttler},
  {Ip}, {Keller}, {Knollenberg}, {Lara}, {Lazzarin}, {L{\`o}pez-Moreno},
  {Marzari}, {Tubiana}, \& {Vincent}}]{Thomas2018}
{Thomas}, N., {El Maarry}, M.~R., {Theologou}, P., {et~al.} 2018, \planss, 164,
  19

\bibitem[{Vincent {et~al.}(2016)Vincent, A\textsc{\char13}Hearn, Lin,
  El-Maarry, Pajola, Sierks, Barbieri, Lamy, Rodrigo, Koschny,
  {et~al.}}]{vincent2016summer}
Vincent, J.-B., A\textsc{\char13}Hearn, M.~F., Lin, Z.-Y., {et~al.} 2016,
  \mnras, 462, S184

\bibitem[{{Vincent} {et~al.}(2016){Vincent}, {A'Hearn}, {Lin}, {El-Maarry},
  {Pajola}, {Sierks}, {Barbieri}, {Lamy}, {Rodrigo}, {Koschny}, {Rickman},
  {Keller}, {Agarwal}, {Barucci}, {Bertaux}, {Bertini}, {Besse}, {Bodewits},
  {Cremonese}, {Da Deppo}, {Davidsson}, {Debei}, {De Cecco}, {Deller},
  {Fornasier}, {Fulle}, {Gicquel}, {Groussin}, {Guti{\'e}rrez},
  {Guti{\'e}rrez-Marquez}, {G{\"u}ttler}, {H{\"o}fner}, {Hofmann}, {Hviid},
  {Ip}, {Jorda}, {Knollenberg}, {Kovacs}, {Kramm}, {K{\"u}hrt}, {K{\"u}ppers},
  {Lara}, {Lazzarin}, {Lopez Moreno}, {Marzari}, {Massironi}, {Mottola},
  {Naletto}, {Oklay}, {Preusker}, {Scholten}, {Shi}, {Thomas}, {Toth}, \&
  {Tubiana}}]{Vincent2016}
{Vincent}, J.-B., {A'Hearn}, M.~F., {Lin}, Z.-Y., {et~al.} 2016, \mnras, 462,
  S184

\bibitem[{Wu {et~al.}(1978)Wu, Phillips, Lee, \& Judge}]{wu1978atomic}
Wu, R.~C., Phillips, E., Lee, L., \& Judge, D. 1978, \jgr: Space Physics, 83,
  4869

\end{thebibliography}

\end{document}